\documentclass{aa}  
\usepackage{graphicx,subfigure}
\usepackage[tableleft]{rotating}
\usepackage{longtable,lscape}
\usepackage{amsmath,amssymb}
\usepackage{txfonts}
\usepackage{xspace}
\usepackage{natbib}
\usepackage{longtable}
\def\um{$\mu$m\xspace}
\def\msun{M$_{\odot}$\xspace}
\def\rsun{R$_{\odot}$\xspace}
\def\lsun{L$_{\odot}$\xspace}
\def\rstar{$R_{\star}$\xspace}
\def\tstar{$T_{\star}$\xspace}
\def\mstar{$M_{\star}$\xspace}
\def\teff{$T_{\mathrm{eff}}$\xspace}
\def\lstar{$L_{\star}$\xspace}

\def\rinner{$R_\mathrm{inner}$\xspace}
\def\router{$R_\mathrm{outer}$\xspace}
\def\mdot{$\dot{M}$\xspace}
\def\mdotinput{$\dot{M}_{\mathrm{input}}$\xspace}
\def\mdotd{$\dot{M}_{\mathrm{D}}$\xspace}
\def\mdotr{$\dot{M}_{\mathrm{R}}$\xspace}

\def\vinfty{$v_{\infty}$\xspace}

\def\ipp3{$I(p)\times p^3$\xspace}
\def\mbol{$M_{\mathrm{bol}}$\xspace}
\def\mbolsun{$M_{\mathrm{bol},\odot}$\xspace}
\def\VmK{\ensuremath{V{\rm -}K}\xspace}
\def\h2o{H$_2$O\xspace}

\def\twco{\element[ ][12]{CO}\xspace}
\def\thco{\element[ ][13]{CO}\xspace}
\def\twc{\element[ ][12]{C}\xspace}
\def\thc{\element[ ][13]{C}\xspace}
\def\twthratio{\twc/\thc}
\def\orich{oxygen-rich\xspace}
\def\crich{carbon-rich\xspace}

\def\fhco{$f_{\mathrm{H,CO}}$\xspace}

\def\ddw{dust-driven wind\xspace}

\def\gastronoom{{\sc GASTRoNOoM}\xspace}

\def\hifi{HIFI \xspace}
\def\pacs{PACS \xspace}

\def\ohir{OH/IR\xspace}
\def\Msun{\hbox{$M_{\odot}$}}               
\def\Lsun{\hbox{$L_{\odot}$}}               
\def\Rstar{\hbox{$R_{\star}$}}              
\def\Tstar{\hbox{$T_{\star}$}}              
\def\Lstar{\hbox{$L_{\star}$}}              
\def\Rinner{\hbox{$R_{\rm inner}$}}              
\def\Mdot{\hbox{$\dot{M}$}}               
\def\vinf{\hbox{$v_\infty$}}               
\def\arcsec{\hbox{$^{\prime\prime}$}\xspace}
\def\isotope{\hbox{$^{12}$C/$^{13}$C}}   
\def\Vinfty{\hbox{$v_{\infty}$}}

\begin{document}
   \title{Probing the mass-loss history of AGB and red supergiant stars from CO rotational line profiles}
    \subtitle{II. CO line survey of evolved stars: derivation of mass-loss rate formulae}
	\titlerunning{CO line survey on evolved stars and \mdot-formulae}
	
   \author{E. De Beck \inst{1}
	\and L. Decin\inst{1,2} 
	\and A. de Koter\inst{2,3}
	\and K. Justtanont \inst{4}
	\and T. Verhoelst \inst{1}
	\and F. Kemper \inst{5}
	\and K. M. Menten \inst{6} 
          }
	\authorrunning{E. De Beck et al.}
   \offprints{E. De Beck}

   \institute{
	Department of Physics and Astronomy, Institute for  Astronomy, K.U.Leuven, Celestijnenlaan 200D, B-3001 Leuven, Belgium\\
        \email{Elvire.DeBeck@ster.kuleuven.be}
	\and Astronomical Institute ``Anton Pannekoek'', University of Amsterdam, Science Park XH, Amsterdam, The Netherlands
	\and Astronomical Institute Utrecht, University of Utrecht, PO Box 8000, NL-3508 TA Utrecht, The Netherlands
	\and Chalmers University of Technology, Onsala Space Observatory, SE-439 92 Onsala, Sweden
	\and Jodrell Bank Centre for Astrophysics, School of Physics and Astronomy, University of Manchester, Manchester, M13 9PL, UK
	\and Max-Planck-Institut f\"ur Radioastronomie, Auf dem H\"ugel 69, D-53121 Bonn, Germany
}

\date{Received ----------; accepted ----------}

\abstract
{The evolution of intermediate and low-mass stars on the asymptotic giant branch is dominated by their strong dust-driven winds. More massive stars evolve into red supergiants with a similar envelope structure and strong wind. These stellar winds are a prime source for the chemical enrichment of the interstellar medium.
}
{We aim to \textit{(1)} set up simple and general analytical expressions to estimate mass-loss rates of evolved stars, and \textit{(2)} from those calculate estimates for the mass-loss rates of the asymptotic giant branch, red supergiant, and yellow hypergiant stars in our galactic sample.
}
{The rotationally excited lines of carbon monoxide (CO) are a classic and very robust diagnostic in the study of circumstellar envelopes. When sampling different layers of the circumstellar envelope, observations of these molecular lines lead to detailed profiles of kinetic temperature, expansion velocity, and density. A state-of-the-art, nonlocal thermal equilibrium, and co-moving frame radiative transfer code that predicts CO line intensities in the circumstellar envelopes of late-type stars is used in deriving relations between stellar and molecular-line parameters, on the one hand, and mass-loss rate, on the other. These expressions are applied to our extensive CO data set to estimate the mass-loss rates of 47 sample stars. 
}
{We present analytical expressions for estimating the mass-loss rates of evolved stellar objects for 8 rotational transitions of the CO molecule and thencompare our results to those of previous studies. Our expressions account for line saturation and resolving of the envelope, thereby allowing accurate determination of very high mass-loss rates. We argue that, for estimates based on a single rotational line, the CO(2--1) transition provides the most reliable mass-loss rate. The mass-loss rates calculated for the asympotic giant branch stars range from $4\times10^{-8}$\,\msun\,yr$^{-1}$ up to $8\times10^{-5}$\,\msun\,yr$^{-1}$. For red supergiants they reach values between $2\times10^{-7}$\,\msun\,yr$^{-1}$ and $3\times10^{-4}$\,\msun\,yr$^{-1}$. The estimates for the set of CO transitions allow time variability to be identified in the mass-loss rate. Possible mass-loss-rate variability is traced for 7 of the sample stars. We find a clear relation between the pulsation periods of the asympotic giant branch stars and their derived mass-loss rates, with a levelling off at $\sim$$3\times10^{-5}$\,\msun\,yr$^{-1}$ for periods exceeding 850\,days. 
}
{
}
\maketitle


\section{Introduction}\label{sect:introduction}
Stars of low and intermediate masses (1\,\msun$\leq\;$\mstar$\leq\;$9\,\msun) expel a large part of their outer layers in a \ddw during the later stages of their evolution, when ascending the asymptotic giant branch (AGB). The expulsion of stellar material creates a cool circumstellar envelope (CSE) containing dust grains and molecular gas phase species. The rates at which mass is lost (\mdot) vary from $\sim$$10^{-8}$\,\msun\,yr$^{-1}$ up to $\sim$$10^{-4}$\,\msun\,yr$^{-1}$. A violent ejection of the circumstellar envelope ends the AGB phase and initiates further evolution into a protoplanetary nebula (PPN) and later into a planetary nebula (PN), when the ejected material is ionised \citep{habing2003}.

High-mass stars (\mstar$>\;$9\,\msun) do not ascend the AGB, but evolve into supergiants and hypergiants. Similar to AGB stars, they are typified by circumstellar envelopes of dusty and molecular material created by strong winds. The mechanisms behind the dense outflows of both types of objects are likely to be different  \citep{josselin2007}, since the central stars have obtained other physical and chemical properties after going through somewhat different phases of evolution.

We want to investigate the physical and chemical properties of these CSEs by analysing molecular-line observations. The most important species for such investigations is carbon monoxide, CO. More specifically, observations of lines resulting from transitions from states with angular momentum quantum number $J$ to $J-1$ can be used to derive density, velocity and kinetic temperature in the envelope. Previous surveys have focussed on the analysis of only a few low-excitation rotational transitions of CO ($J=1-0,\;2-1,\;3-2$), or have used simple analytical expressions to derive the mass-loss rate from the parameters of these transitions \citep{knapp1982,knapp1985,olofsson1993,loup1993,neri1998,ramstedt2008}. Since we have access to an extensive data set covering both low and high-$J$  molecular lines, we want to provide mass-loss estimators based on multiple CO lines, including those of high excitation levels. These analytical expressions will link stellar and CO-line parameters to the (variable) mass-loss rates of the central stars. 

The presence of both \twco and \thco line transitions in the data set allows us to estimate the \twthratio isotope ratio for many stars in the sample. This abundance ratio provides a measure for the chemical evolution and hence of the evolutionary state of the stars. In this paper we give first order estimates for \twthratio, derived from line intensity ratios.
\\

The sample of evolved stars and the CO-line observations are presented in Sect.~\ref{sect:data} and in the (online) appendix. We discuss some of the sample stars individually based on the observations. Sect.~\ref{sect:radtrananalysis} deals with the radiative transfer analysis and the parameter study that will lead to the construction of analytical expressions to estimate mass-loss rates. The constructed formalism is then compared to those presented in the literature. In Sect.~\ref{sect:stellarparameters} and Appendix~\ref{sect:basicstellarparameters}, we discuss the determination of the basic stellar parameters needed to derive mass-loss rates. The results of our study in terms of \mdot, \twthratio and wind driving efficiency are presented in Sect.~\ref{sect:resultsanddiscussion}. Conclusions are given in Sect.~\ref{sect:conclusions}.


\section{Observations}\label{sect:data}
\subsection{The sample}\label{subsect:sampledescription}
The data for our sample of 69 galactic objects have been assembled over many years and mainly consists of targets originally selected as potential candidates to be included in guaranteed time key programs (GTKP) of the Herschel-\hifi and Herschel-\pacs instruments. Different evolutionary  and chemical types are covered, but there is a rather strong bias towards oxygen-rich (C/O$<1$) AGB stars. Several categories of AGB stars, differing in pulsational and mass-loss properties, are represented in the sample, i.e. Mira-variables, semi-regular (SR) variables and \ohir-type stars. SRs and Miras have low to intermediate mass-loss rates ($10^{-8} - 10^{-5}$\,\msun\,yr$^{-1}$), while \ohir stars have very dusty and optically opaque envelopes, formed by intense mass loss that can reach up to a few $10^{-4}$\,\msun\,yr$^{-1}$. Typical for these objects are the strong OH masering at 1612\,MHz and a very large infrared excess. Miras have very regular pulsations, with a fixed period of a few 100~days (average $\sim$400~days) and large amplitudes ($\geq$2.5\,mag in V-band). Semi-regulars, on the other hand, have smaller amplitudes and can exhibit irregularities in their pulsations. Their pulsation periods can range from 20 up to 2000~days \citep[GCVS][]{gcvs}. \ohir stars have pulsation periods ranging from a few 100~days up to more than 1000~days with an average of $\sim$1000~days and they could be considered the evolutionary successors of Miras \citep{vassiliadis1993}. Their thick envelopes, obscuring the central star in the optical, are produced by the so-called \textit{superwind phases} \citep{iben1983,vassiliadis1993}, which occur during the late parts of the quiescent hydrogen burning phase, just before the central star goes through a thermal pulse (TP). During these phases, the pulsation period increases with increasing mass loss (up to a few $10^{-5}$\,\msun\,yr$^{-1}$) and luminosity. The enhanced mass loss leads to a very optically thick and dusty CSE, characterising an \ohir star. When a superwind phase is halted, the pulsation period decreases again, the expelled material diffuses outwards, and a less intense mass-loss process is initiated \citep{vassiliadis1993}. The combination of these factors can lead to the central star being again visible in the optical, and being again observed and classified as a Mira or SR variable. This scenario is more plausible for the less massive stars (\mstar$\leq 2.5$\,\msun), as the higher-mass stars (\mstar$>2.5$\,\msun) experience relatively modest variations in their high mass-loss rates, luminosities, and pulsation periods due to the TPs \citep[see Figs. 3 to 9 in][]{vassiliadis1993}. Low-mass stars experience only very short periods of enhanced mass loss. Moreover, during these periods the absolute value of the mass-loss rate is significantly lower than for their massive counterparts. Therefore, the chance to observe them in their superwind phases, i.e. as \ohir, is far smaller than for the higher-mass stars.

Supergiants and hypergiants and some post-AGB objects are also included in the sample. The latter fit in this study as the progeny of AGB stars. Studying their extended envelopes will lead to an improved knowledge of the envelopes of AGB-type stars and the late stages of the AGB evolution. Two young stellar objects (YSOs) --- AFGL\,5502 and the Gomez Nebula --- were observed together with the sample of evolved stars. The obtained data on these YSOs  were not yet published and are presented in this paper. We will not further discuss these objects, since the focus of this paper is on evolved stars.

\subsection{The observations}\label{subsect:datadescriptionandreduction}
The data set presented in this paper consists of observations of multiple rotationally excited lines of both \twco and \thco in 69 stars. The bulk of the presented lines were first published by \cite{kemper2003}. Transitions from $J=1-0$ up to $J=7-6$ for \twco and from $J=2-1$ up to $J=6-5$ for \thco are covered. Since the high-$J$ lines have formation regions deeper within the CSEs, this multitude of rotational lines can sample a much larger part of the CSE than only the low-excitation transitions. Also, because of slightly different molecular properties, the \thco-data probe regions somewhat different from  those sampled with the \twco-rotational lines. Moreover, since the abundance of \thco is lower than that of \twco, the circumstellar layers are optically thin(ner) for the rotational lines of the former, implying that \thco-rotational lines provide good diagnostics for the envelope's density structure. In this respect, we should be able to obtain a more complete view on the structure and properties of the envelopes and the mass-loss history of the objects, than was possible before with single-dish data only sampling low-excitation \twco-lines.

All previously unpublished data in our sample were obtained with APEX\footnote{This publication is based on data acquired with the Atacama Pathfinder Experiment (APEX). APEX is a collaboration between the Max-Planck-Institut fur Radioastronomie, the European Southern Observatory, and the Onsala Space Observatory.} (Atacama Pathfinder EXperiment) and JCMT\footnote{The James Clerk Maxwell Telescope is operated by The Joint Astronomy Centre on behalf of the Science and Technology Facilities Council of the United Kingdom, the Netherlands Organisation for Scientific Research, and the National Research Council of Canada.} (James Clerk Maxwell Telescope). Other data were obtained via private communication or retrieved from the literature. 

APEX is a 12m single-dish telescope with a frequency range from 210 up to 1500\,GHz located at Llano Chajnantor, Chile. CO-line data were obtained with three heterodyne SIS-receivers mounted on the Nasmyth-A focus: APEX-2A, FLASH-I and FLASH-II. Instrument specifics are listed in Table~\ref{tbl:instruments}, together with the CO-lines observable in the respective frequency ranges. All observations were performed in beam-switching mode.

Reduction of the APEX data was done with CLASS, part of the GILDAS-package. The process mainly consisted of combining different scans of the same molecular line towards one object into a single spectrum and subsequently removing baselines and spikes in the obtained spectra. Correcting the data for telescope efficiencies  was carried out by the APEX online calibrator.

JCMT data were obtained with heterodyne receivers A, B, HARP, RxW (bands C and D), and E in beam-switching mode. Data reduction was performed with the reduction package SPECX. Spectra of standard stars taken during the observing runs were compared to the standard spectra available on the web site of JCMT. In case of large deviations in the measured line intensities of these standard stars ($>10\,\%$), the scientific-programme data obtained around the time of the measurement of these standard stars were corrected for these discrepancies. In case of multiple available standard spectra with other magnitudes of deviation, the correction factor was determined via linear interpolation in time. As for APEX data, most corrections and conversions of the data are performed online. The correction for the main-beam efficiency $\eta_{\mathrm{MB}}$ had to be done manually in the course of the reduction process using the values given in Table~\ref{tbl:instruments}. $\eta_{\mathrm{MB}}$ allows transforming antenna temperatures, $T_{\mathrm{A}}$, into main-beam-brightness temperatures, $T_{\mathrm{MB}}$. The latter is the equivalent of the brightness temperature of measurements performed with a perfect antenna outside the earth's atmosphere. The main advantage of a $T_{\mathrm{MB}}$-based temperature scale is that the data are no longer dependent of any of the instrumental properties, apart from the beam width. All data in this paper are presented on the $T_{\mathrm{MB}}$-scale.

The uncertainties on the JCMT data, so-called absolute errors, are fairly well established for all receivers \citep{kemper2003}. For APEX this is not so clear. No standard spectra were available to check the performance of the instruments at the time of the execution of the scientific programme and no standard values for the uncertainties on the data were found in the literature. \cite{ramstedt2006} mention an uncertainty of 20\,\% on the absolute intensity scale for the $J=3-2$ transition, but give no reference for this number.

\begin{table}\centering
\caption{Frequency ranges, half power beam widths $\theta_{\rm b}$, main beam efficiencies $\eta_{\mathrm{MB}}$, and observable CO rotational excitation lines for APEX and JCMT instruments.}
\label{tbl:instruments}
\setlength{\tabcolsep}{1.25mm}
 \begin{tabular}{lcccc}\hline\hline\\[-2ex]
 Instrument 	&Frequency&	$\theta_{\mathrm{b}}$	&$\eta_{\mathrm{MB}}$	& Observable\\
		&(GHz)	  &(arcsec)	&	& CO lines \\\hline\\[-2ex]
\multicolumn{5}{l}{\textit{APEX}}\\
APEX-2A		&$279 - 381$ 	&17.3	&	0.73	&\twco, \thco($3-2$)\\
FLASH-I		&$430 - 492$ 	&13.3	&	0.60	&\twco, \thco($4-3$)\\
FLASH-II	&$780 - 887$	&7.7	&	0.43	&\twco($7-6$)\\
\hline\\[-2ex]
\multicolumn{5}{l}{\textit{JCMT}}\\
A 		&  $211-279$		& 20	&0.69		& \twco, \thco($2-1$)\\
B 		&  $315-375$		& 14	&0.63		& \twco, \thco($3-2$)\\
HARP 		&  $325-375$		& 14	&0.63		& \twco, \thco($3-2$)\\
RxW(C)		&  $430-510$		& 11	&0.52		& \twco, \thco($4-3$)\\
RxW(D) 		&  $626-710$ 		& 8	&0.30		& \twco, \thco($6-5$)\\
E		&  $790-840$		& 7	&0.25		& \twco($7-6$)\\
\hline
\end{tabular}
\end{table}

\subsection{Results}\label{subsect:dataresults}
Tables~\ref{tbl:12CO} and \ref{tbl:13CO} in Appendix~\ref{sec:app_data} (available online) contain main-beam-brightness temperatures at the centre of the line profile ($T_{\mathrm{MB,c}}$), velocity integrated main-beam intensities ($I_{\mathrm{MB}}$), and expansion velocities (\vinfty) for all CO-data of the sample stars. Among the high-quality CO-data for over fifty stars, there are \thco data for 29 targets.

Some of the observed lines towards AGB or RSG stars in the sample were not detected, e.g. \twco($3-2$) for $\alpha$\,Sco (=IRAS\,16262-2619), and some others were too noisy or heavily contaminated by interstellar lines, e.g. V1360\,Aql (=IRAS\,18432-0149). In these cases, the data are presented in the appendix, but are not used in the further analysis of the sample. The respective targets are listed in Table~\ref{tbl:overview_no_estimates} together with Post-AGB objects and YSOs. 

Characterisations of the sample targets in terms of chemical type (\orich, \crich or S-type) and pulsational and/or evolutionary type (e.g. Mira, SR, \ohir) are listed in Table~\ref{tbl:fundamentalparameters}.

\begin{figure}
\begin{center}
 \includegraphics[height=\linewidth,angle=90]{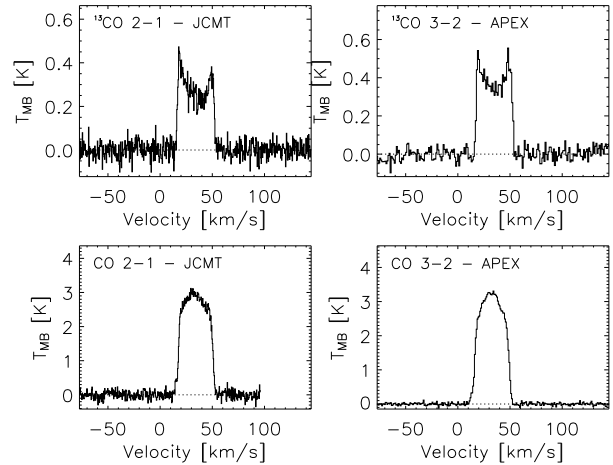}
\caption{\twco and \thco data of IK\,Tau, obtained with JCMT and APEX. It is clear from the graphs that the \twco lines are more parabolic in shape than the \thco lines. This effect is caused by the lower abundance of \thco, leading to optically thinner line profiles. \label{fig:iktau}}
\end{center}
\end{figure}

\begin{figure*}
\begin{center}
\subfigure[R\,Cas]{\includegraphics[angle=90,width=.45\linewidth]{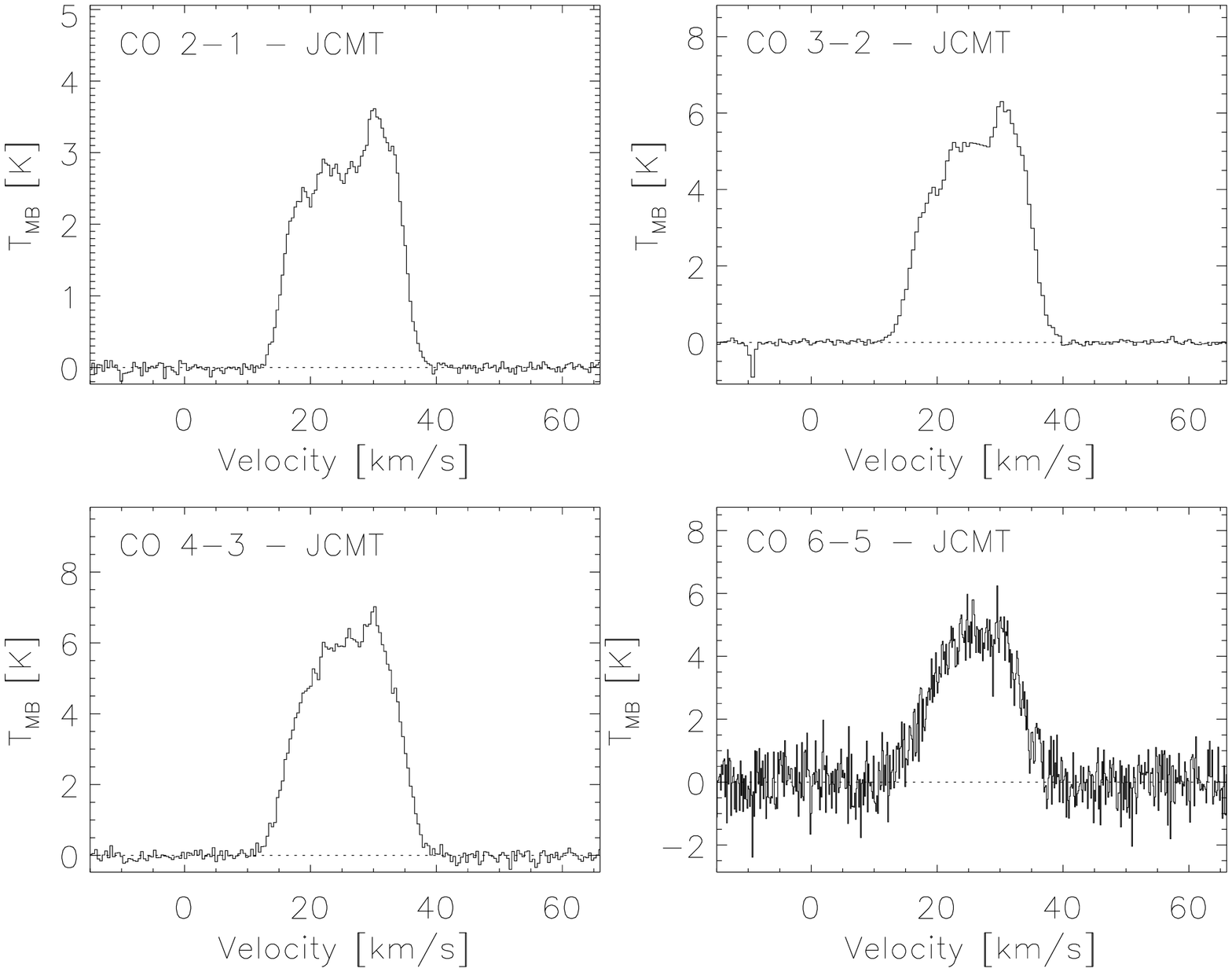}}
\subfigure[V\,Hya]{\includegraphics[angle=90,width=.45\linewidth]{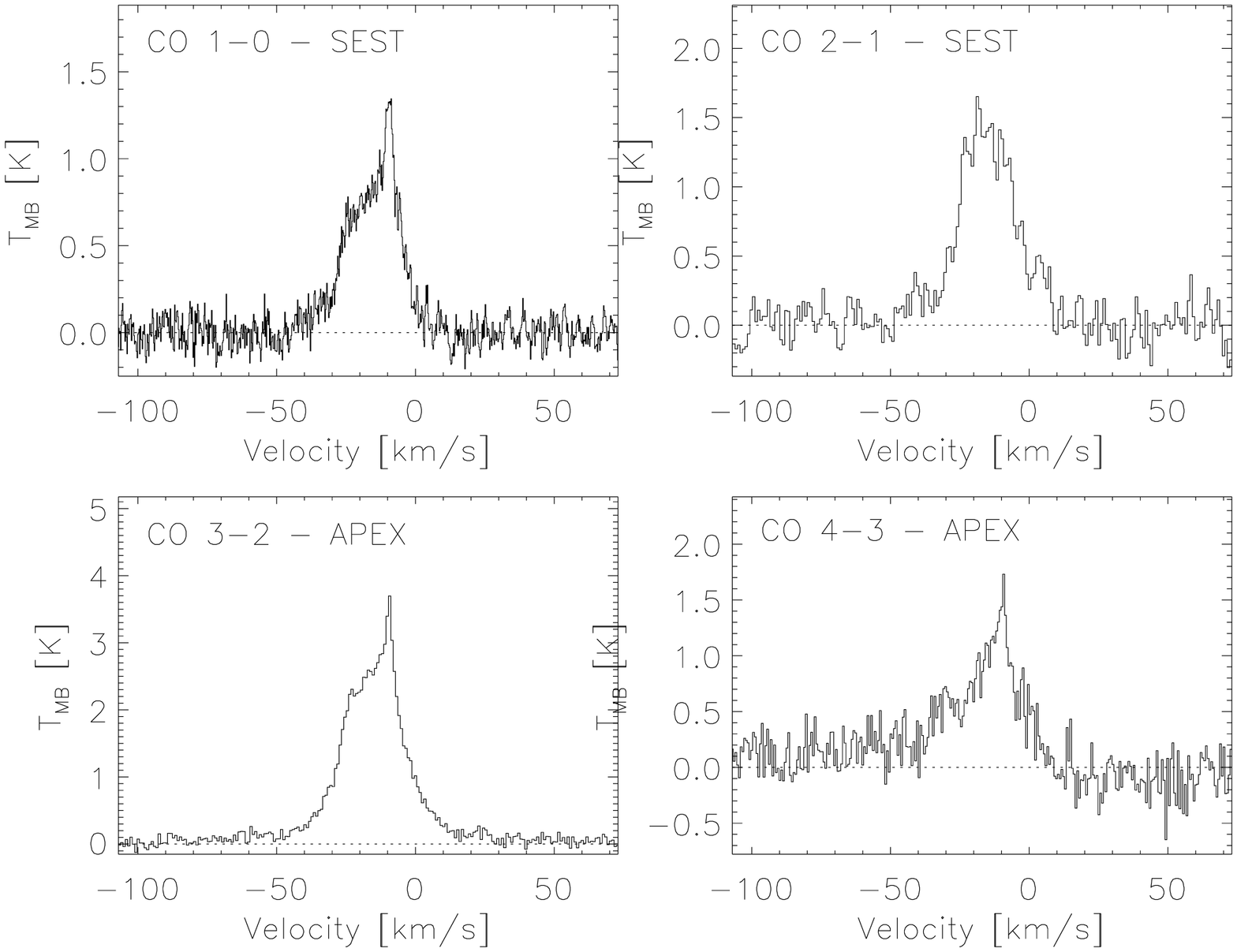}}
\subfigure[VY\,CMa]{\includegraphics[angle=90,width=.45\linewidth]{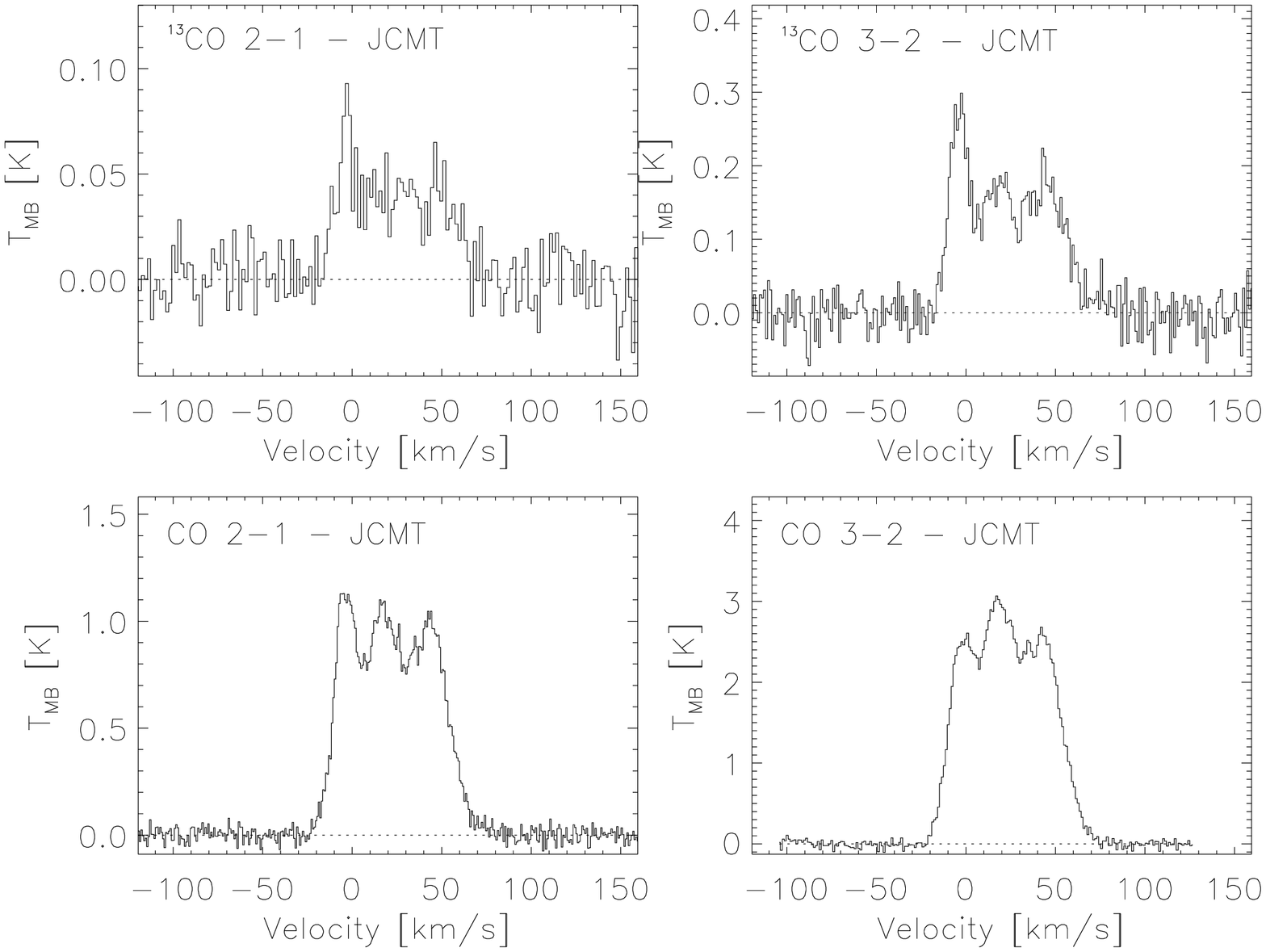}}
\subfigure[U\,Ant]{\includegraphics[angle=90,width=.45\linewidth]{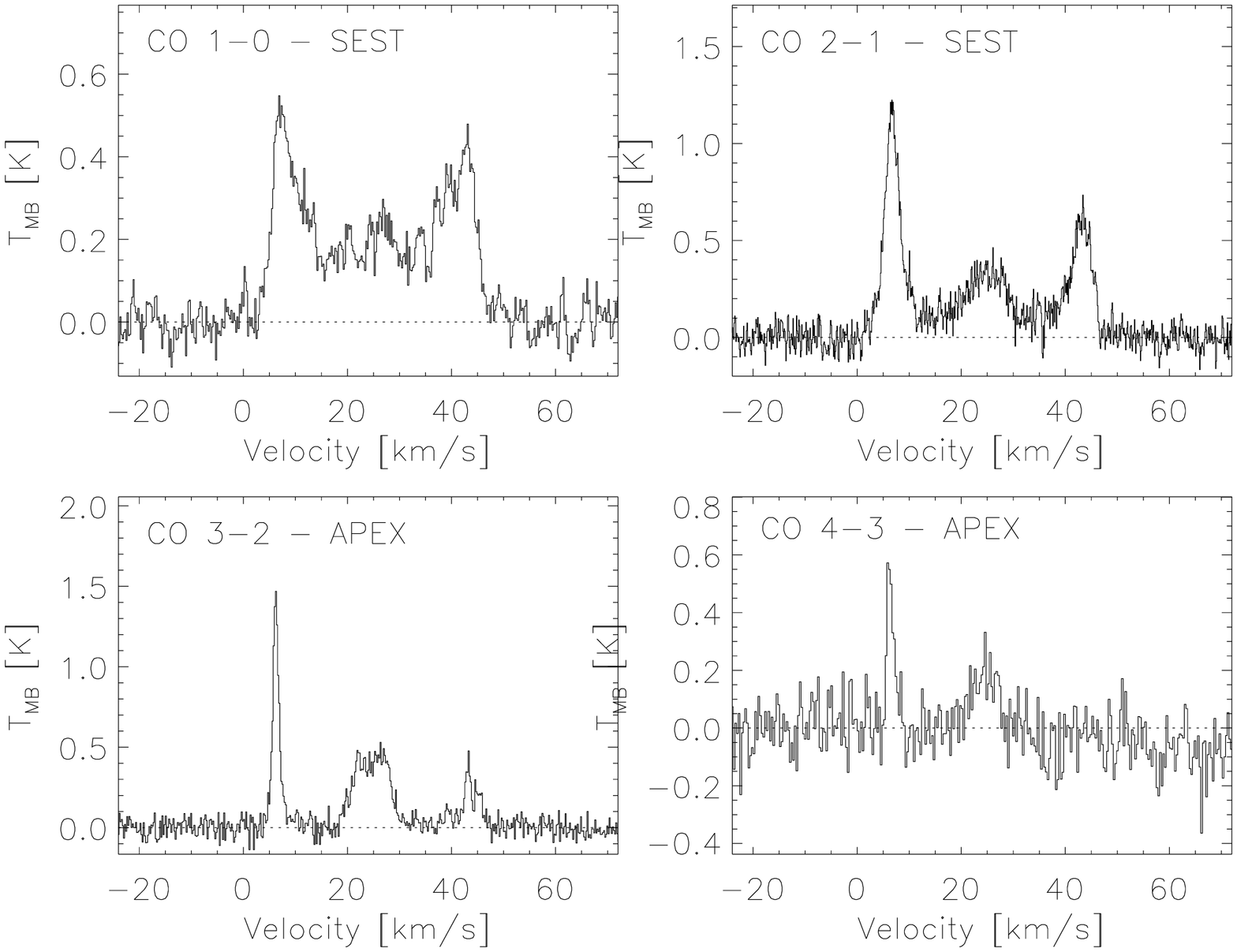}}
\subfigure[S\,Sct]{\includegraphics[angle=90,width=.45\linewidth]{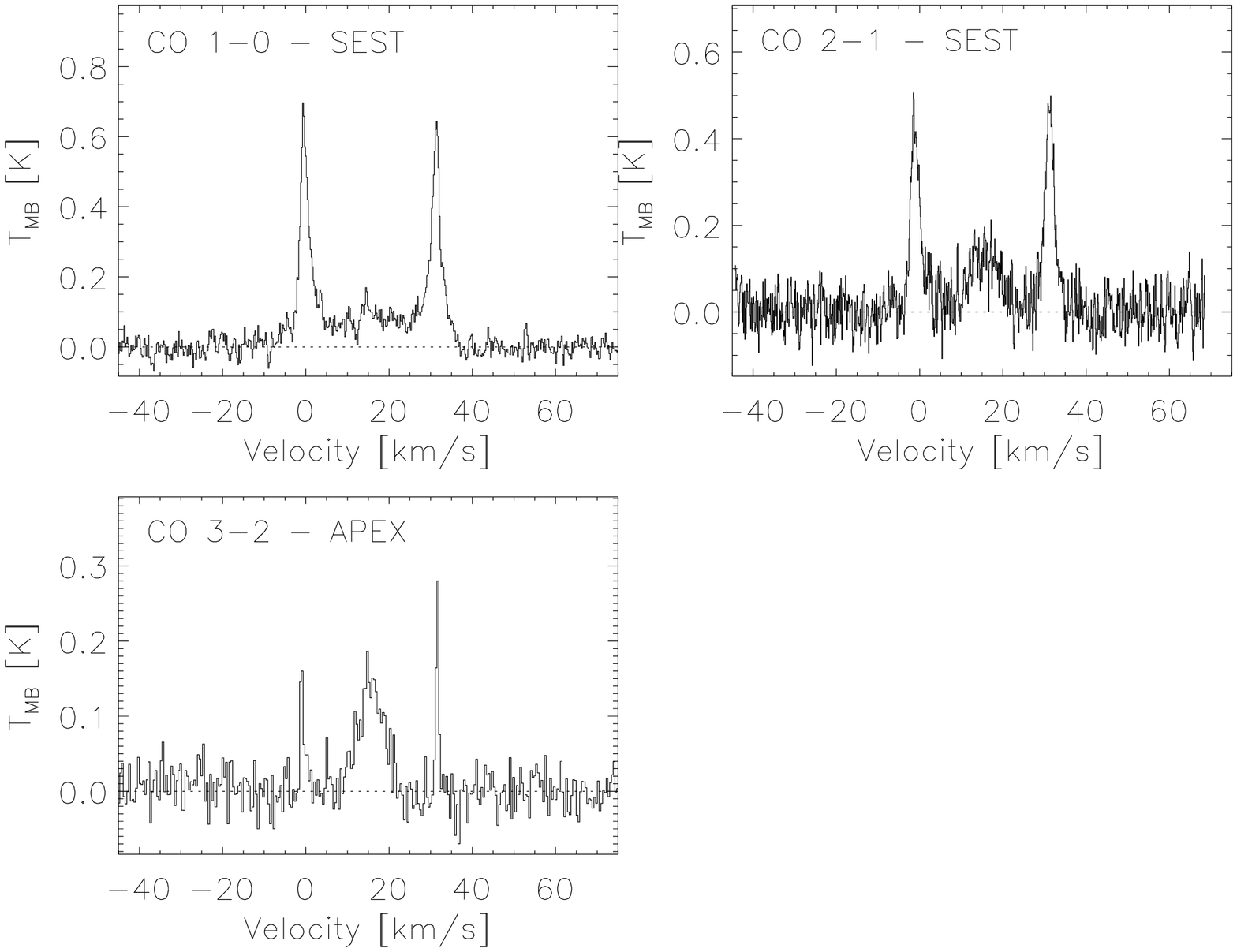}}
\subfigure[R\,Scl]{\includegraphics[angle=90,width=.45\linewidth]{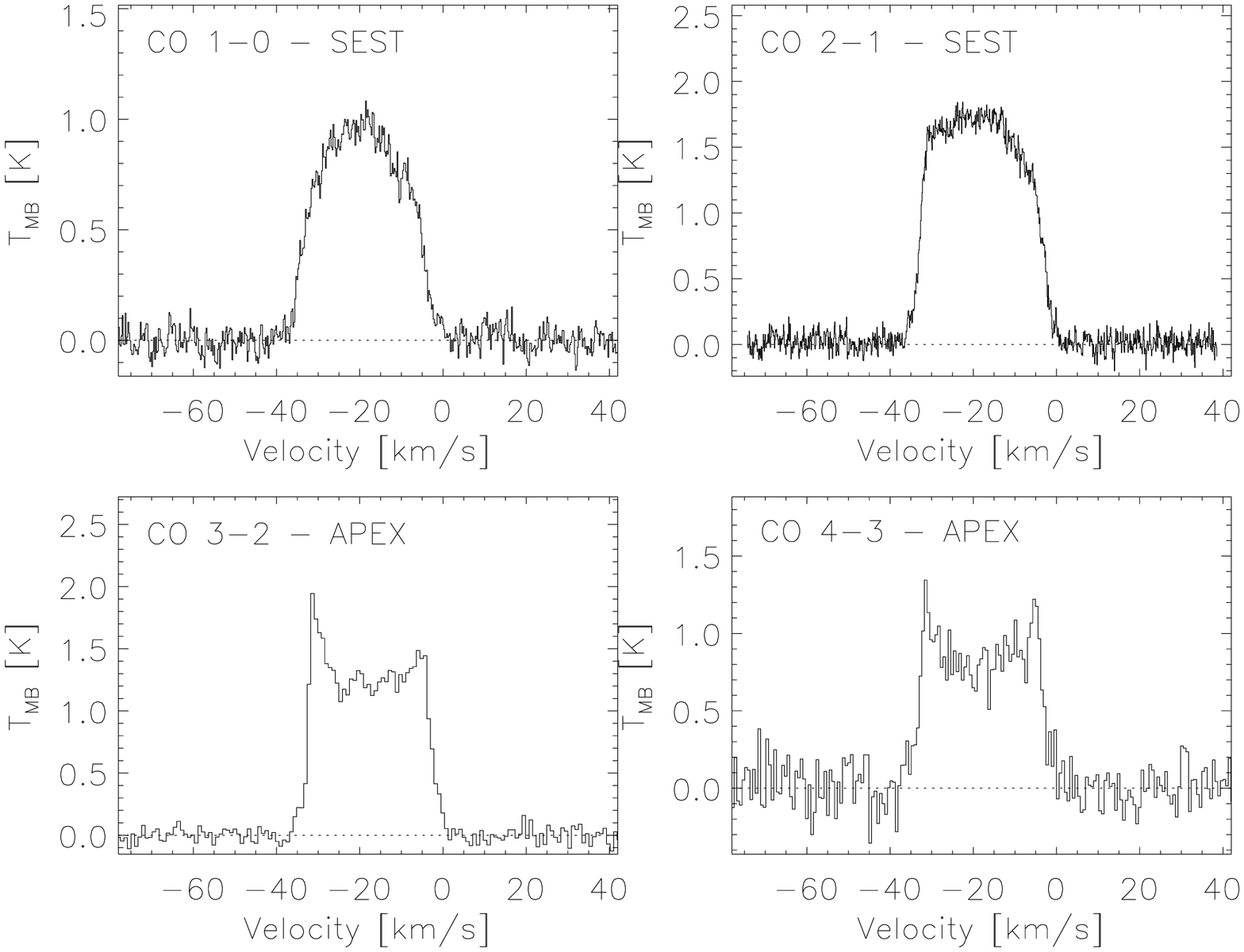}}
\caption{Panels a to c show the multiple superimposed peaks that are easily detected in the CO emission lines of R\,Cas, V\,Hya, and VY\,CMa. Panels d and e show the two separate velocity components seen in the CO lines of U\,Ant and S\,Sct. These are clear evidence for detached shells around these targets. It has also been shown that R\,Scl (panel f) has a detached shell. See text for further comments. \label{fig:superimposed_detached} }
\end{center}
\end{figure*}

\begin{figure}
\begin{center}
\includegraphics[angle=90,width=\linewidth]{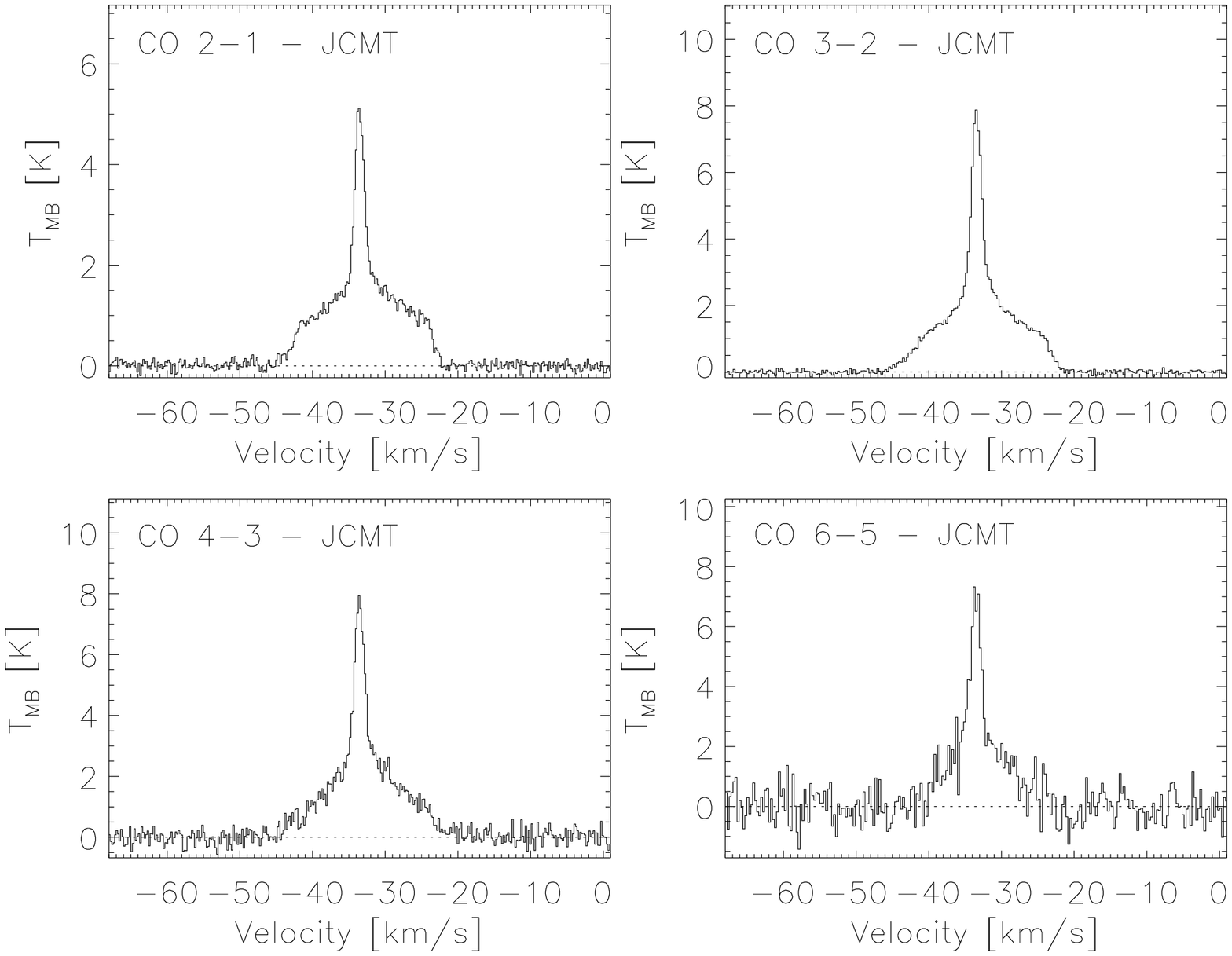}
\caption{The CO line profiles measured towards the \orich SRb variable EP\,Aqr bear clear evidence of a composite circumstellar environment. See Sect.~\ref{subsubsect:individualtargets} for further details. \label{fig:composite}}
\end{center}
\end{figure}

\subsubsection{Line profile diagnostics}\label{subsubsect:softparabola}
Because of its high stability against photodissociation and its large molecular abundance, carbon monoxide exists throughout the largest part of the CSE. Rotational transitions of CO are easily excited in the cool envelopes and can therefore trace many properties of all different layers, e.g. density and temperature. These properties are linked to the optical thickness of the envelope layers and the mass-loss rate of the central star.

In case of optically thick, unresolved envelopes, the observed line profiles have parabolic shapes. Changing to the optically thin and/or spatially resolved case, the molecular lines are more flat-topped or even display two-horned profiles \citep{knapp1985}. Considering this simple rule, the shape of the line profiles is a useful first diagnostic for the (density) properties and the geometric extent of a star's CSE relative to the telescope beam. The data of, e.g., IK\,Tau shown in Fig.~\ref{fig:iktau} clearly reflect that the envelope layers are optically thinner for \thco than for \twco, since the line profiles of the latter are consistently more parabolic in shape. The mass loss of CW\,Leo is very strong \citep[$\sim$$2\times 10^{-5}$\,\msun\,yr$^{-1}$;][]{mauron2000}, causing optically thick CO emission lines. These are, however, flattened due to the large angular size of the envelope \citep[$\sim$200\,arcsec;][]{mauron2000}, causing resolution by the telescope beams with half power beam widths between 7 and 20\,arcsec (see Table~\ref{tbl:instruments}).

A first analysis of the data consists in fitting a so-called \textit{soft parabola} \citep{olofsson1993} to the observed line profiles. Every rotational line is specified by several line parameters: the main-beam-brightness temperature at the line centre, $T_{\mathrm{MB,c}}$, the velocity at the line centre, i.e. the velocity of the star with respect to the Local Standard of Rest, $v_{\mathrm{LSR}}$, and the half width of the line profile, i.e. the expansion velocity of the CSE in the outermost regions where the studied molecule is present, \vinfty. These parameters are obtained by fitting the data with the soft parabola line profile function, given in  Eq.~\ref{eq:softparabola}, where $\beta$ is a measure for the shape of the profile function:
\begin{eqnarray}
\label{eq:softparabola}
T(v)&=&T_{\mathrm{MB,c}}\left[1-\left(\frac{v-v_{\mathrm{LSR}}}{v_{\infty}}\right)^2\right]^{\beta/2},  \,|v-v_{\mathrm{LSR}}|\leq v_{\infty}\\
&=&0,\hspace{.5\linewidth}|v-v_{\mathrm{LSR}}|> v_{\infty}\nonumber
\end{eqnarray}
When $\beta=2$, the line profile has a parabolic shape, representative for optically thick lines, observed towards spatially unresolved CSEs. Smaller, positive values for $\beta$ lead to more flat topped profiles. Negative $\beta$ values can be used to fit two-horned profiles observed towards optically thin envelopes.  The best fit to the line profile was determined through minimising the total absolute difference between the data and the soft-parabola fit with the IDL-routine AMOEBA.

Figs.~\ref{fig:12CO_parabola} and \ref{fig:13CO_parabola} show all \twco and \thco observations and the soft-parabola fits to these data. The $\beta$-values per line profile are listed in Tables~\ref{tbl:12CO} and \ref{tbl:13CO}. The parameter $\beta$ provides information on the optical thickness of the envelope (and hence on \Mdot) and the resolving power of the instrument. Its value will be used to improve the accuracy of the mass-loss determination (see Sect.~\ref{grid_results}). 

\subsubsection{Discussion on individual targets}\label{subsubsect:individualtargets}
If strong deviations from the soft-parabola fit are found, this fitting procedure could reveal significant deviations from the assumptions of, e.g., spherical symmetry or constant \mdot, since detached shells, variability in the mass loss, vast clumps in the envelope, or jets can affect the symmetry and the overall shape of the CO lines. Some of the targets in the sample and their deviations from the soft-parabola fits are discussed individually in this section.

It is obvious from Figs.~\ref{fig:12CO_parabola} and \ref{fig:13CO_parabola} that indeed not all profiles in the data set can be fitted with the simple type of line profile function presented in Eq.~\ref{eq:softparabola}. R\,Cas (Mira) and V\,Hya (SRa) have line profiles exhibiting superimposed peaks --- see Fig.~\ref{fig:superimposed_detached} --- that could be linked to bipolar outflows \citep{olofsson1993,sahai2003}. \cite{olofsson1993} suggest that these bipolar outflows could be related to the possible presence of a binary companion in case of both targets. \cite{sahai2003} suggest that V\,Hya could already be transitioning to the Post-AGB phase. The high-velocity bipolar outflow would be linked to this transition process.

U\,Ant and S\,Sct are discussed by \cite{olofsson1993} as targets with convincing evidence in the CO data for highly episodic mass loss, possibly due to a thermal pulse. The shapes of their line profiles --- see Fig.~\ref{fig:superimposed_detached} --- indeed imply the presence of a detached shell. The inner parts of the outflow have terminal velocities of respectively 7\,km\,s$^{-1}$ and 8\,km\,s$^{-1}$, while the outflow velocities of the outer parts of the envelopes reach 16\,km\,s$^{-1}$ and 20\,km\,s$^{-1}$. The \crich semi-regular R\,Scl also has a detached shell \citep{olofsson1996}, albeit not directly visible in the CO line profiles.

The CO line profiles of the SRb variable EP\,Aqr (Fig.~\ref{fig:composite}) reveal the composite nature of its circumstellar environments. Except for the low-S/N observation of \element[ ][13]{CO}(4--3), all measured line profiles towards EP\,Aqr exhibit a composite profile with (a) a broad, low-$T_{\mathrm{MB}}$ profile and (b) a very narrow, high-$T_{\mathrm{MB}}$ profile centrally superposed on this broad profile. Only very little is known about the origin of these types of line profiles. The broad plateau emission has been suggested to originate from episodic mass loss, bipolar outflows, or circumstellar disks \citep{knapp1998,kerschbaum1999}. \cite{winters2007} mention that no obvious departure from circular symmetry can be seen, but that there is evidence for a ringed structure in the \element[ ][12]{CO}(2--1) map implying variation of \mdot in time.

R\,Hya is well-known for its declining period and mass-loss rate \citep{decin2008_rhya}, which \cite{wood1981} attributed to a possible recent thermal pulse. The detached shell that has been detected at 60\,\um can be explained by a slowing down of the wind. \cite{decin2008_rhya} have modelled the wind of R\,Hya in detail, checking models produced with the radiative transfer code \gastronoom against data of both rotational and vibration-rotation transitions of CO. WX\,Psc is a second object for which \mdot could be considered to be under influence of TPs. This extreme oxygen-rich star could currently be going through a superwind phase \citep{decin2007_wxpsc}, causing the large infrared excess. For both targets, however, these presumed deviations from constancy in mass-loss rate are not directly visible in the rotational CO line profiles.

The triply peaked low-$J$ transition profiles in both \twco and \thco towards the supergiant VY\,CMa --- see Fig.~\ref{fig:superimposed_detached} --- are most probably a superposition of an optically thin component (red-wing and blue-wing peaks) and an optically thick one (middle peak), as shown by \cite{ziurys2007}. For the higher-$J$ transitions ($J=4-3$ and up), the extra parabolic feature is no longer clearly present. This, in combination with the asymmetry of the profile, led to the assumption of a variable mass-loss history as described by \cite{decin2006_vycma}. NML\,Cyg and $\alpha$\,Ori both have spiked line profiles, somewhat similar to those of VY\,CMa. In the case of NML\,Cyg, the lines could be contaminated with interstellar CO detection. The line profiles of IRC\,+10420 and $\mu$\,Cep exhibit asymmetries, with a blue-wing feature superimposed on a more or less parabolic profile. The extra feature is far more pronounced for $\mu$\,Cep and is also further out to the blue part of the line profile.

For some of the more evolved objects in our sample, i.e. Post-AGB stars, we also see deviations from the soft-parabola shape. The \element[ ][12]{CO}($3-2$) line profile towards GSC\,08608-00509 exhibits a sharp red-wing peak. For the Cotton Candy Nebula, we see a similar sharp feature in the blue wing. Since we have only few observations for these targets, it is hard to draw well-founded conclusions about the properties or geometry of their extended envelopes. 

All three line profiles towards the Gomez Nebula are two-horned, supporting the hypothesis of a CO emitting disk in keplerian rotation, as mentioned by \citet[their Fig.~3]{bujarrabal2008}. The CO line profiles towards the other young stellar object, AFGL\,5502, bear evidence of asymmetry, since the higher-excitation lines have superimposed features in their blue wings.
\\

We note here that T\,Cet was not included in the sample of red supergiants, contrary to what is often found in literature \citep[e.g. ][]{sloan1998}, but in the sample of AGB stars. The low luminosity and the dust features in the ISO spectrum make classification of T\,Cet as AGB star more plausible \citep{speck2000,verhoelst2009}.
\begin{figure*}
\begin{center}
  \includegraphics[angle=0,width=.95\textwidth]{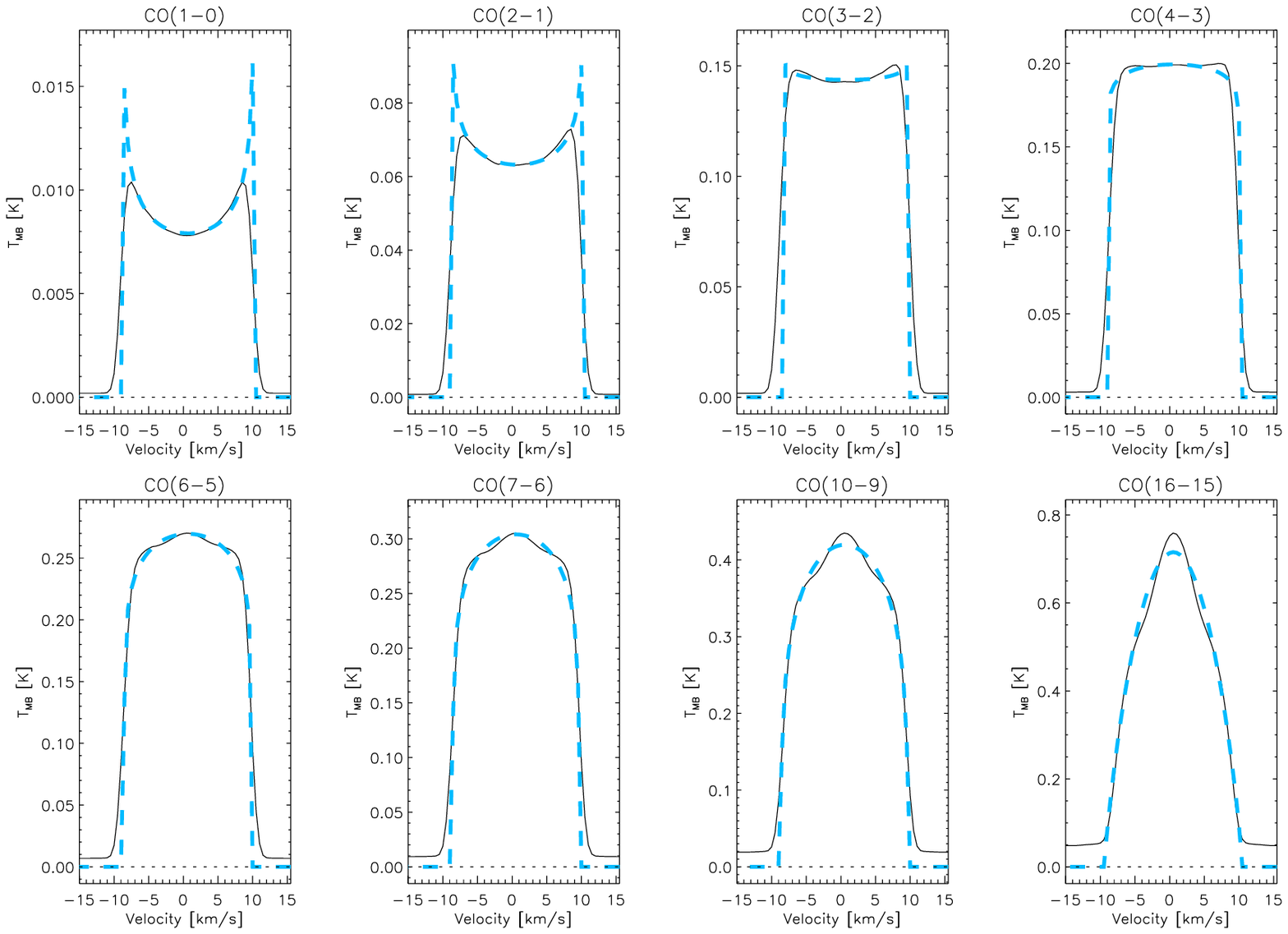} 
\caption{Line profile predictions for \Tstar\,=\,2000\,K, \Lstar\,=\,3000\,\Lsun, $d$\,=\,300\,pc, \fhco\,=\,$1\times10^{-4}$, \isotope\,=\,10, \Rinner\,=\,3\,\Rstar, \Mdot\,=\,$1\times10^{-7}$\,\Msun\,yr$^{-1}$, \vinfty\,=\,10\,km\,s$^{-1}$ for a beam width of 10\arcsec. The soft-parabola fit is plotted in dashed grey/light blue lines. For the high-$J$ excitation lines one can see both the stellar and dust continuum and the acceleration zone (the narrow peaks around the line centres). \label{fig_profiles}}
\end{center}
\end{figure*}


\section{Radiative transfer analysis}\label{sect:radtrananalysis}
Physical information on the envelope structure of evolved stars can be extracted from the observed line fluxes using a non-local thermodynamic equilibrium (non-LTE) radiative transfer code (see Sect.~\ref{subsect:radtranmodel} for a short description). The availability of a large \twco\ and \thco\ line flux data base, as  presented in Sect.~\ref{subsect:dataresults}, offers interesting possibilities for statistical analyses. However, this large database also forces us to construct simplified analytical expressions relating the mass-loss rate and the observed integrated CO-line intensities to deduce the mass-loss history of each target.  We therefore derived new mass-loss rate formulae based on a grid computation. Similar formalisms that have been presented in the literature are summarised in Sect.~\ref{subsect:comp}.  The grid construction and the resulting analytical expressions are discussed in Sect.~\ref{subsect:grid} and~\ref{grid_results}. A discussion on the quality of the formalism can be found in Sect.~\ref{subsect:formuladiscussion}.

\subsection{Radiative transfer model}\label{subsect:radtranmodel}
The structure of circumstellar envelopes is thought to be quite complex, due to the possible presence of shocks, bipolar outflows, or episodic mass-loss events. The complex heating and cooling processes due to excitation of molecules, and the unknown grain-size distributions contribute to the complexity of the envelopes. The often taken approach to model these envelope structures is therefore to choose a simplified approach with reasonable assumptions. We use the non-LTE radiative code \gastronoom, which was presented and discussed by \cite{decin2006_vycma, decin2007_wxpsc,decin2008_rhya}  and benchmarked according to the method described by \cite{vanzadelhoff2002}, considering a simple 2-level molecule (their cases 1a and 1b). The CSE is assumed to be spherically symmetric and formed by a constant mass-loss rate. The code calculates the kinetic temperature and velocity structure in the shell by solving the equations of motion of gas and dust and the energy balance simultaneously \cite{justtanont1994}. Adopting a value for the gas mass-loss rate, the dust-to-gas ratio is adjusted until the required terminal velocity is obtained \citep{decin2006_vycma}. The statistical equilibrium equations and the radiative transfer equation are solved in the co-moving frame using the Approximated Newton-Raphson operator as developed by \cite{schoenberg1986}. Finally, a ray tracing program uses the computed non-LTE level populations to calculate the line profiles. For a full description of the \gastronoom-code, we refer to \cite{decin2006_vycma}. 

As described by \cite{decin2006_vycma,decin2007_wxpsc,decin2008_rhya}, a detailed radiative transfer modelling of one target often results in the computation of many ($>10^5$) models. Some prior knowledge on the input parameters helps to restrict the extensive parameter space. The main input parameters for this research are the stellar temperature \tstar, the stellar radius \rstar, the distance $d$, the CO fractional abundance $f_{\mathrm{H,CO}}=\mathrm{CO/H}$, the \twthratio isotope abundance ratio, the terminal velocity \vinfty, the mass-loss rate \mdot, and the dust condensation radius \rinner. When each of these eight parameters would only be scanned over five different values, almost 400\,000 models should be computed. Performing an in-depth line profile analysis for each target presented in this study is therefore too time-consuming and beyond the scope of this paper. The mass-loss rates of some twenty targets of the current sample will be individually modelled in a forthcoming paper (Lombaert et al., in prep.). 

Before describing the grid computation in Sect.~\ref{subsect:grid}, we give a short overview of mass-loss rate formulae presented in the literature. This overview is for sure not complete, but provides a few often applied formulae, relevant to this study. These formalisms are also compared with the one we present in this paper.

\subsection{Comparison with other studies} \label{subsect:comp}
Several theoretical mass-loss rate formulae based on observed line intensities or fluxes were already presented in the literature. \cite{knapp1985} derived a formula to estimate \mdot in terms of the CO($J=1-0$)\footnote{When no isotope number is mentioned, CO is equivalent to \twco.} line intensity. Their formula relates the mass-loss rate to an assumed CO abundance and a number of easily determined observables: the main beam temperature $T_{\mathrm{MB,c}}$ at the line centre (in K), the terminal velocity \vinfty (in km\,s$^{-1}$), and distance $d$ (in pc). A distinction was made between optically thick and optically thin envelopes. The model for the CO emission assumed a spherically symmetric envelope expanding at constant velocity. The kinetic temperature profile throughout the envelope was assumed in all cases to be that derived by \cite{kwan1982} for CW\,Leo. The statistical equilibrium equations were solved for the lowest 15 rotational levels. \cite{olofsson1993} transformed the formula of \cite{knapp1985} for an optically thick envelope to the slightly more general form
\begin{equation}
\label{eq:mdotknapp1985}
\dot{M}_{\mathrm{KM}}=5.7\times10^{-20}\frac{T_{\mathrm{MB,c}}v_{\infty}^2 d^2 \theta_{\rm b}^2}{s(J) f_{\mathrm{H,CO}}^{0.85}}\; \mathrm{M}_{\odot}/\mathrm{yr,}
\end{equation}
where $\theta_{\rm b}$ is the beamwidth of the instrument (in arcsec) used for the observation of the line under study. The factor $s(J)$ depends on the rotational transition $J\rightarrow J-1$.
For the $J=1-0$ transition $s(J) =1$, and for the $J=2-1$ transition $s(J) = 0.5$ \citep[see the discussion in][]{olofsson1993}. In this equation --- valid in the optically thick case --- the CO envelope size was fixed at $3\times10^{17}$\,cm, appropriate for high-\mdot targets. \cite{schoeier2001} compared mass-loss rates derived using their Monte Carlo non-LTE radiative transfer model to those estimated using the formulae of \cite{knapp1985}. They found that the formula underestimated the mass-loss rates when compared to those obtained from their radiative transfer analysis on average by a factor of about four, and the discrepancies were found to increase for lower mass-loss rate objects \citep[][see their Fig.~8]{schoeier2001}. In the case of optically thin envelopes, the exact envelope size matters in estimating \mdot, so Eq.~\ref{eq:mdotknapp1985} is indeed of lesser value. A correction for the envelope size \cite[see][and references therein]{neri1998} would decrease the discrepancy found by \cite{schoeier2001}. However, considering the simplicity of Eq.~\ref{eq:mdotknapp1985}, the agreement is still remarkably good.

\cite{loup1993} calculated the envelope size $R_{\mathrm{CO}}$ and the mass-loss rate $\dot{M}_{\mathrm{L}}$ by numerically solving the system of Eqs.~\ref{eq:loup1993_1} and \ref{eq:loup1993_2}.
\begin{eqnarray}
 \label{eq:loup1993_1}
\dot{M}_{\mathrm{L}}&=&\frac{v_{\infty}^2 d^2 T_{\mathrm{MB,c}}}{2\times10^3 (2f_{\mathrm{H,CO}})^{0.85}}\times F^{-1}(R_{\mathrm{CO}}) \;\;  \mathrm{M}_{\odot}\,\mathrm{yr}^{-1}\\
 \label{eq:loup1993_2}
\dot{M}_{\mathrm{L}}&=&10^{-6}\left(\frac{R_{\mathrm{CO}}}{7.3\times10^{16}}\right)^{1/0.58} \left(\frac{v_{\infty}}{10}\right)^{1/1.45} \\
&&\quad\times\left(\frac{4\times10^{-4}}{2f_{\mathrm{H,CO}}}\right)^{1/1.16}\;\;  \mathrm{M}_{\odot}\,\mathrm{yr}^{-1} \nonumber
\end{eqnarray}
They assumed  \fhco of $2.5\times10^{-4}$ for \orich stars and $5\times10^{-4}$ for both \crich and S-type stars. The function $F(R_{\mathrm{CO}})$ introduces a correction factor for the envelope size and is normalised to unity for $R_{\mathrm{CO}}=3\times10^{17}$\,cm.

\cite{ramstedt2008} derived a formula relating the mass-loss rate to the integrated intensity $I_{{\rm CO},J}$ of the \twco($J\rightarrow J-1$) lines,  the terminal velocity \vinfty, the abundance \fhco, the distance $d$ to the object, and the telescope's beam size $\theta_{\rm b}$. Using the Monte Carlo radiative transfer code described by \cite{schoeier2001}, the estimator was derived from a grid of 60 models varying in \mdot (1, 3, 10, 30, and 100\,$\times10^{-7}$\,\msun\,yr$^{-1}$), \vinfty (5, 10, 15, and 20\,km\,s$^{-1}$), and \fhco (1, 3, and 10\,$\times 10^{-4}$).  The distance was taken to be 1000\,pc. For each model, the velocity integrated intensities in CO ($J=1-0,\;2-1,\;3-2,\; 4-3$) were calculated for a 20\,m telescope (with corresponding beam sizes of 33$\arcsec$, 16.5$\arcsec$, 11$\arcsec$, and 8.5$\arcsec$).  They use the so-called $h$-parameter to describe the free parameters of the dust \citep{schoeier2001}:
\begin{equation}
\label{h}
h = \left( \frac{\Psi}{0.01} \right) \left( \frac{2.0 \  \mathrm{g \ cm^{-3}}}{\rho_{\mathrm{d}}} \right) \left( \frac{0.05 \ \mu \mathrm{m}}{a_{\mathrm{d}}} \right),
\end{equation}
\noindent
where $\Psi$ is the dust-to-gas mass ratio, and $\mathrm{\rho_{d}}$ and $a_{\mathrm{d}}$ are the average density and the radius of an individual dust grain, respectively. The value of $h$ was assumed to be 0.2 for \mdot in the range up to $3\times10^{-7}$\,\msun\,yr$^{-1}$, 0.5 for $1 - 3 \times10^{-6}$\msun\,yr$^{-1}$, and 1.5 for $10^{-5}$\,\msun\,yr$^{-1}$. The stellar temperature \tstar, stellar radius \rstar or inner dust condensation radius \rinner were not specified. An estimate for the  mass-loss rate was determined from a minimisation procedure applied to Eq.~\ref{eq:mdotramstedt2008}:
\begin{equation}
\label{eq:mdotramstedt2008}
\dot{M}_{\mathrm R}=s_J(I_{\mathrm{CO},J} \theta_{\mathrm{b}}^2 d^2/10^6)^{a_J} v_{\infty}^{b_J} f_{\mathrm{H,CO}}^{-c_J} \mathrm{M}_{\odot}\,\mathrm{yr}^{-1},
\end{equation}
with $I_{{\rm CO},J}$ in K\,km\,s$^{-1}$, \vinfty in km\,s$^{-1}$, $d$ in pc, and $\theta_{\rm b}$ in arcsec. A comparison between their input mass-loss rates and the values found from Eq.~\ref{eq:mdotramstedt2008} was not shown, nor were estimates of the uncertainties on the derived \mdot-values provided. The comparison between mass-loss rates derived using Eq.~\ref{eq:mdotramstedt2008} and mass-loss rates derived using their Monte Carlo code for a sample of C-, M- and S-stars shows quite a good agreement. Close inspection, however, reveals that for low-$J$ transitions ($J=1-0,\,2-1$)  Eq.~\ref{eq:mdotramstedt2008} tends to underestimate the mass-loss rate for $\dot{M}\lesssim10^{-6}$\,\msun\,yr$^{-1}$ and to overestimate it for $\dot{M}\gtrsim10^{-5}$\,\msun\,yr$^{-1}$ \citep[see Fig A.1. in][]{ramstedt2008}. 
\\

Since \textit{(1)} our data set holds observations up to the high excitation $J=6-5$ and $J=7-6$ rotational line transitions of CO, and we want \textit{(2)} to include the shape of the line profiles, represented by parameter $\beta$, in the estimates of \mdot, and \textit{(3)} to extend the computations to higher mass-loss rate values (\Mdot $> 1 \times 10^{-5}$\,\Msun\,yr$^{-1}$), we elaborated on the studies discussed above by performing a grid computation covering more than 15\,000 models.

\begin{table}[t]
\setlength{\tabcolsep}{1.mm}
\caption{Values for the input parameters used in the grid computations. \label{table:parameters_grid}}
\begin{tabular}{ll|ll}
 \hline \hline
Param. & Grid values & Param. & Grid values \\
\hline\\[-2ex]
\Tstar & 2000, 2500, 3000\,K & \Rinner& 3, 10, 30\,\rstar\\
\Lstar & 3000, 10\,000, 30\,000\,\lsun &\isotope & 10, 30, 100 \\
$d$ & 300,\, 1000,\, 3000\,pc &  \Mdot & 1, 3, 10, 30,  \\
\vinf & 10, 15, 20\,km\,s$^{-1}$ &    & 100, 300, 1\,000 \\
\fhco & 1, 3, 5$\times10^{-4}$     &          &$\times10^{-7}$\msun\,yr$^{-1}$\\
$\theta$&10, 20, 30\arcsec								&	&\\
\hline
\end{tabular}
 \end{table}

\subsection{Description of the grid}\label{subsect:grid}
To study the influence of different stellar and envelope parameters --- like stellar temperature, stellar radius (or stellar luminosity) and dust condensation radius --- on the CO line intensities, the input parameters were varied to cover the relevant parameter space for evolved low-mass targets (see Table~\ref{table:parameters_grid}). The values for the \fhco were chosen to be $1\times10^{-4}$ for O-rich stars \citep{kahane1994},  $3\times10^{-4}$ for  S-type stars \citep{ramstedt2006}, and  $5\times10^{-4}$ for C-rich stars \citep{zuckerman1986}. All computations were performed for a 15\,m class telescope, with beam sizes of 10\arcsec, 20\arcsec and 30\arcsec as grid input values. This grid spans 15\,309 star models, for which each rotational line intensity was computed for three values of the beamwidth.

As was done for the observational line profiles, the theoretical line profiles were fitted with a soft parabola. In the case of the theoretical line profiles, however, an extra \textit{offset} $T_0^g$ should be added to the equation, resulting in
\begin{eqnarray}
\label{eq:softparabolatheory}
T^g(v)&=&T^g_{\mathrm{MB,c}}\left[1-\left(\frac{v^g-v^g_{\mathrm{LSR}}}{v^g_{\infty}}\right)^2\right]^{\beta/2}  +T_0^g \nonumber \\
         &   & \qquad \qquad \qquad \qquad \qquad {\rm for\ }|v^g-v^g_{\mathrm{LSR}}|\leq v^g_{\infty} \\
&=&T_0^g  \ \ \qquad \qquad \qquad \qquad \,{\rm for\ } |v^g-v^g_{\mathrm{LSR}}|> v^g_{\infty} \nonumber
\end{eqnarray}
where $T$ is the line intensity\footnote{Throughout the paper, we denote line intensity, expressed as a temperature, with $T$, and (velocity-)integrated line intensity with $I$.} in main-beam temperature. The superscript '$g$' designates that we are dealing with theoretical profiles that were computed using the \gastronoom-code. The reason for the offset is that, in case of optically thin envelopes, one still detects the almost unattenuated stellar continuum (see Fig.~\ref{fig_profiles}). This offset parameter is not present in Eq.~\ref{eq:softparabola} that was used to fit the data. During the data-reduction process, the offset is removed from the spectra when baselines are subtracted, since the observer does not know if and how much stellar light penetrates the envelope. When confronting observational data with theoretical integrated line intensities, one therefore has to compare $I^g-2v_{\infty}T^g_0$ to $I$, where $I$ is the integrated line intensity $\int T(v) dv$ in K\,km\,s$^{-1}$. The quantity $I^g-2v_{\infty}T^g_0$ will from now on be referred to as $I^{g,cc}$, with the superscript '$cc$' denoting the continuum correction. Optically thin lines allow one to trace the wind acceleration zone of which the emission is visible as a superimposed small peak around the line centre (see Fig.~\ref{fig_profiles}). 

The soft-parabola fit is not ideal to reproduce optically thin resolved emission lines (see upper panels of Fig.~\ref{fig_profiles}). The $\beta$-value was determined in a way that the soft parabola should yield a good representation of the line depression. The integrated intensity of the fit was not required to match the integrated intensity of the \gastronoom-predictions. This method provides information about the degree to which the line profiles are resolved by the telescope beam.

\begin{figure}
 \includegraphics[angle=90,width=.45\textwidth]{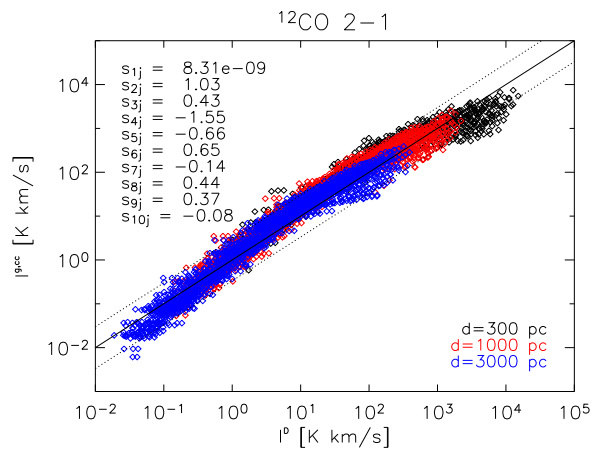}
\caption{The continuum corrected integrated intensities  $I^{g,cc}$ predicted with \gastronoom are compared to the values $I^{\mathrm{D}}$ calculated with Eq.~\ref{eq:minimisationfunction}. Values for the exponents $s_{i,J}$ in Eq.~\ref{eq:minimisationfunction} are specified in the upper left corner. The full line represents equality of $I^{g,cc}$ and $I^{\mathrm{D}}$, the dotted lines show a factor 3 difference w.r.t. this relation. For all three distance values assumed in the model grid, the estimates are plotted in a different color: black for $d=300$\,pc, red for $d=1000$\,pc, and blue for $d=3000$\,pc. See Sect.~\ref{grid_results} for further details. \label{intensity_CO21}}
\end{figure}

\begin{figure*}
\subfigure{\includegraphics[height=.5\textwidth,angle=90]{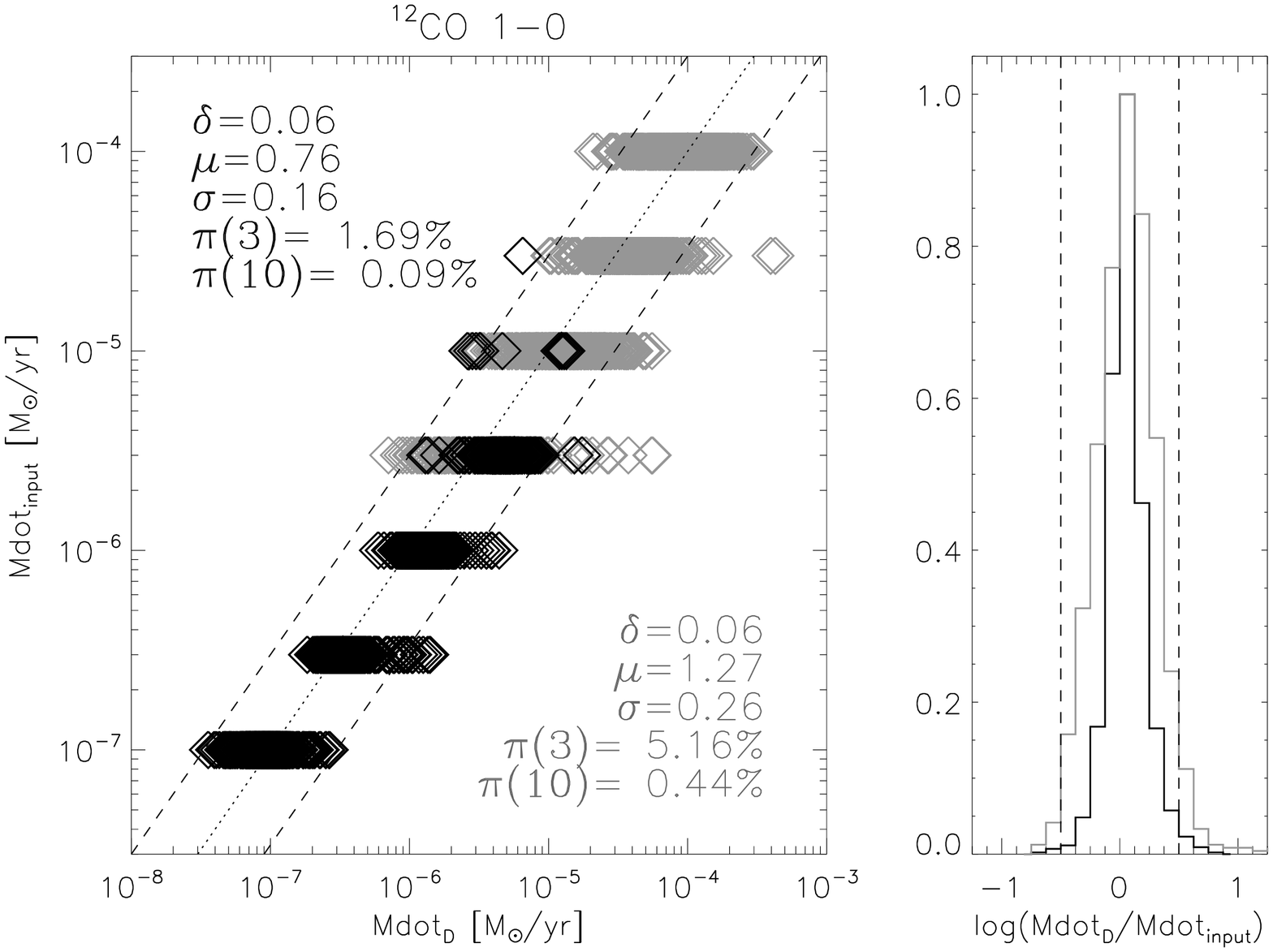}} \hspace{.1\linewidth} \hspace{-0.1\linewidth}\subfigure{\includegraphics[height=.5\textwidth,angle=90]{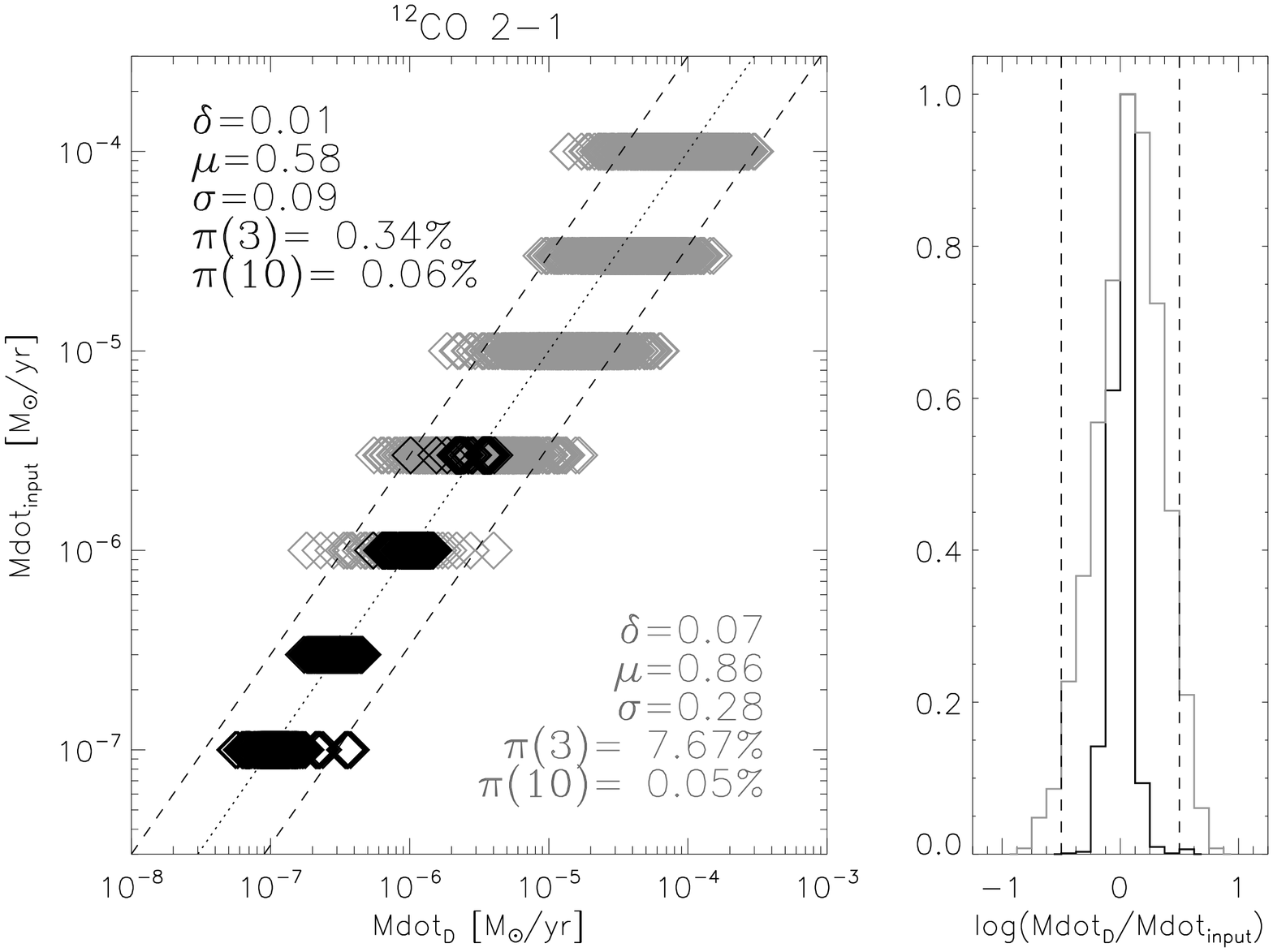}}
\subfigure{\includegraphics[height=.5\textwidth,angle=90]{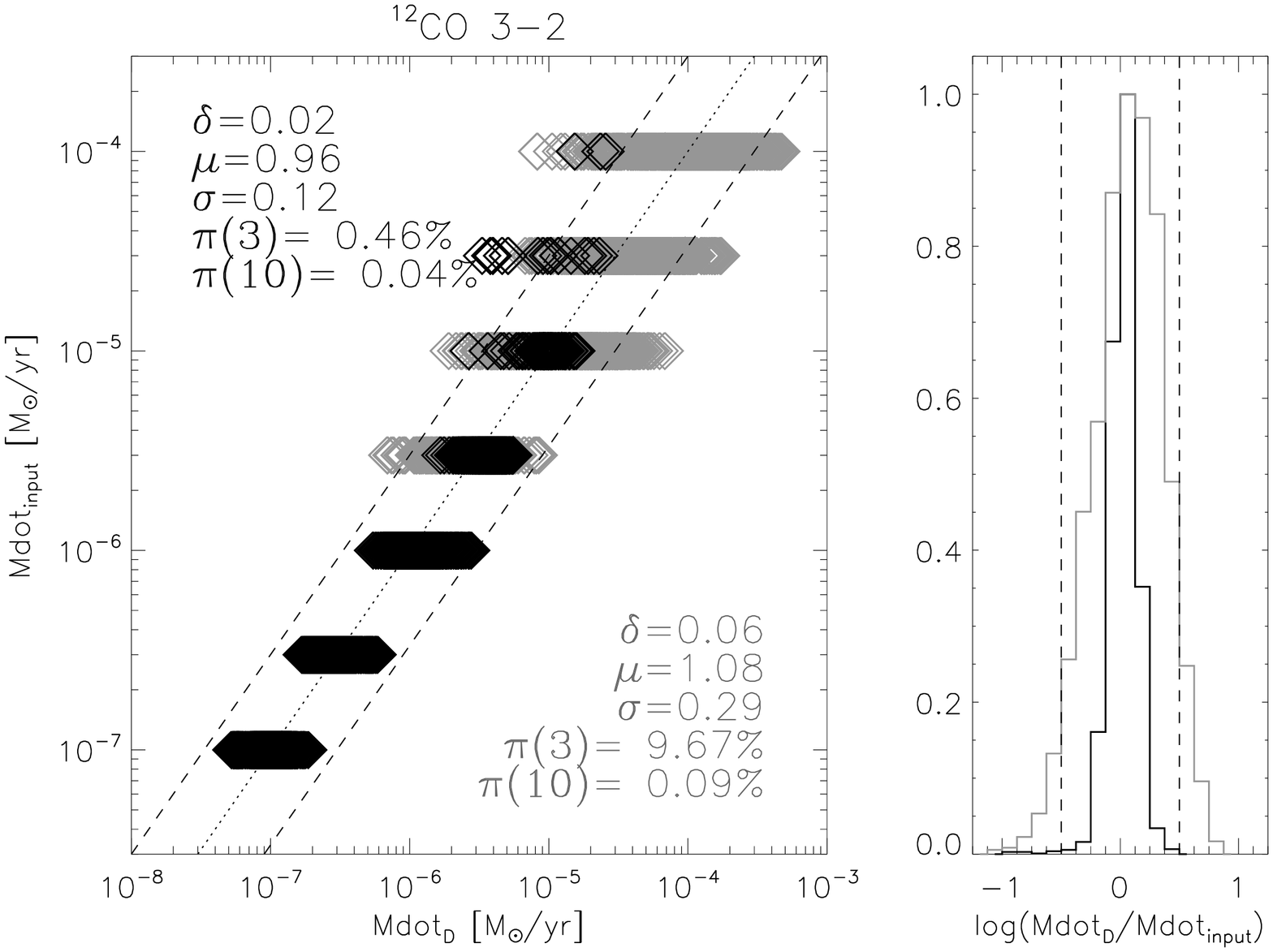}} \hspace{.1\linewidth} \hspace{-0.1\linewidth}\subfigure{\includegraphics[height=.5\textwidth,angle=90]{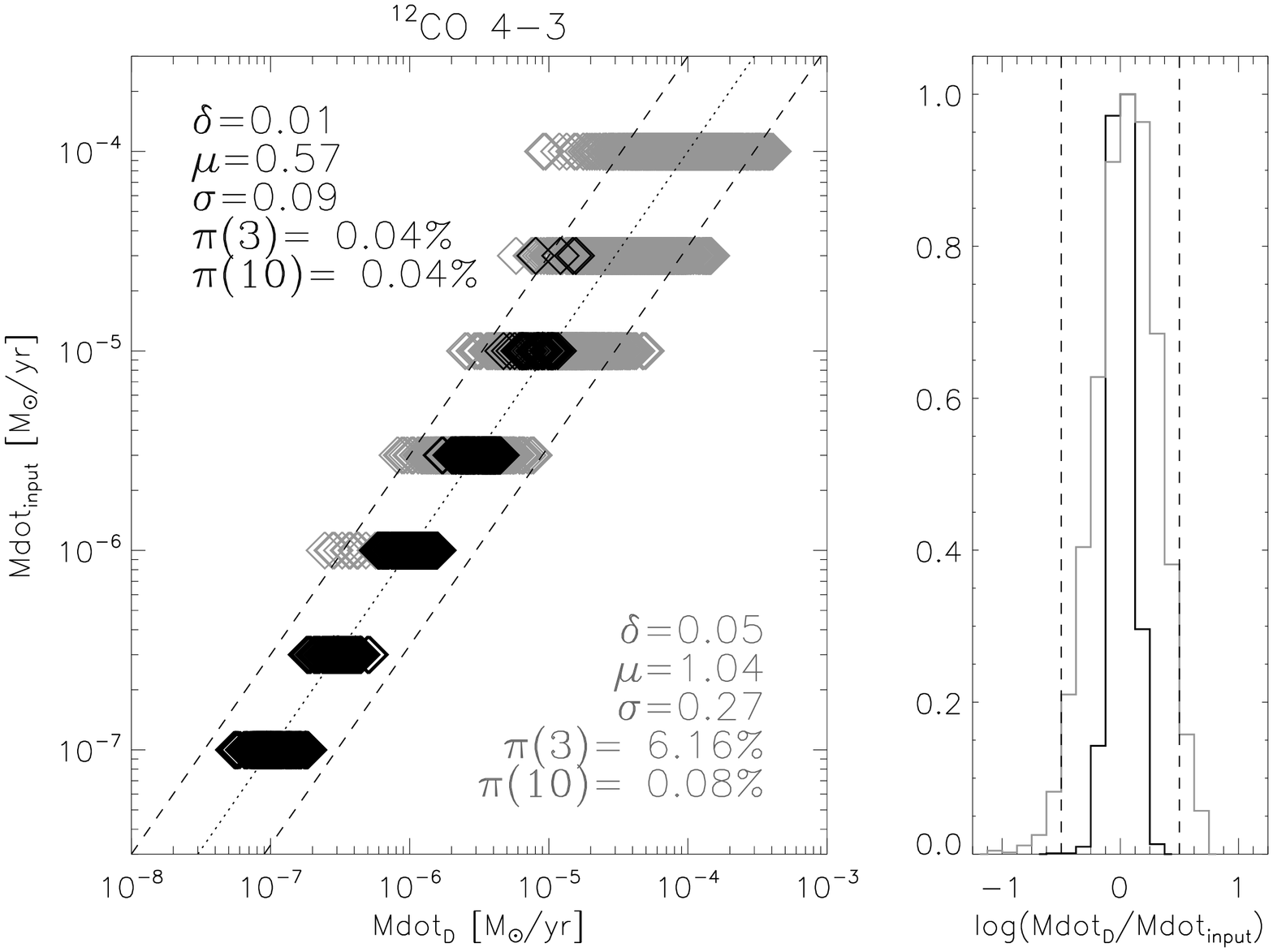}}
\subfigure{\includegraphics[height=.5\textwidth,angle=90]{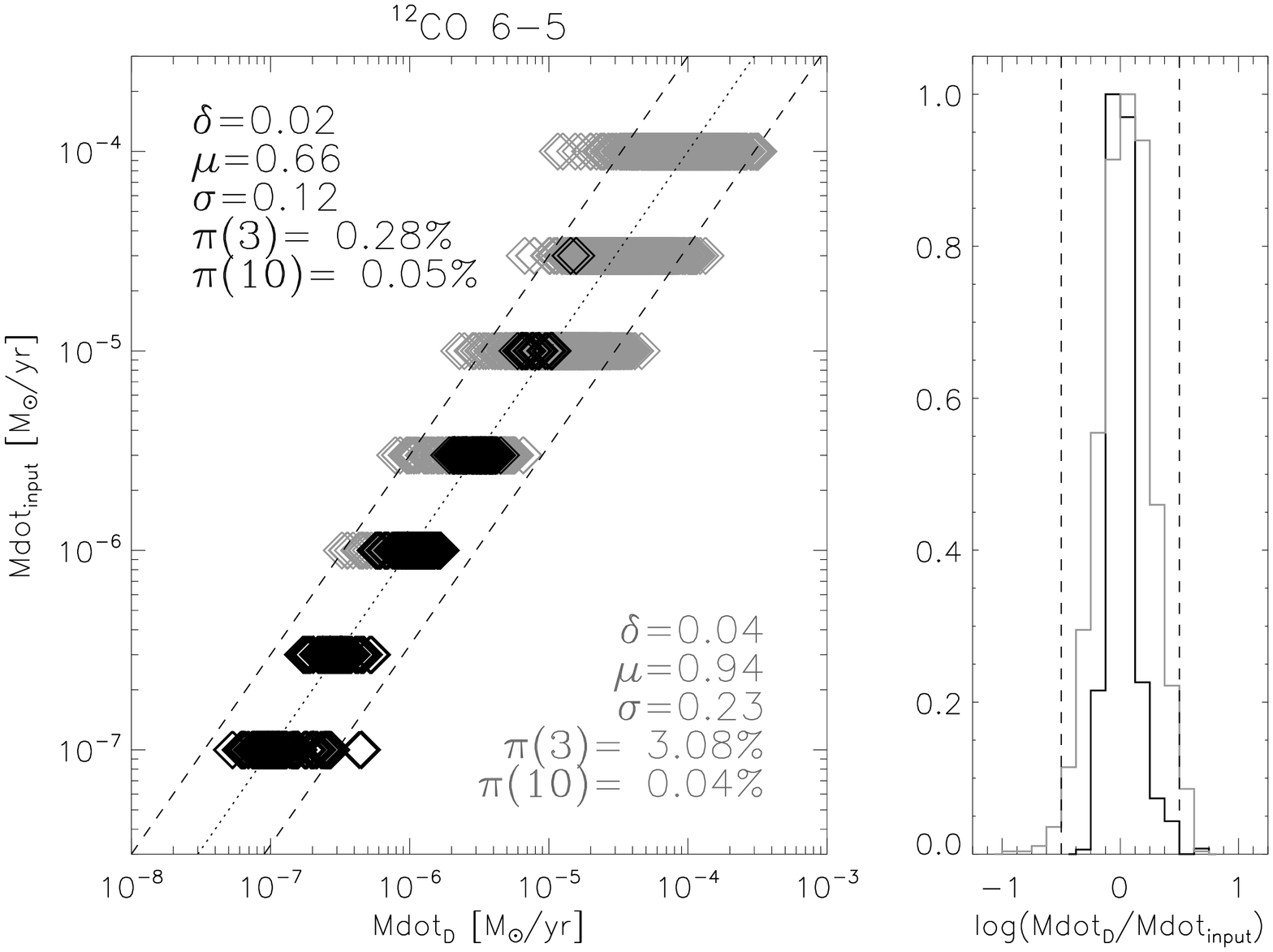}} \hspace{.1\linewidth} \hspace{-0.1\linewidth}\subfigure{\includegraphics[height=.5\textwidth,angle=90]{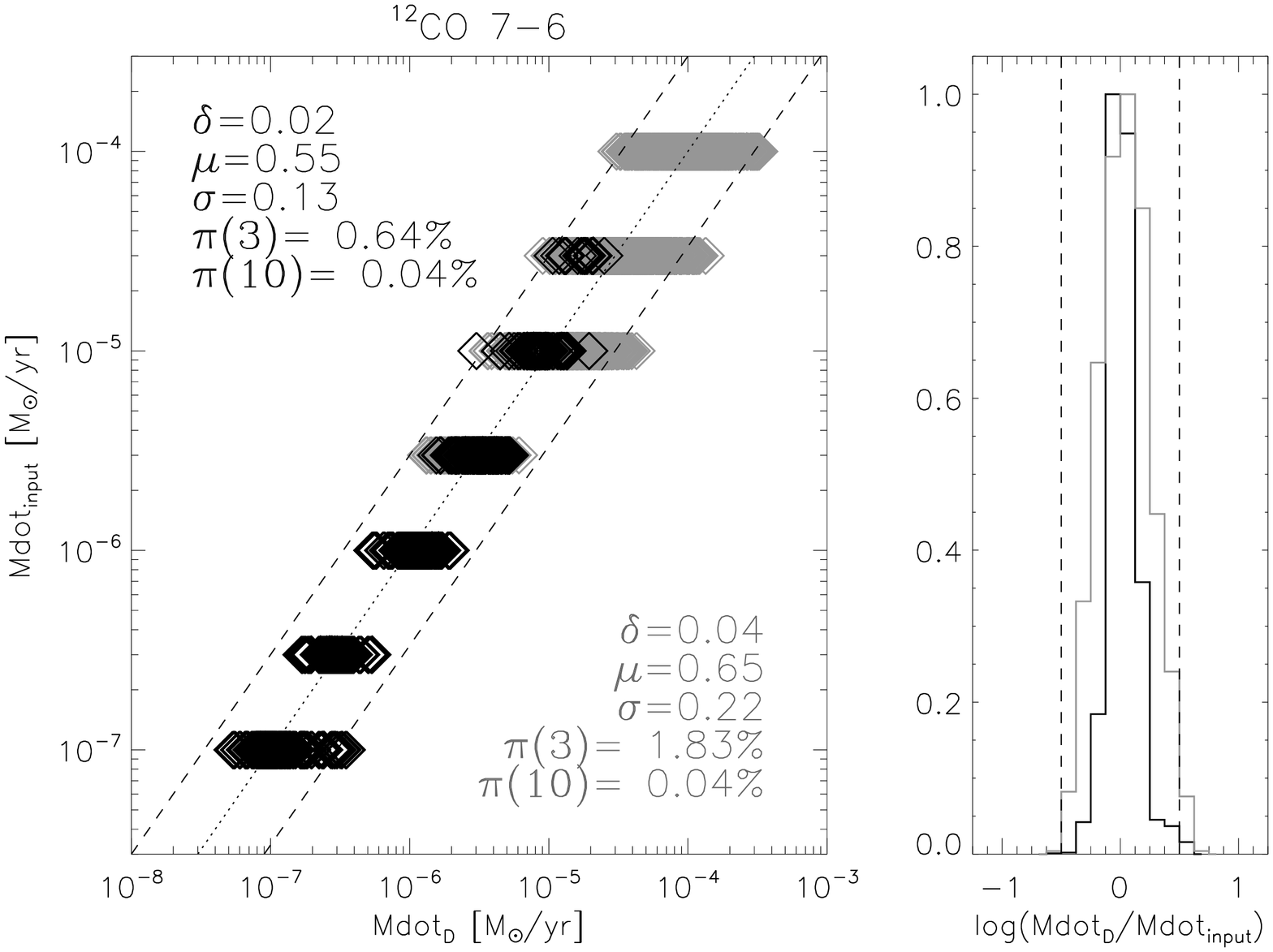}}
\caption{Comparison between the input value \mdotinput and \mdotd resulting from using Eq.~\ref{eq:ourmdotformula} for different rotational CO line transitions.  Estimates for the non-saturated regime are plotted in black, for the saturated regime in grey. The dotted line represents equality of the input and estimated mass-loss rates, the dashed lines show a factor 3 difference w.r.t. this relation. The listed $\delta$, $\sigma$, and $\mu$ are the mean, the standard deviations, and the maximum absolute values of $\log(\dot{M}_{\mathrm{D}}/\dot{M}_{\mathrm{input}})$, respectively. $\pi(3)$ and $\pi(10)$ show the percentage of grid models that produce values of $\dot{M}_{\mathrm{D}}$ deviating more than a factor 3, resp. 10, from $\dot{M}_{\mathrm{input}}$. The histogram on the right of each panel shows the peak-normalised distribution of $\log(\dot{M}_{\mathrm{D}}/\dot{M}_{\mathrm{input}})$. The vertical dashed lines in these histograms again show a factor three difference between  $\dot{M}_{\mathrm{D}}$ and $\dot{M}_{\mathrm{input}}$. The black and grey histograms represent estimates for the non-saturated, and saturated regimes, respectively. \label{Fig:Mdoteq}}
\end{figure*}

\begin{table*}\setlength{\tabcolsep}{1.6mm}
\centering
\caption{Values for the exponents $s_{i,J}$ in Eq.~\ref{eq:ourmdotformula} for each \twco($J \rightarrow J-1$) transition. The boundary values for saturation (see text for further details) are given in the third column. The $s_{i,J}$ are calibrated such that the mass-loss rate \mdotd is in units of \msun\,yr$^{-1}$. Standard deviations $\sigma$, maximum absolute values $\mu$, and mean values $\delta$ of $\log(\Mdot_{\mathrm{D}}/\dot{M}_{\mathrm{input}})$ are given in the last three columns.\label{exp_results}}
\begin{tabular}{rc|c|rrrrrrrrrr|ccc}
\hline
\hline\\[-2ex]
Regime & Transition & $I_{\mathrm{MB}} \times d^2$& $s_{1,J}$ & $s_{2,J}$ & $s_{3,J}$ & $s_{4,J}$ & $s_{5,J}$ & $s_{6,J}$ & $s_{7,J}$ & $s_{8,J}$ & $s_{9,J}$ & $s_{10,J}$  & $\sigma$ &  $\mu$ & $\delta$\\
           &              &   [K\,km\,s$^{-1}$\,$\times$\,pc$^2$]   &               &               &                &                &             &                &               &               &              &           		& 		&  &\\
\hline\\[-2ex]
  unsat. &   $1-0$ & $<1\times10^{7}$ &   $9.73\times10^{-16}$ &  1.25 &  0.31 & -1.78 & -1.29 &  1.03 &  0.33 &  0.10 &  0.21 & -0.01&0.16&0.76 &0.06\\
    sat. &  $1-0$ & $\geq1\times10^{7}$ &   $9.16\times10^{-9}$ &  0.78 &  0.50 & -1.22 & -0.34 &  0.44 & -0.66 &  0.28 &  0.37 & -0.03 &0.26&1.27&0.06\\
 unsat. &   $2-1$ & $<2\times10^{7}$ &   $1.00\times10^{-15}$ &  1.26 &  0.05 & -1.96 & -1.26 &  0.79 &  0.05 &  0.24 &  0.10 &  0.07 &0.09&0.58&0.01\\
    sat. &  $2-1$ & $\geq2\times10^{7}$ &   $9.04\times10^{-10}$ &  0.75 &  0.43 & -1.47 & -0.08 &  0.39 & -1.00 &  0.35 &  0.41 & -0.00 &0.28&0.86&0.07\\
 unsat. &   $3-2$ & $<9\times10^{7}$ &   $9.93\times10^{-14}$ &  1.14 &  0.13 & -1.92 & -1.39 &  0.78 & -0.25 &  0.40 &  0.29 & -0.11 &0.12&0.96&0.02\\
    sat. &  $3-2$ & $\geq9\times10^{7}$ &   $9.34\times10^{-10}$ &  0.61 &  0.31 & -1.67 & -0.00 &  0.36 & -1.00 &  0.37 &  0.43 &  0.01 &0.29&1.08&0.06\\
 unsat. &   $4-3$ & $<9\times10^{7}$ &   $9.69\times10^{-15}$ &  1.17 &  0.04 & -1.98 & -1.26 &  0.79 & -0.38 &  0.39 &  0.32 &  0.02 &0.09&0.57&0.01\\
    sat. &  $4-3$ & $\geq9\times10^{7}$ &  $ 9.30\times10^{-10}$ &  0.63 &  0.25 & -1.77 & -0.06 &  0.38 & -1.00 &  0.42 &  0.45 &  0.03&0.27&1.04&0.05 \\
 unsat. &   $6-5$ & $<1\times10^{8}$ &   $9.43\times10^{-14}$ &  1.12 &  0.01 & -1.99 & -1.19 &  0.75 & -0.48 &  0.46 &  0.42 & -0.04 &0.12&0.66&0.02\\
    sat. &  $6-5$ & $\geq1\times10^{8}$ &  $ 9.52\times10^{-10}$ &  0.69 &  0.15 & -1.86 & -0.22 &  0.49 & -1.00 &  0.52 &  0.46 & -0.11 &0.23&0.94&0.04\\
 unsat. &   $7-6$ & $<2\times10^{8}$ &   $9.44\times10^{-12}$ &  1.09 &  0.01 & -1.99 & -1.06 &  0.77 & -0.42 &  0.57 &  0.52 & -0.02 &0.13&0.55&0.02\\
    sat. &  $7-6$ & $\geq2\times10^{8}$ &  $ 9.44\times10^{-10}$ &  0.64 &  0.12 & -1.87 & -0.21 &  0.47 & -1.00 &  0.51 &  0.47 & -0.15 &0.22&0.65&0.04\\
 unsat. &  $10-9$ & $<2\times10^{8}$ &   $9.73\times10^{-10}$ &  0.98 &  0.00 & -2.00 & -0.72 &  0.68 & -0.47 &  0.63 &  0.60 &  0.01&0.13&0.45&0.02 \\
    sat. & $10-9$ & $\geq2\times10^{8}$ &  $ 9.41\times10^{-10}$ &  0.58 &  0.03 & -1.93 & -0.08 &  0.49 & -1.00 &  0.51 &  0.47 & -0.15 &0.24&0.67&0.05\\
 unsat. &  $16-15$ & $<2\times10^{8}$ &   $9.84\times10^{-10}$ &  0.87 &  0.01 & -1.98 & -0.57 &  0.61 & -0.85 &  0.65 &  0.61 &  0.02 &0.16&0.52&0.03\\
    sat. & $16-15$ & $\geq2\times10^{8}$ & $  9.20\times10^{-10}$ &  0.48 &  0.02 & -1.92 &  0.00 &  0.47 & -1.00 &  0.46 &  0.43 & -0.09 &0.31&0.92&0.10\\
\hline
 \end{tabular}
\end{table*}

The values in Table~\ref{exp_results} show that the dependence of the integrated intensity on the distance is strongest. For the unsaturated regime, the integrated intensity is quite sensitive on the terminal velocity and mass-loss rate, while this dependence is, as expected, much lower in the saturated regime. The intensity is moderately dependent on $\beta$, the stellar temperature, stellar radius, and dust condensation radius. 

\subsection{Grid results}\label{grid_results}
To estimate the minimisation function to the integrated intensities, the dependence of the integrated intensity $I^g$ on the different input parameters was inspected, and a power-law relation between the integrated intensity and all parameters was assumed. To estimate the outcome variable $I^{g,cc}$ we therefore have minimised the function
\begin{eqnarray}
\label{eq:minimisationfunction}
I^{\mathrm{D}}&=&\frac{1}{s_{1,J}} \dot{M}^{s_{2,J}} \theta_{\rm b}^{s_{3,J}} d^{s_{4,J}} v_{\infty}^{s_{5,J}} f_{\mathrm{H,CO}}^{s_{6,J}} \\\nonumber
		      &  &\qquad\times \; T_{\star}^{s_{7,J}} R_{\star}^{s_{8,J}} R_{\mathrm{inner}}^{s_{9,J}} \beta^{s_{10,J}}.
\end{eqnarray}
The minimisation was carried out for each \twco  line transition, using the generalized reduced gradient method \citep{lasdon1978}. The correlation between the \gastronoom output variable $I^{g,cc}$ and the integrated intensity $I^{\mathrm{D}}$  derived from Eq.~\ref{eq:minimisationfunction} is not bad, but a clear saturation effect is visible for all transitions when $I^{g,cc}$ exceeds a certain boundary value (see Fig.~\ref{intensity_CO21}). These boundary values are higher for higher $J$. A higher mass-loss rate not only leads to an increase of the cooling of the circumstellar gas, but also to a decrease in drift velocity between the dust and the gas \citep{decin2006_vycma}. This leads to a weaker dependence of the integrated intensity of the CO lines on \mdot, visible as saturation. Not accounting for this saturation effect typically leads to underestimating \mdot by a factor of $3-10$ for $\dot{M}\geq3\times10^{-5}$\,\msun\,yr$^{-1}$.
\\
Moreover, it is visible from Fig.~\ref{intensity_CO21} that the boundary values for saturation are distance dependent, with lower boundary values for larger distances. This dependence was removed by minimising Eq.~\ref{eq:minimisationfunction} again, but with both the left-hand and the right-hand side multiplied with a factor $d^2$. The coefficients $s_{i,J}$ for $J=1-0,\;2-1,\;3-2,\;4-3,\;6-5,\;7-6,\;10-9,\;16-15$ and the boundary values of $I_{\mathrm{MB}}\times d^2$ for saturation are listed in Table~\ref{exp_results}. 

Replacing $I^{\mathrm{D}}$ in Eq.~\ref{eq:minimisationfunction} with the observed integrated line intensity $I_{\mathrm{MB}}$ yields an estimate for the mass-loss rate, \mdotd, provided that the other parameters are known:
\begin{equation}
 \label{eq:ourmdotformula}
\dot{M}_{\mathrm{D}}=\left[\frac{  s_{1,J} \times I_{\mathrm{MB}}}{\theta_{\rm b}^{s_{3,J}} d^{s_{4,J}} v_{\infty}^{s_{5,J}} f_{\mathrm{H,CO}}^{s_{6,J}}T_{\star}^{s_{7,J}} R_{\star}^{s_{8,J}} R_{\mathrm{inner}}^{s_{9,J}} \beta^{s_{10,J}}}\right]^{\frac{1}{s_{2,J}}}.
\end{equation}

In Fig.~\ref{Fig:Mdoteq} we show a comparison between the mass-loss rate $\dot{M}_{\mathrm{input}}$ used as input for the grid models, and the mass-loss rate $\Mdot_{\mathrm{D}}$, calculated from the CO line intensities that result from the models, using Eq.~\ref{eq:ourmdotformula}. In Table~\ref{exp_results} and Fig.~\ref{Fig:Mdoteq} we listed the mean values $\delta$, standard deviations $\sigma$, and maximum absolute values $\mu$ of the quantity $\log(\Mdot_{\mathrm{D}}/\dot{M}_{\mathrm{input}})$. All listed $\delta$-values are positive, meaning that on average we slightly overestimate the input-mass-loss rate. This is however limited to overestimates of a few percents only, except for the cases of $1-0$ (15\,\%, $\delta$=0.06), and $2-1$ in the saturated regime (17\,\%, $\delta$=0.07). These deviations are still well within the standard deviation $\sigma$ and therefore not significant. In the panels of Fig.~\ref{Fig:Mdoteq} we also mention the quantities $\pi(3)$ and $\pi(10)$. These are the percentages of models in the grid that result in a value of $\Mdot_{\mathrm{D}}$ that deviates by more than a factor 3, resp. 10, from the input mass-loss rate.  The $\sigma$-values indicate uncertainties of a factor $\sim$1.3 for the unsaturated regime, and a factor $\sim$1.9 for the saturated regime when comparing $\Mdot_{\mathrm{D}}$ to $\dot{M}_{\mathrm{input}}$. When applying Eq.~\ref{eq:ourmdotformula} to data, the error bars on the measured parameters, e.g. $T_{\star}$, add to these uncertainties, increasing the typical uncertainty to a factor of three. 

The spread on the estimates reflects the interplay between the different stellar parameters and the line intensities, and the large parameter space covered by the grid. In the saturated regime the spread is somewhat larger than in the unsaturated regime, with standard deviations $\sigma$ ranging up to 0.16 and 0.31 in the unsaturated and saturated regimes, respectively. These values are still rather low, and are also reflected in the values of $\pi(3)$ and $\pi(10)$ that do not exceed $10\,\%$ and $0.45\,\%$ in the saturated regime, clearly showing the very small number of outliers present in our results.  We point out that we found no obvious relations between the (few) grid points showing the larger values of $|\log(\Mdot_{\mathrm{D}}/\dot{M}_{\mathrm{input}})|$. Therefore, we do not attach larger intrinsic uncertainties to specific input parameters. The quantities $\sigma,\,\mu,\,\delta,\,\pi$ indicate a high accuracy of the estimator for the complete range of mass-loss rates included in the grid (Table~\ref{table:parameters_grid}).

\subsection{Discussion on the derived mass-loss rate formulae}\label{subsect:formuladiscussion}
\subsubsection{Varying the number of free parameters}\label{subsubsect:varyingnumberofparameters}
\begin{figure*}
\centering
\subfigure{\includegraphics[height=.45\textwidth,angle=90]{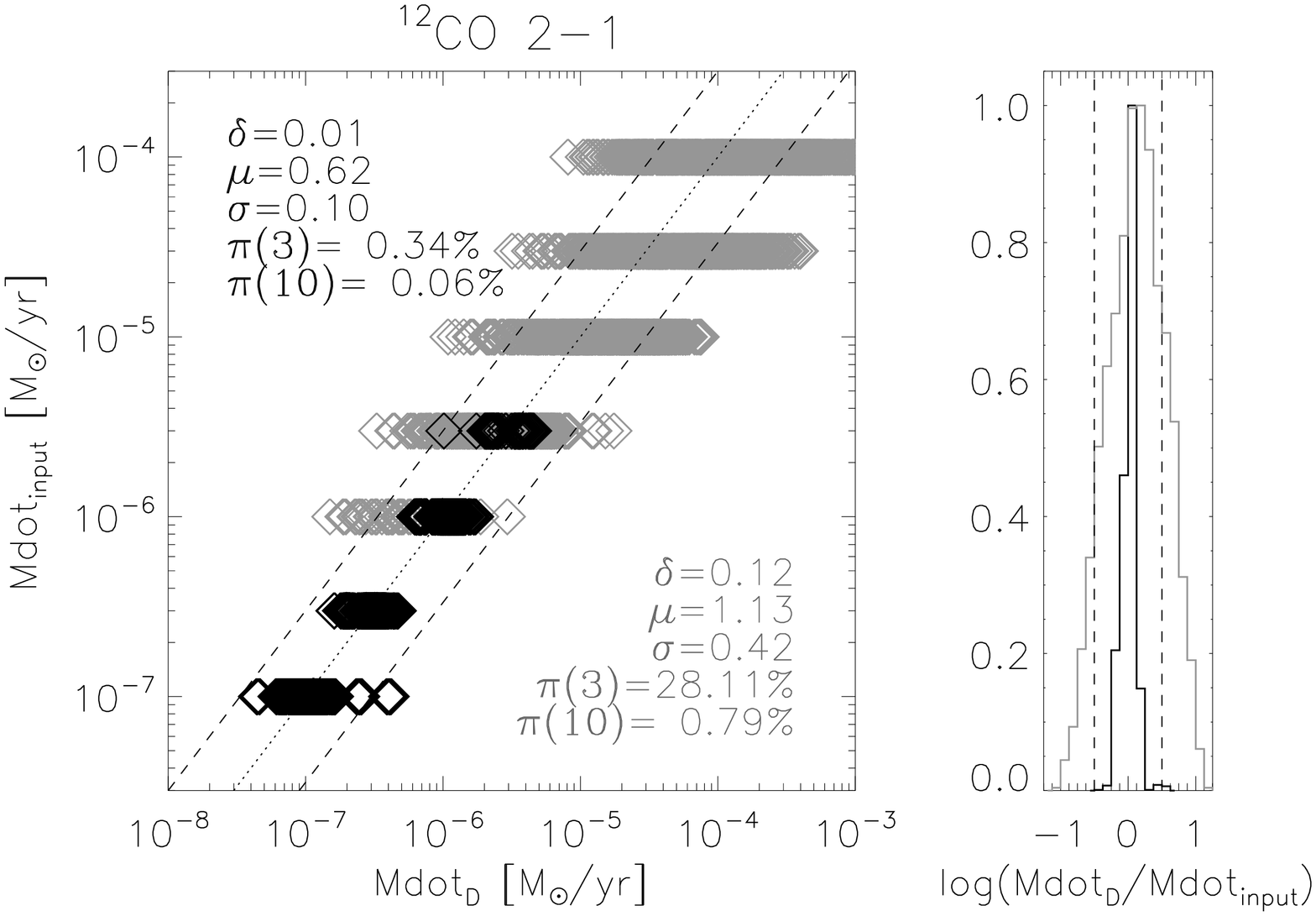}}\hspace{.3cm}\subfigure{\includegraphics[height=.45\textwidth,angle=90]{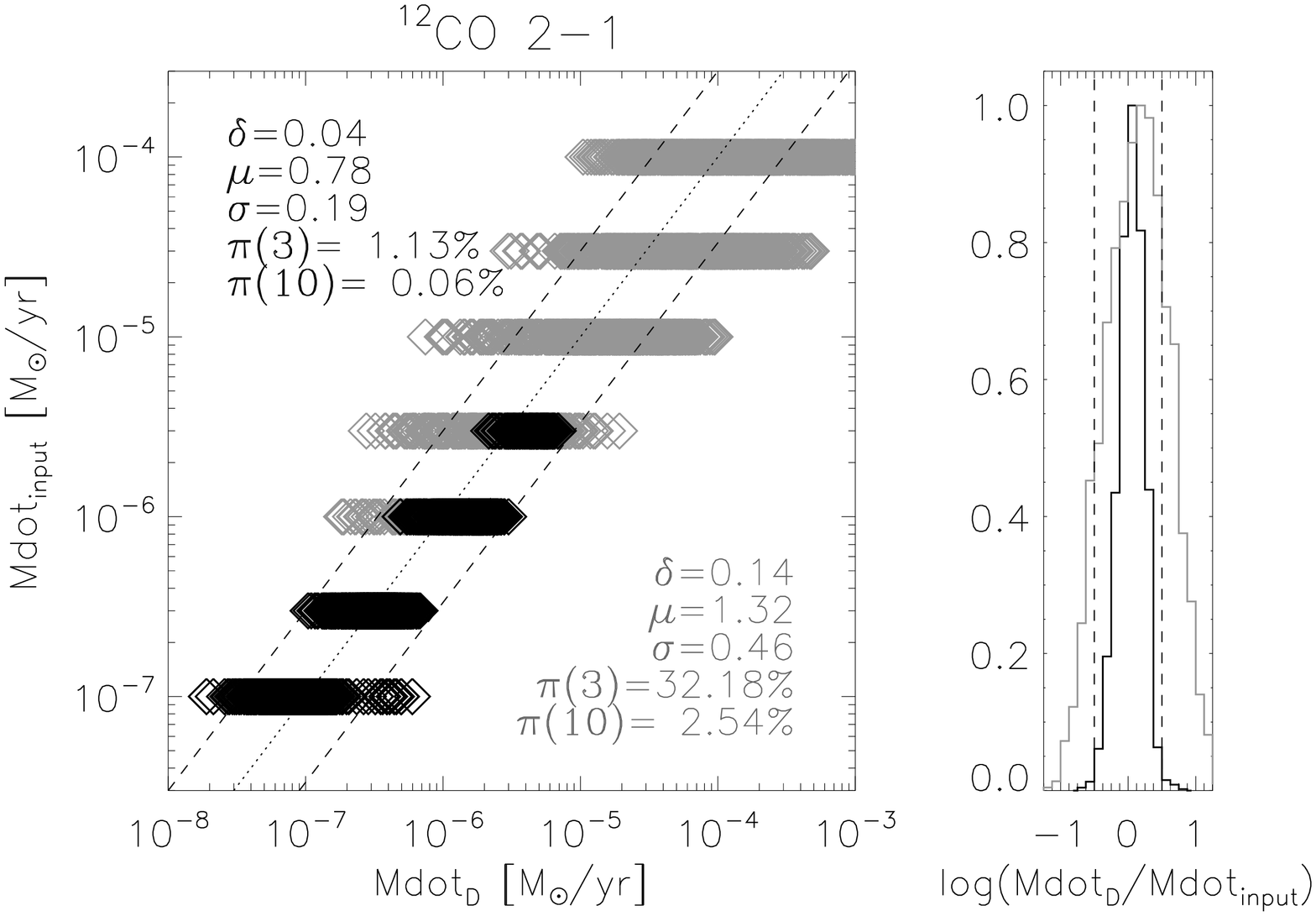}}\caption{Comparison between the input value \mdotinput (ordinate) and \mdotd resulting from using a formula analogous to  Eq.~\ref{eq:ourmdotformula} (abscissa) for different rotational CO line transitions, but omitting the dependence of the mass-loss rate on \Rinner\ and $\beta$ (left panel), and on \Rinner, $\beta$, \Tstar\,, and \Rstar\ (right panel). Estimates for the non-saturated regime are plotted in black, for the saturated regime in grey. $\delta,\,\sigma,\,\mu,\,\pi(3)$ and $\pi(10)$ are as in Fig.\ref{Fig:Mdoteq}. The dotted line represents equality of the input and estimated mass-loss rates, the dashed lines show a factor 3 difference w.r.t. this relation. The histogram on the right of each panel shows the peak-normalised distribution of $\log(\dot{M}_{\mathrm{D}}/\dot{M}_{\mathrm{input}})$. The vertical dashed lines in these histograms again show a factor three difference between  $\dot{M}_{\mathrm{D}}$ and $\dot{M}_{\mathrm{input}}$.
\label{Fig:lessparam}}
\end{figure*}

As was done in other studies \citep[e.g.][]{ramstedt2008}, one can lower the number of free parameters in Eq.~\ref{eq:minimisationfunction}. This is illustrated in Fig.~\ref{Fig:lessparam} for the $J=2-1$ transition. We show the estimates in two cases, omitting the dependence of the integrated intensity on $\beta$ and \rinner (left panel) and on $\beta$, \rinner, \tstar, and \rstar (right panel). Mainly for the mass-loss rates exceeding $10^{-5}$\,\msun\,yr$^{-1}$, the estimates exhibit larger uncertainties.  An overview of the values for the standard deviations $\sigma$, and maximum absolute values $\mu$ of $\log(\dot{M}_{\mathrm{D}}/\Mdot_{\mathrm{input}})$ is given in Table~\ref{tbl:lessparamstats} for the different cases. Both $\mu$ and $\sigma$ increase significantly when lowering the number of parameters. This implies that it is better to include a well-founded estimate for some parameter, like $\beta$ or \rinner, than to omit it from Eq.~\ref{eq:minimisationfunction}.

\begin{table}
\centering
\caption{Standard deviations $\sigma$ and maximum deviations $\mu$ of $\log(\dot{M}_{\mathrm{D}}/\dot{M}_{\mathrm{input}})$ for the different numbers of parameters included in the formalism. Values for both the unsaturated and saturated regimes for \twco($J=2-1$) are shown. \label{tbl:lessparamstats}}
 \begin{tabular}{l|cc|cc}
\hline
\hline\\[-2ex]
\twco($J=2-1$)					&\multicolumn{2}{c}{$\sigma$}	&\multicolumn{2}{c}{$\mu$}\\
						&unsat.	&sat.		&unsat.	&sat.\\
\hline\\[-2ex]
all parameters				&0.09	&0.28	&0.58		&0.86\\
no $\beta$, \rinner			&0.10		&0.42		&0.62	&1.13\\
no $\beta$, \rinner, \tstar,\rstar	&0.19		&0.46		&0.78	&1.32\\
\hline
\end{tabular}
\end{table}

\subsubsection{Sensitivity analysis}\label{subsubsect:sensitivityanalysis}
To further assess the quality of our estimator, we focus on the influence of \textit{(1)} input parameters for Eq.~\ref{eq:ourmdotformula}, such as effective temperature, luminosity, dust condensation radius, photospheric CO abundance, and distance, \textit{(2)} the envelope's outer radius, and \textit{(3)} the gas kinetic temperature.

\begin{figure*}
\includegraphics[angle=90,width=\linewidth]{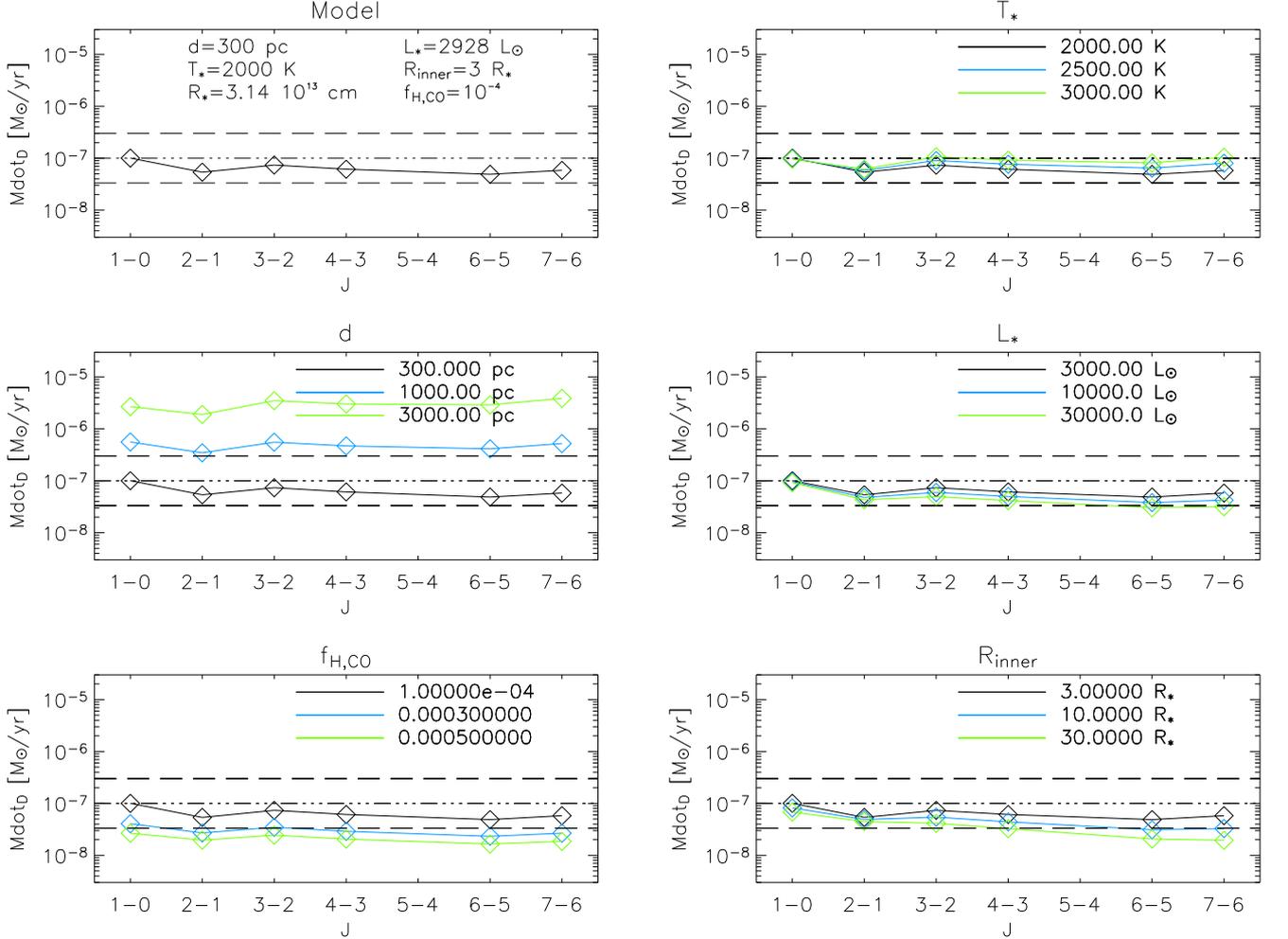}\caption{Sensitivity check of the used formalism for five parameters: $d$, \fhco, \tstar, \lstar (\rstar), and \rinner. The upper left panel gives the estimates based on the input parameters of the model and the output values of the integrated intensities $I^{\mathrm{g,cc}}$. The grid values for the different parameters and the colour coding are specified in the legend of every panel. The dash-dotted line in all panels marks the model's input-\mdot, $10^{-7}$\,\msun\,yr$^{-1}$, the dashed lines mark a factor three difference w.r.t. this value.  \label{fig:sensitivity_parameters}}
\end{figure*}

\paragraph{Input parameters:}
We tested the influence of $d$, \tstar, \lstar (\rstar), \fhco and \rinner on the \mdot-estimates for $J=1-0,\,2-1,\,3-2,\,4-3,\,6-5,\,7-6$. We assumed a standard stellar model with \mdotinput$=10^{-7}$\,\msun\,yr$^{-1}$, \tstar$=2000$\,K, \rstar$=3.14\times10^{13}$\,cm, \rinner$=3$\,\rstar, \fhco$=10^{-4}$, $d=300$\,pc, and a fixed beamwidth $\theta_{\mathrm{b}}=10$\,\arcsec. The integrated intensities $I_{\mathrm{MB}}$ are those calculated with the \gastronoom-code for this model. In Fig.~\ref{fig:sensitivity_parameters} we show the influence of the different parameters on the \mdot-estimates by varying $d$, \fhco, \tstar, \lstar and \rinner one by one while keeping all others constant at the model values. We point out that a slight gradient is present in the \mdot-estimates for the standard model, i.e. for higher $J$ we seem to estimate somewhat lower values of \mdot. The deviations and the spread of the estimates are, however, well within a factor three from the input mass-loss rate. Since the latter is used as a criterion to decide upon constancy of \mdot, we can state that the estimator can very well reproduce the input-\mdot. 

As expected, the distance has the strongest influence on the estimated \mdot-values, with a similar increase or decrease in mass-loss rate for all transitions. Altering \fhco has a pure scaling effect on the estimates, with a lower \mdot for a higher \fhco. Changing \tstar, \lstar, or \rinner leads to the largest differences for the higher-$J$ transitions, since changing these parameters affects the layers closest to the star more strongly than it does the more outward layers. This effect is reflected in the larger absolute values of exponents $s_{7,J},\,s_{8,J}$ and $s_{9,J}$ in Eq.~\ref{eq:ourmdotformula}, and by the fact that the mentioned gradient is slightly shallower or steeper when considering other values for \tstar, \lstar, or \rinner. This implies that the estimates based on the higher-$J$ transitions suffer from larger uncertainties and that more value can be attached to the estimates based on the lower-$J$ transitions.

Inspecting the different panels in Fig.~\ref{fig:sensitivity_parameters}, we find no indications for changes in the \mdot-estimates with an opposite character for the different transitions, in the sense that, e.g., the estimates for some $J$ would go lower, while those for other transitions go higher. For a constant mass-loss rate we will therefore only see a gradient (if any) and not a random distribution of the estimates. This implies that a faulty parameter assumption will only lead to a general shift of the estimates and is unlikely to change the interpretation of possible variability or constancy of the mass-loss rate.

\paragraph{Outer envelope radius:}
The influence of the outer radius of the CSE, \router, can not be tested in the same way since it is not an input parameter in Eq.~\ref{eq:ourmdotformula}. This variable has therefore no direct influence on the estimates. It did however influence the derivation of the coefficients $s_{i,J}$, since the geometrical extent of the envelope affects the integrated intensities and the shapes of the emission lines. The transition most sensitive to changes in \router is $J=1-0$, implying that it is a safer choice to use \mdot-estimates based on the $J=2-1$ transition.

Clumping of CSE material causes a larger \router than a smooth mass outflow because of shielding properties of the clumps. This implies that higher integrated intensities can be reached with the same mass-loss rate. Therefore, if clumping is present, we could be overestimating the mass-loss rate. This will again be most strongly reflected in the $J=1-0$ estimates.

\begin{figure}
\centering\subfigure[Gas kinetic temperature $T$ versus radius $r$ for \mdot=$10^{-7}$\,\msun\,yr$^{-1}$ (full line), and $10^{-5}$\,\msun\,yr$^{-1}$ (dotted line).
\label{subfig:sensitivity_tkin}]{\includegraphics[width=\linewidth]{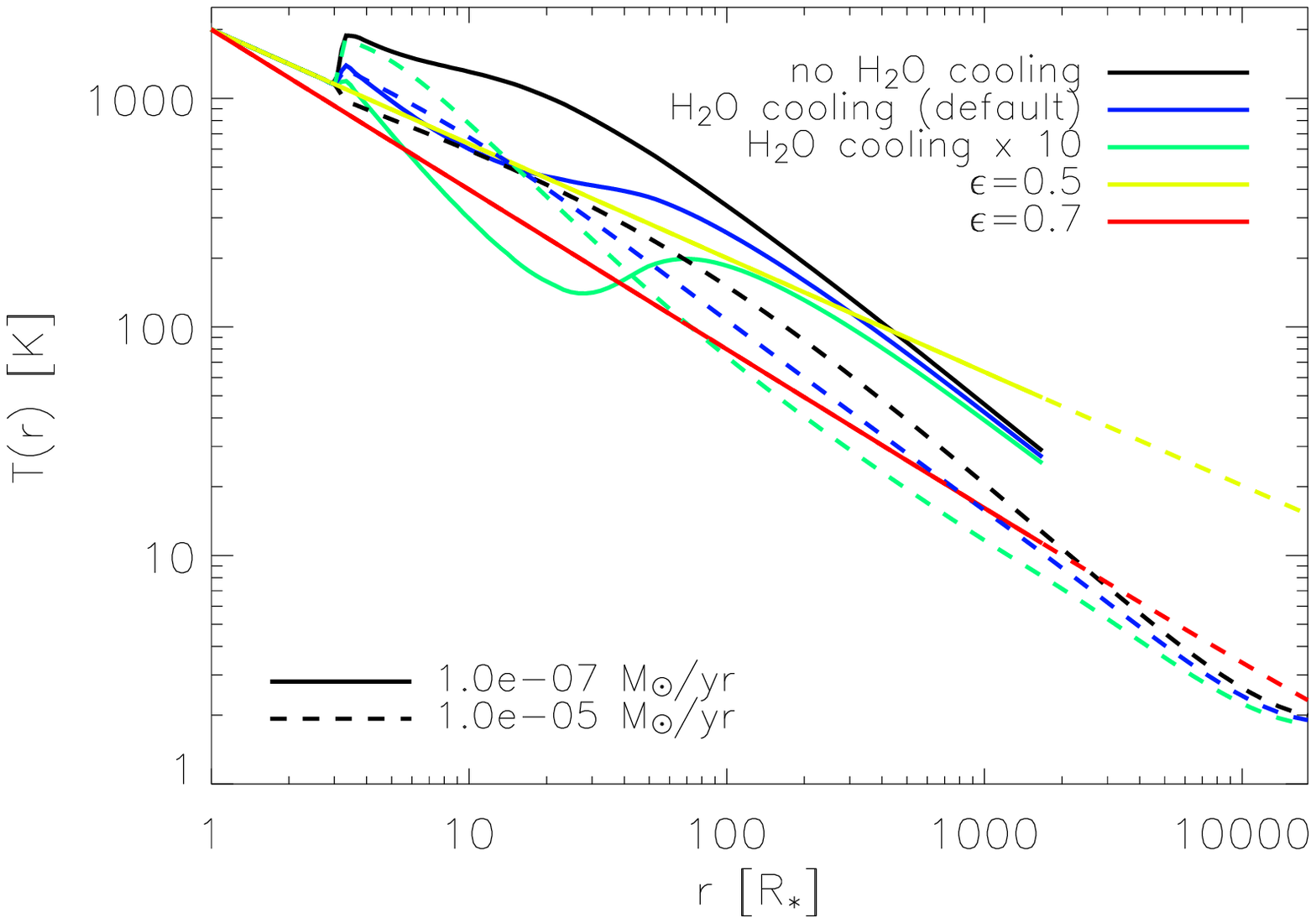}}
\centering\subfigure[Integrated intensities $I^{g,cc}$ for transitions $J=1-0$ up to $J=7-6$.\label{subfig:sensitivity_intensity}]{\includegraphics[width=\linewidth]{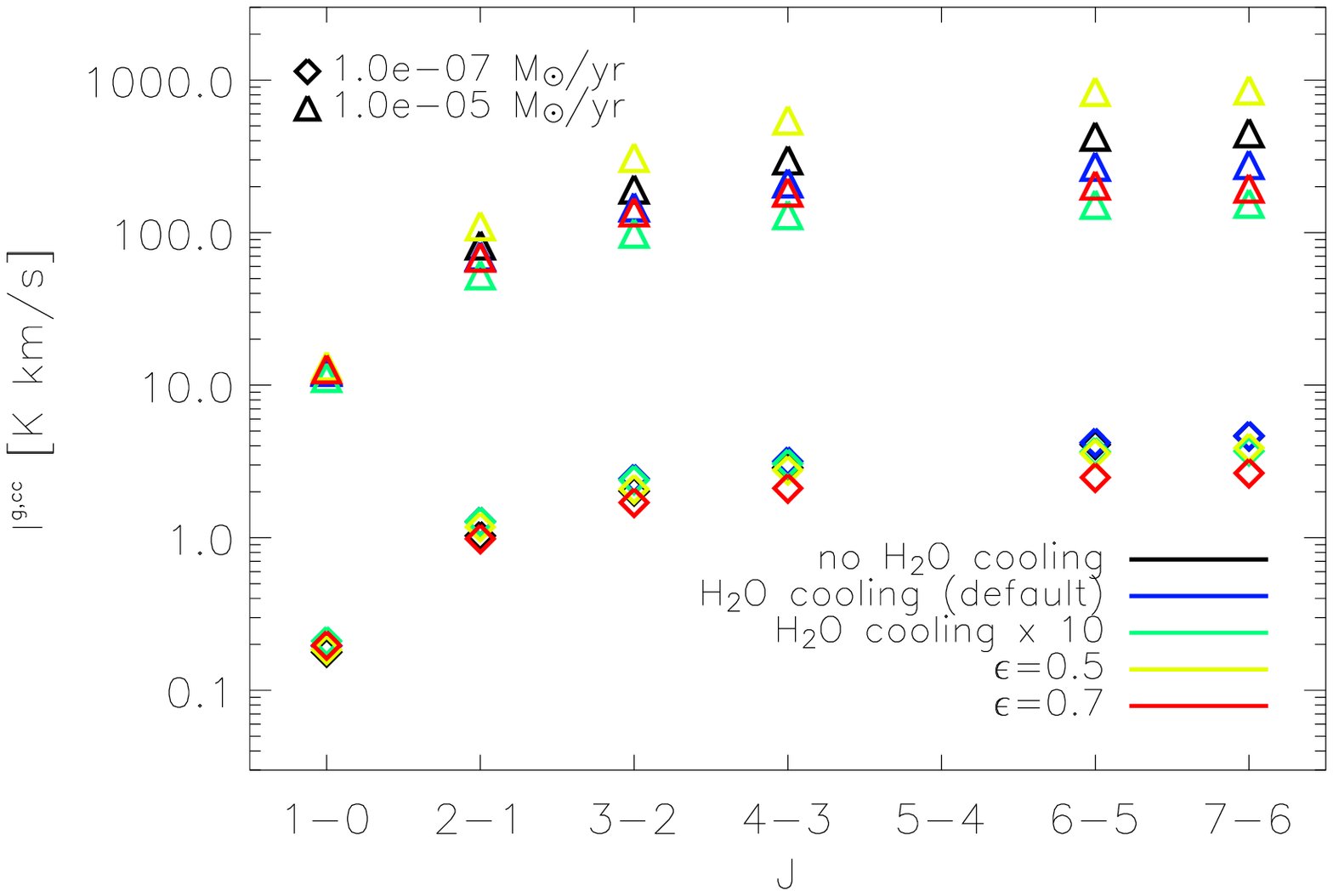}}
\caption{Gas kinetic temperatures (a) and integrated intensities (b) for different \mdotinput-values and different cooling mechanisms. The colour coding in both panels is as follows: black was used for omission of \h2o-cooling, blue for inclusion of \h2o-cooling, green for inclusion of \h2o-cooling ten times stronger than by default, yellow for a temperature power law, as in Eq.~\ref{eq:temppowerlaw}, with $\epsilon=0.5$, and red for $\epsilon=0.7$. In the lower panel, the different values of the input-mass-loss rate are indicated with diamonds and triangles for \mdot=$10^{-7}$\,\msun\,yr$^{-1}$ and $10^{-5}$\,\msun\,yr$^{-1}$, respectively. See discussion in Sect.~\ref{subsubsect:sensitivityanalysis}. \label{fig:sensitivityanalysis}}
\end{figure}

\paragraph{Gas kinetic temperature:}
In Fig.~\ref{fig:sensitivityanalysis} we compare gas kinetic temperatures $T$, which have been consistently computed with the \gastronoom-code, and integrated intensities $I^{g,cc}$ of transitions $J=1-0$ up to $J=7-6$ for two different values of \mdotinput ($10^{-7}$\,\msun\,yr$^{-1}$ and $10^{-5}$\,\msun\,yr$^{-1}$) and five different effects in cooling. All other model parameters are the same as those mentioned above for the standard model. The default cooling scheme, which was used to derive the formalism presented in this paper, includes among others cooling due to rotational excitation of \h2o. The cooling schemes we considered \textit{(1)} ignore the \h2o-cooling, \textit{(2)} include \h2o-cooling (default), \textit{(3)} include \h2o-cooling ten times stronger than assumed by default, \textit{(4)} assume a power law for the gas kinetic temperature 
\begin{equation}
 \label{eq:temppowerlaw}
T(r)=\Tstar\left(\Rstar/r\right)^{\epsilon}
\end{equation}
with $\epsilon=0.5$, and \textit{(5)} assume a power law with $\epsilon=0.7$.

We point out that, except for the cases in which a power law was used to describe $T(r)$, the temperature can increase or decrease with increasing radius $r$, i.e., instead of cooling there can be heating of the envelope; see Fig.~\ref{fig:sensitivityanalysis}. 

We focus on the \h2o rotational cooling since this part of the cooling is still subject to large uncertainties. In \gastronoom, \h2o is modelled as a three-level system, which is a gross simplification of the actual structure of the molecule. The rotational rate coefficients used here are based on the results of \cite{green1993}, as are those of \cite{neufeld1993}. \cite{faure2008} recently calculated new rotational rates and compared them to those of \cite{neufeld1993}, finding higher cooling rates for temperatures $\lesssim 1500$\,K and lower cooling rates for temperatures $\gtrsim 1500$\,K. They also found that the rotational rate coefficients can be uncertain by factors of a few up to an order of magnitude, which is reflected in the uncertainties on the cooling rates.

The contribution of \h2o-cooling is most important in the inner wind regions, since farther out in the envelope CO-rotational cooling and especially cooling due to adiabatic expansion dominate. This implies that the gas kinetic temperature in the inner wind regions is more uncertain than in the rest of the envelope. 

As is visible in Fig.~\ref{subfig:sensitivity_intensity}, $I^{g,cc}$ decreases for increasing \h2o-cooling. Again, this effect is more explicit for higher $J$. Overestimating the \h2o-cooling therefore leads to an underestimate of the mass-loss rate, considering the relation between the integrated intensities and the \mdot-estimates.

Since the excitation regions of the higher-$J$ transitions are partly situated in the inner wind regions, the integrated intensities of these transitions will be subject to larger uncertainties than those of the low-$J$ transitions $J=1-0$ and $2-1$. This different degree of influence on the integrated intensities for the different $J$-values is very clear in Fig.~\ref{subfig:sensitivity_intensity}, where the spread on the integrated intensities of the transitions $J\rightarrow J-1$ increases for higher $J$. A second implication is that the mentioned gradient of (slightly) lower \mdot for higher $J$-values is possibly linked to the uncertainties on the \h2o-cooling in the inner regions. If this is indeed the case, then the \h2o-cooling rates used in our models are overestimates of the actual rates.

Considering the current models, the uncertainties on the intensities of higher-$J$ transitions (as discussed above), and taking into account that the intensities of the $J=1-0$ transition can be affected by e.g. masering and clumping, we put forward that the $J=2-1$ transition is likely the most reliable transition to use in estimating mass-loss rates.

\begin{figure*}
\subfigure{\includegraphics[height=.5\textwidth,angle=90]{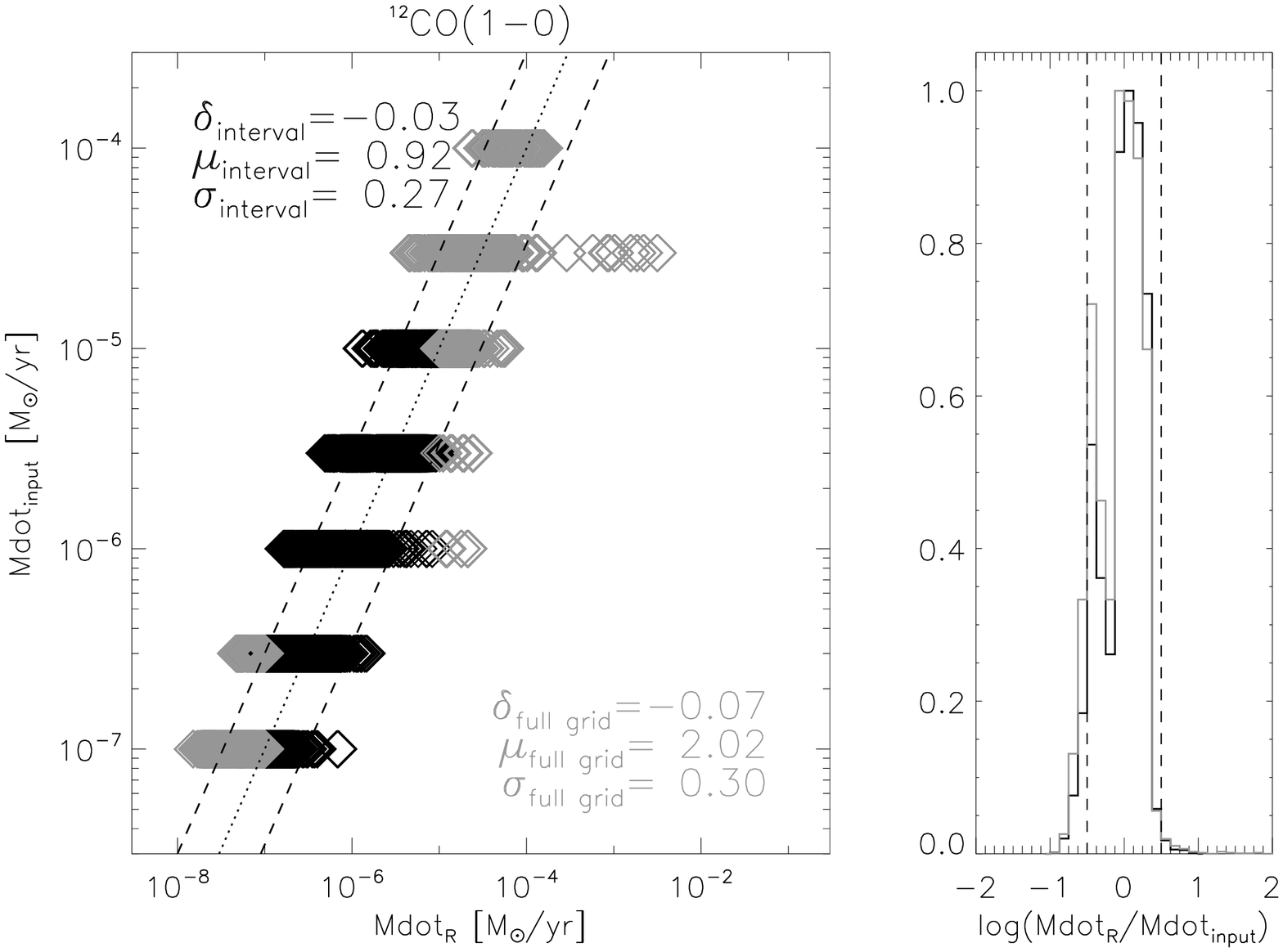}} \hspace{.1\linewidth} \hspace{-0.1\linewidth}\subfigure{\includegraphics[height=.5\textwidth,angle=90]{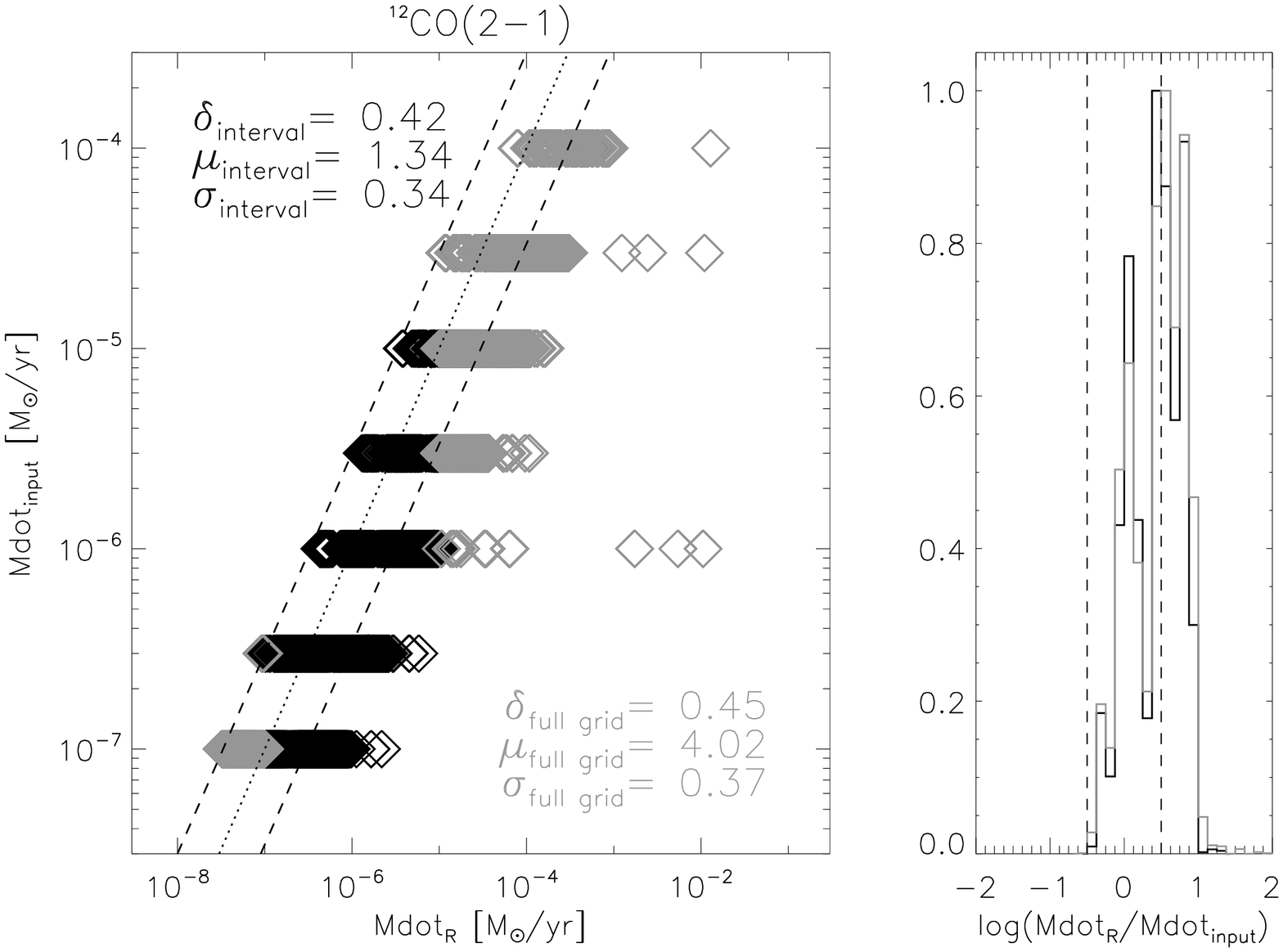}}
\subfigure{\includegraphics[height=.5\textwidth,angle=90]{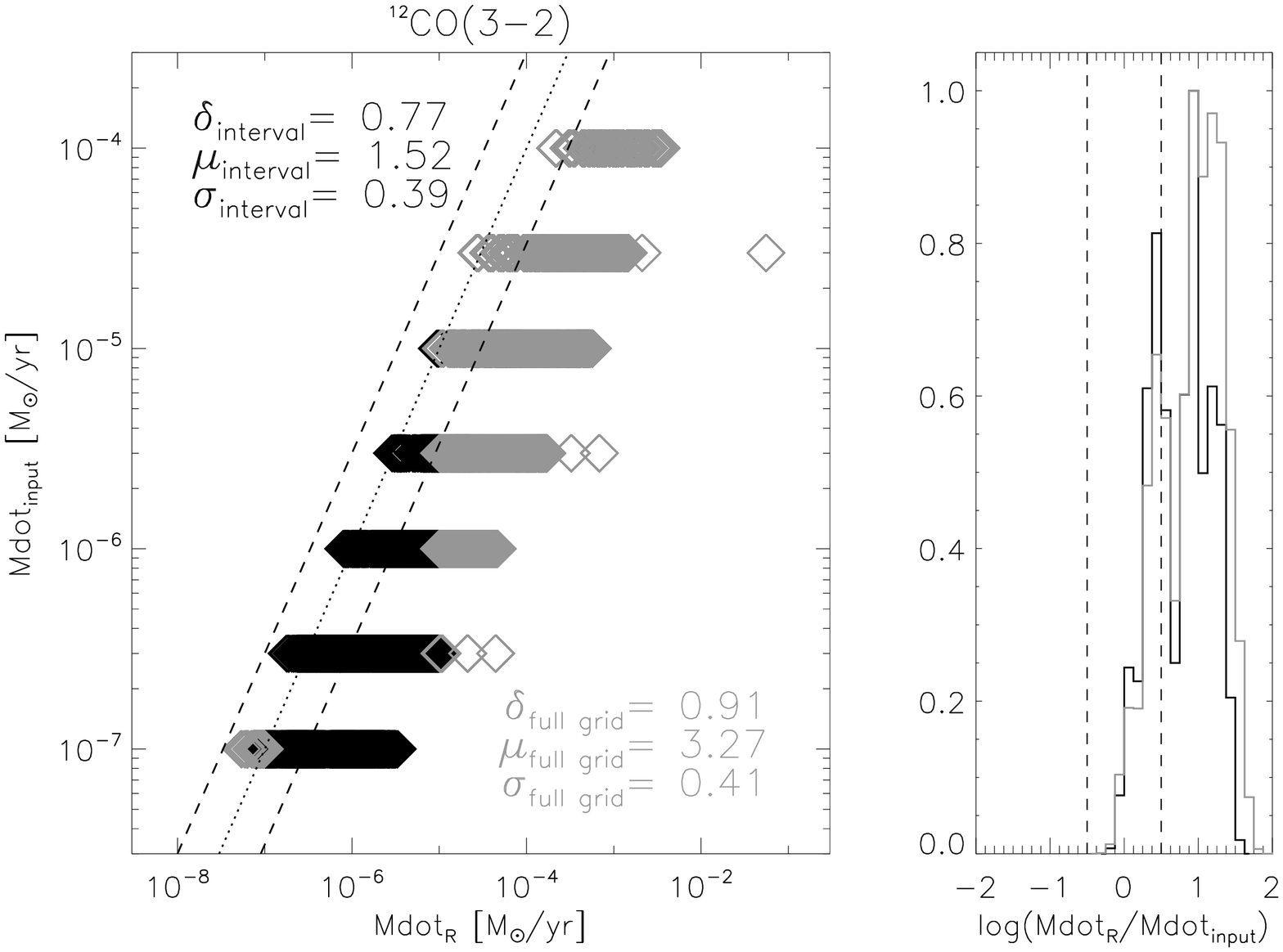}} \hspace{.1\linewidth} \hspace{-0.1\linewidth}\subfigure{\includegraphics[height=.5\textwidth,angle=90]{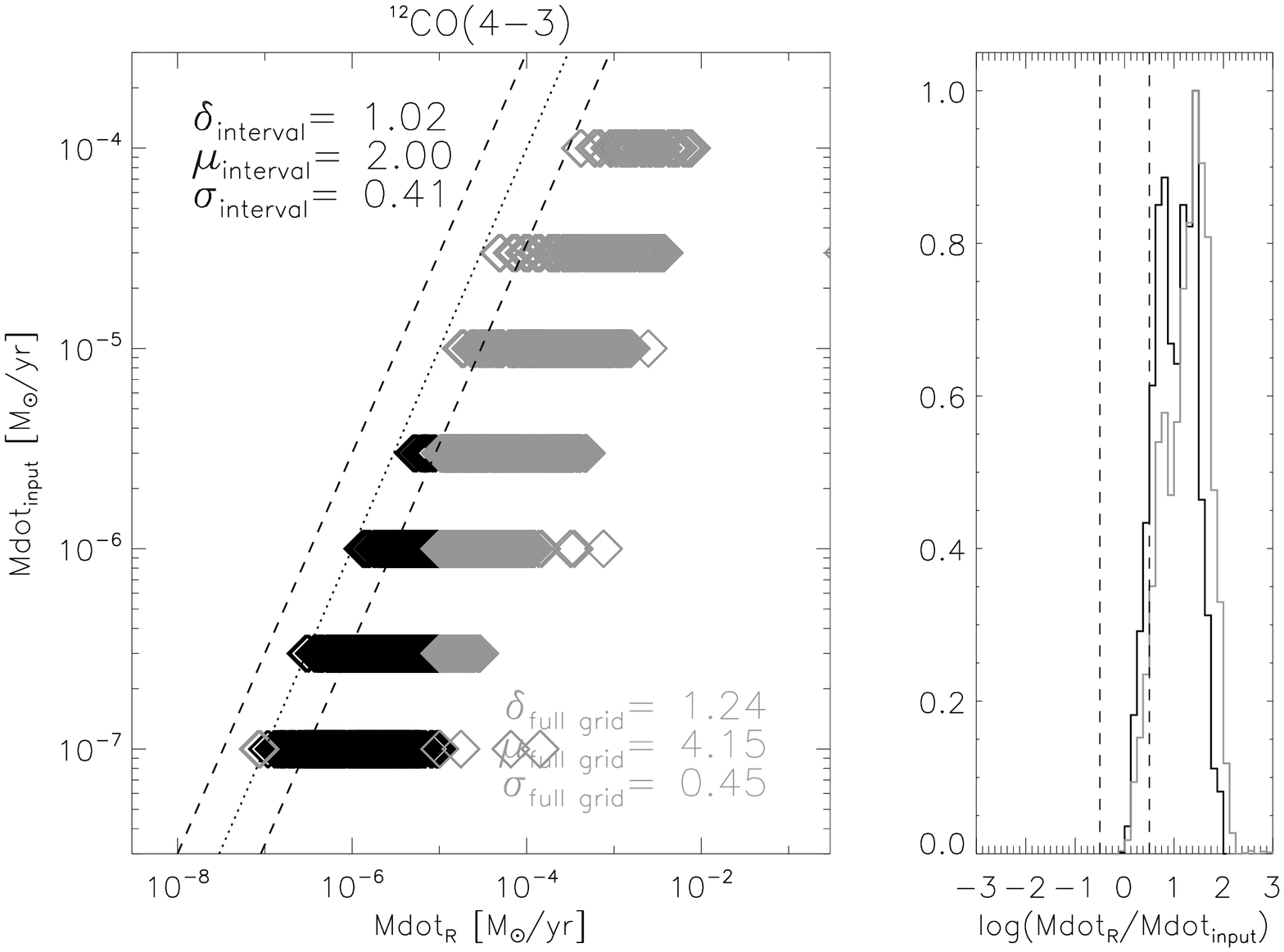}}
\caption{\mdotinput versus \mdotr calculated from the grid also used to construct Fig.~\ref{Fig:Mdoteq} and Eq.~\ref{eq:mdotramstedt2008}. The dotted line indicates where the quantities are equal, the dashed lines show a factor three difference w.r.t. this relation. The listed $\delta,\,\mu$, and $\sigma$ are, respectively, the mean values, the standard deviations, and the maximum absolute values of $\log(\dot{M}_{\mathrm{R}}/\dot{M}_{\mathrm{input}})$. We specify these quantities for the full grid (plotted in grey), and for the grid points inside the interval $[10^{-7};\,10^{-5}]$\,\msun\,yr$^{-1}$ (plotted in black). The histogram in each panel shows the peak-normalised distribution of $\log(\dot{M}_{\mathrm{R}}/\dot{M}_{\mathrm{input}})$. \label{fig:ramstedt_v_input}}
\end{figure*}

\subsubsection{Comparison to \cite{ramstedt2008}} \label{subsubsect:compwithramstedt}
In benchmarking our formalism against mass-loss rate estimators presented in the literature, we focus on the one derived by \cite{ramstedt2008}. A direct comparison to the estimators of \cite{knapp1985} or \cite{loup1993} is not possible, since these have the main-beam temperature at the line centre, $T_{\mathrm{MB,c}}$, as a fundamental parameter, while we consider the integrated intensity $I_{\mathrm{MB}}$ and the shape of the lines. 

In Fig.~\ref{fig:ramstedt_v_input} we show the mass-loss rate estimates derived for transitions $J=1-0,\; 2-1,\; 3-2$ and $4-3$ via Eq.~\ref{eq:mdotramstedt2008}, using the same parameter grid as was used to derive the formalism presented in this paper. \cite{ramstedt2008} mention that their estimator $\dot{M}_{\mathrm{R}}$ is only valid for mass-loss rates  between  $10^{-7}$ and $10^{-5}$\,\msun\,yr$^{-1}$, so we marked those outside this interval in grey, for the sake of the clarity. \cite{ramstedt2008} also note the requirement of unresolved envelopes to use their estimator to high accuracy, and we therefore eliminated the grid points that result in resolved envelopes by determining the CO-photodissociation radius, $R_{\mathrm{CO}}$, and subsequently the outer radius, $R_{\mathrm{outer}}$, of the envelope according to \cite{stanek1995} and \cite{schoeier2001}: 
\begin{eqnarray}
\label{eq:photodissociationradius}
 R_{\mathrm{CO}}&=&5.4\times10^{16}\left(\frac{\Mdot}{10^{-6}}\right)^{0.65} \nonumber\\
&&\times\left(\frac{15}{\Vinfty}\times\frac{f_{\mathrm{H,CO}}(\Rstar)}{4\times10^{-4}}\right)^{0.55} \nonumber\\
&&+7.5\times10^{15}\left(\frac{\Vinfty}{15}\right)\mathrm{\,cm}\\
\alpha&=&2.79\times\left(\frac{\Mdot}{10^{-6}}\times\frac{15}{\Vinfty}\right)^{0.09}\\
f_{\mathrm{H,CO}}(r)&=&f_{\mathrm{H,CO}}(\Rstar)\times\exp\left[-\ln(2)\times\left(\frac{r}{R_{\mathrm{CO}}}\right)^{\alpha}\right]
\end{eqnarray}
The CO-photodissociation radius and the outer radius of the envelope are the radii at which the CO abundance has dropped to, respectively, 50\,\% and 1\,\% of its photospheric value, $f_{\mathrm{H,CO}}(\Rstar)$. $\alpha$ describes the abundance profile throughout the envelope. We adopted the \mdotinput-value to calculate $R_{\mathrm{CO}}$ and $R_{\mathrm{outer}}$. When the angular size of the envelope with radius \router exceeded the beam size $\theta_{\rm b}$, we excluded the respective grid point.

The mean values $\delta$, standard deviations $\sigma$, and maximum absolute values $\mu$ of $\log(\dot{M}_{\mathrm{R}}/\dot{M}_{\mathrm{input}})$ are printed in each panel of Fig.~\ref{fig:ramstedt_v_input}, both for the full grid and for only those grid points inside the interval $[10^{-7};\,10^{-5}]$\,\msun\,yr$^{-1}$. Very high values of $\mu$ are reached when the complete grid is considered. Confining the statistics to the interval $[10^{-7};\,10^{-5}]$\,\msun\,yr$^{-1}$, we find that these values are much lower. For all transitions $\mu$ and $\sigma$ are higher than the values mentioned in Table~\ref{exp_results}, derived for our estimator. There is also a significant shift to higher \mdot-values, for all but the $J=1-0$ transition, using the estimator of \cite{ramstedt2008} in Eq.\,\ref{eq:mdotramstedt2008}. This shift is indicated by $\delta$-values significantly deviating from zero ($\delta>\sigma$) and is expected due to the differences in the temperature structure adopted by us and by \cite{ramstedt2008}. The uncertainties on the cooling, especially by rotational excitation of H$_2$O, are still rather large, especially so in the inner wind region, as discussed in Sect.~\ref{subsubsect:sensitivityanalysis}. Furthermore, our approach utilises a higher number of parameters and expands the validity range of the \mdot-estimator by correcting for saturation of the CO lines, ensuring that we get small intrinsic uncertainties for our estimator.


\section{Stellar parameters} \label{sect:stellarparameters}
\subsection{Temperature, luminosity and distance} Since we want to obtain empirical relations between wind parameters (e.g. \mdot, \vinfty), molecular-line parameters (e.g. line intensity), and basic stellar parameters (e.g. \tstar, \lstar), the latter need to be well determined. To draw meaningful conclusions from sample statistics, a highly homogeneous determination of basic stellar parameters is needed. Effective temperatures were derived from the dereddened \VmK colour, and luminosities from period-luminosity relations. A description of the methods used to obtain these parameters is given in Appendix~\ref{sect:basicstellarparameters}. The results for the sample are given in Table~\ref{tbl:fundamentalparameters}, together with the evolutionary stage, chemical type and pulsational type of the objects.

\subsection{Inner envelope radius}
As discussed in Sect.~\ref{sect:radtrananalysis}, the inner radius of the circumstellar envelope, \rinner, is used in determining the mass-loss rate. In estimating \mdot for the sample targets, \rinner is calculated assuming that the dust temperature $T_{\mathrm{d}}$ at this radius is equal to the condensation temperature, and that
\begin{equation}
\label{eq:dusttemp}
T_{\mathrm{d}}(r)=\Tstar\left(\frac{\Rstar}{2r}\right)^{2/(4+s)}
\end{equation}
with $s\approx1$ \citep[Olofsson in ][]{habing2003}. The condensation temperature is taken to be 1500\,K for O-rich and S-type stars and 1200\,K for C-rich stars.

\begin{table}\centering
\caption{Overview of sample targets for which no mass-loss rate estimates were made. The third column lists the evolutionary stage of each target.\label{tbl:overview_no_estimates}}
 \begin{tabular}{llc}
\hline\hline\\[-2ex]
  IRAS & Other identifier & Evolutionary stage\\
\hline \\[-2ex]
06176-1036	&	Red Rectangle		&	P-AGB \\
07399-1435	&	Calabash Nebula		&	P-AGB\\
10197-5750	&	GSC 08608-00509		&	P-AGB\\
13428-6232	&	GLMP 363		&	P-AGB\\
16262-2619	&	Alpha Sco		&	RSG\\
17150-3224	&	Cotton Candy Nebula	&	P-AGB \\
17443-2949	&	PN RPZM 39		&	P-AGB\\
17501-2656	&	V4201 Sgr		&	AGB\\
18059-3211	&	Gomez Nebula		&	YSO\\
18100-1915	&	OH 11.52 -0.58		&	AGB\\
18257-1000	&	V441 Sct		&	AGB\\
18308-0503	&	AFGL 5502		&	YSO\\
18327-0715	&	OH 24.69 +0.24		&	AGB\\
18361-0647	&	OH 25.50 -0.29		&	AGB\\
18432-0149	&	V1360 Aql		&	AGB\\
18460-0254	&	V1362 Aql		& 	AGB\\
18488-0107	&	V1363 Aql		&	AGB\\
18498-0017	&	V1365 Aql		& 	AGB\\
19067+0811	&	V1368 Aql		&	AGB\\
19110+1045	&	KJK G45.07		&	HII-region\\
\hline
 \end{tabular}
\end{table}

\longtabL{6}{\small
\begin{landscape}
\begin{longtable}{llrcrrrrrcrcrrr}
\caption[Fundamental stellar parameters.]{Overview of the evolutionary stage, atmospheric chemistry, pulsational type, spectral type, and stellar parameters of the sample stars. Spectral types were taken from (i) the SIMBAD database, (ii) \cite{kwok1997}, (iii) \cite{feast2000}, (iv) \cite{justtanont1996}, (v) \cite{suarez2006}, and (vi) \cite{castro-carrizo2007}. For each star the following fundamental parameters are specified: distance $d$, period of pulsation $P$, effective temperature \teff, luminosity \lstar, and stellar radius \rstar. \teff and \lstar were determined according to the methods presented in Appendix~\ref{sect:basicstellarparameters}. \\References to the values adopted from the literature are:
 (1)	\cite{massey2006},
 (2)    \cite{schoeier2007},
 (3)    \cite{levesque2007},    
 (4)    \cite{matsuura2002},     
 (5)  \cite{decin2008_parent}, 
 (6)   \cite{teyssier2006},       
 (7)    	\cite{whitelock1985},   
 (8)   \cite{levesque2005},     
 (9)   	\cite{schoeier2005},    
 (10)  	\cite{faundez2004},      
 (11)  	\cite{dejager1998},    
 (12)  	\cite{likkel1991}, 
 (13)  	\cite{groenewegen1999}.    
$(V-K)_0$ is the $V-K$ colour that was dereddened according to the method discussed in Appendix~\ref{subsect:dereddening}. $K_{\mathrm{mean}}$ is the mean K-band magnitude. }\label{tbl:fundamentalparameters}\\
\hline \hline \\[-2ex]
IRAS 	&OTHER 	&Evol.  	&Chem. 	&Puls. 	&Sp. Type 	&$d$ 	&$P$ 	&\teff 	&Ref.&\lstar 	&Ref.		&\rstar 	&$(V-K)_0$ 	&$K_{\mathrm{mean}}$ \\
 	& 		&Type 	& 		&Type	& 		&(pc) 	&(days) 	&(K) 		&	&(\lsun) 	&		&(\rsun) 	&\multicolumn{2}{c}{(mag)} \\
\hline \\[-2ex]
\endfirsthead
\multicolumn{3}{c}{\textbf{{\tablename} \thetable{}.} Continued} \\[0.5ex]
\hline \hline \\[-2ex]
IRAS 	&OTHER 	&Evol.  	&Chem. 	&Puls. 	&Sp. Type 	&$d$ 	&$P$ 	&\teff 	&Ref.&\lstar 	&Ref.		&\rstar 	&$(V-K)_0$ 	&$K_{\mathrm{mean}}$ \\
 	& 		&Type 	& 		&Type	& 		&(pc) 	&(days) 	&(K) 		&	&(\lsun) 	&		&(\rsun) 	&\multicolumn{2}{c}{(mag)} \\
\hline \\[-2ex]
\endhead
\hline\\[-2ex]
\multicolumn{3}{l}{{Continued on Next Page\ldots}} \\
\endfoot
\\[-1.8ex]
\endlastfoot
-			&	 NML Cyg			&	 RSG 		&	O			&	SRc			&	M6I			&	1220			&	1280			&	3834			&	-			&	272035		&	-		&	1183			&	3.9				&	   12.3 \\
00192-2020		&			 T Cet			&			 AGB 			&			OS 			&			SRb 			&			M5/M6Ib/II 			&			237			&			158			&			2788			&			-			&			4900			&			-			&			312			&			6.5			 	&	  -0.8 \\
01037+1219		&			 WX Psc			&			 AGB 			&			O 			&			OH/IR 			&			M8/10II 			&			740			&			660			&			2750			&			-			&			13914			&			-			&			520			&			13.3			&		    2.3 \\
01246-3248		&			 R Scl			&			 AGB 			&			C 			&			SRb 			&			CII 			&			475			&			370			&			2295			&			-			&			9527			&			-			&			617			&			6.9			   	&	-0.1 \\
01304+6211		&			 V669 Cas			&			 AGB 			&			O 			&			OH/IR 			&			M9III 			&			6210			&			1525			&			2750			&			-			&			38012			&			-			&			859			&			-2.3			&		   13.7 \\
02168-0312		&			 o Cet			&			 AGB 			&			O 			&			MIRA 			&			M7e 			&			107			&			331			&			2193			&			-			&			6099			&			-			&			541			&			8.7			  	&	  2.4 \\
03507+1115		&			 IK Tau			&			 AGB 			&			O 			&			MIRA 			&			M8/10IIe 			&			260			&			470			&			2667			&			-			&			9258			&			-			&			451			&			13.9			&		   -0.5 \\
04566+5606		&			 TX Cam			&			 AGB 			&			O 			&			MIRA 			&			M8.5 			&			380			&			557			&			2779			&			-			&			11360			&			-			&			460			&			15.4			  	&	 -0.6 \\
05073+5248		&			 NV Aur			&			 AGB 			&			O 			&			MIRA 			&			M10 			&			1200			&			635			&			2500			&			-			&			13284			&			-			&			615			&			1.6			    	&	2.9 \\
05524+0723		&			 Alpha Ori			&			 RSG 			&			O 			&			SRc 			&			M2Iab 			&			131			&			2335			&			3546			&			-			&			508887			&			-			&			1891			&			4.8			&		   -4.3 \\
07209-2540		&			 VY CMa			&			 RSG 			&			O 			&			SRc 			&			M2/4II 			&			1500			&			2000			&			3605			&		1		&			428303			&			-			&			1679			&			-0.4			  	&	  8.1 \\
09448+1139		&			 R Leo			&			 AGB 			&			O 			&			MIRA 			&			M8IIIe 			&			82			&			309			&			2890			&			-			&			5617			&			-			&			299			&			11		&			    2.5 \\
09452+1330		&			 CW Leo			&			 AGB 			&			C 			&			MIRA 			&			C9,5 			&			120			&			630			&			2000			&	2	&			9820			&			-			&			826			&			15.1			   	&	 1.1 \\
10131+3049		&			 RW LMi			&			 AGB 			&			C 			&			SRa 			&			Ce 			&			440			&			640			&			2000			&	2	&			5912			&			-			&			641			&			11.8			 	&	   1.2 \\
10329-3918		&			 U Ant			&			 AGB 			&			C 			&			LB 			&			C5,3 			&			256			&			365			&			2775			&			-			&			5640			&	9	&			325			&			5.9			  	&	 -0.5 \\
10350-1307		&			 U Hya			&			 AGB 			&			C 			&			SRb 			&			C6,3 			&			161			&			450			&			2982			&			-			&			11123			&			-			&			395			&			5.5			&		   -0.7 \\
10491-2059		&			 V Hya			&			 AGB 			&			C 			&			SRa 			&			C6, 5e 			&			2160			&			530			&			2007			&			-			&			5414			&			-			&			609			&			8		&			   -0.1 \\
12427+4542		&			 Y CVn			&			 AGB 			&			C 			&			SRb 			&			C7Iab 			&			217			&			157			&			2750			&			-			&			4853			&			-			&			307			&			5.9		&			   -0.7 \\
13269-2301		&			 R Hya			&			 AGB 			&			O 			&			MIRA 			&			M7IIIe 			&			118			&			388			&			2128			&			-			&			7375			&			-			&			631			&			9	&				    2.4 \\
13462-2807		&			 W Hya			&			 AGB 			&			O 			&			SRa 			&			M7e 			&			77			&			361			&			3129			&			-			&			4525			&			-			&			229			&			10.7				&	    3.1 \\
14219+2555		&			 RX Boo			&			 AGB 			&			O 			&			SRb 			&			M7.5e 			&			155			&			340			&			3010			&			-			&			8912			&			-			&			347			&			-1.7			&		    1.9 \\
15194-5115		&			 II Lup			&			 AGB 			&			C 			&			MIRA 			&			C 			&			500			&			575			&			2400			&	2	&			8933			&			-			&			547			&			4.4			&		    1.7 \\
16269+4159		&			 G Her			&			 AGB 			&			O 			&			SRb 			&			M6III 			&			310			&			89			&			3297			&			-			&			3121			&			-			&			171			&			-5.5		&			   11.4 \\
17123+1107		&			 V438 Oph			&			 AGB 			&			O 			&			SRb 			&			M8e 			&			416			&			169			&			2890			&			-			&			5163			&			-			&			286			&			14.8		&			    0.6 \\
17411-3154		&			 AFGL 5379			&			 AGB 			&			O 			&			OH/IR 			&			(OH) 			&			1190			&			1440			&			2750			&			-			&			35484			&			-			&			830			&			5.7	&				    9.5 \\
18050-2213		&			 VX Sgr			&			 RSG 			&			O 			&			SRc 			&			M4eIa 			&			1570			&			732			&			3535			&	3	&			102294			&			-			&			853			&			0.4				&	    7.7 \\
18308-0503		&			 AFGL 5502				&			 YSO 			&			O 			&			YSO 			&			- 			&			3100			&			- 			&			- 			&			-			&			21000			&	10	&			- 			&			1.5			  	&	 13.2 \\
18333+0533		&			 NX Ser			&			 AGB 			&			O 			&			MIRA 			&			(CO) 			&			2480			&			795			&			3300			&			-			&			17395			&			-			&			403			&			4.9		&			    3.6 \\
18348-0526		&			 OH 26.5+0.6			&			 AGB 			&			O 			&			OH/IR 			&			(OH) 			&			1370			&			1570			&			2750			&			-			&			39362			&			-			&			874			&			-0.4		&			    8.1 \\
18397+1738		&			 IRC +20370			&			 AGB 			&			C 			&			MIRA 			&			Ce 			&			600			&			637			&			2200			&	2	&			9933			&			-			&			686			&			4.6			   	&	 1.8 \\
18448-0545		&			 R Sct			&			 AGB 			&			O 			&			Rva 			&			K0Ibpv 			&			431			&			146			&			5000			&	4	&			4000			&	4	&			84			&			5.6			   	&	12.1 \\
18476-0758		&			 S Sct			&			 AGB 			&			C 			&			SRb 			&			C6, 4 			&			398			&			148			&			2425			&			-			&			4634			&			-			&			386			&			6.6			&		    0.5 \\
19114+0002		&			 AFGL 2343			&			 HYPERGIANT 			&			O 			&			SRd 			&			G5Ia 			&			4080			&			200			&			5418			&			-			&			199526			&	11	&			507			&			0.3			  	&	 14.8 \\
19126-0708		&			 W Aql			&			 AGB 			&			S 			&			MIRA 			&			S6e 			&			680			&			490			&			2800			&	5	&			9742			&			-			&			419			&			15.4			   	&	 0.5 \\
19192+0922		&			 OH 44.8-2.3			&			 AGB 			&			O 			&			OH/IR 			&			(OH) 			&			1130			&			552			&			2750			&			-			&			11228			&			-			&			467			&			4.2		&			   13.6 \\
19244+1115		&			 IRC +10420			&			 HYPERGIANT 			&			O 			&			SRd 			&			A5Ia 			&			5000			&			- 			&			4442			&			-			&			630957			&	11	&			1342			&			2.2			&		   13.9 \\
19283+1944		&			 AFGL 2403			&			 AGB 			&			O 			&			OH/IR 			&			(OH) 			&			2300			&			- 			&			2750			&			-			&			1574			&	12	&			174			&			3.2			  	&	 13.3 \\
19486+3247		&			 Chi Cyg			&			 AGB 			&			S 			&			MIRA 			&			S6 +/1e 			&			149			&			408			&			2000			&	6	&			7813			&			-			&			737			&			16				&	    1.9 \\
20075-6005		&			 X Pav			&			 AGB 			&			O 			&			SRb 			&			M8III 			&			270			&			199			&			2046			&			-			&			5849			&			-			&			608			&			9.3			  	&	 -1.1 \\
20077-0625		&			 IRC -10529			&			 AGB 			&			O 			&			OH/IR 			&			M 			&			620			&			680			&			2750			&			-			&			14421			&			-			&			529			&			2.3				&	    2.1 \\
20120-4433		&			 RZ Sgr			&			 AGB 			&			S 			&			SRb 			&			Se 			&			730			&			223			&			2710			&	7	&			6396			&			-			&			363			&			12.7			    	&	1.2 \\
20396+4757		&			 V Cyg			&			 AGB 			&			C 			&			MIRA 			&			C5, 3e 			&			271			&			421			&			2581			&			-			&			6472			&			-			&			402			&			6.3			&		    0.4 \\
21419+5832		&			 Mu Cep			&			 RSG 			&			O 			&			SRc 			&			M2Ia 			&			390			&			730			&			3660			&	8	&			111215			&			-			&			830			&			-5.9			    	&	8.5 \\
21439-0226		&			 EP Aqr			&			 AGB 			&			O 			&			SRb 			&			M8IIIv 			&			135			&			55			&			2302			&			-			&			2145			&			-			&			291			&			8.2		&			   -1.6 \\
21554+6204		&			 GLMP 1048			&			 AGB 			&			O 			&			OH/IR 			&			(OH) 			&			2030			&			- 			&			2750			&			-			&			4984			&	13	&			311			&			3.2			  	&	 14.8 \\
22177+5936		&			 OH 104.9+2.4			&			 AGB 			&			O 			&			OH/IR 			&			(OH) 			&			2300			&			1620			&			2750			&			-			&			40871			&			-			&			891			&			1.1		&			   13.9 \\
22196-4612		&			 pi1 Gru			&			 AGB 			&			S 			&			SRb 			&			S5, 7e 			&			152			&			150			&			2257			&			-			&			4683			&			-			&			447			&			8.4			&		    5.8 \\
23166+1655		&			 LL Peg			&			 AGB 			&			C 			&			MIRA 			&			C 			&			980			&			696			&			2000			&	2	&			10887			&			-			&			869			&			2			   	&	10.5 \\
23320+4316		&			 LP And			&			 AGB 			&			C 			&			MIRA 			&			C 			&			630			&			614			&			2000			&	2	&			9561			&			-			&			815			&			1.6			   	&	 3.5 \\
23558+5106		&			 R Cas			&			 AGB 			&			O 			&			MIRA 			&			M7IIIe 			&			106			&			430			&			3129			&			-			&			8331			&			-			&			310			&			16.8	&				    1.8 \\
\hline
\end{longtable}
\end{landscape}}
\normalsize

\longtabL{7}{\small\setlength{\tabcolsep}{1.5mm}
\begin{landscape}
\begin{longtable}{lccclcrr}
 \caption[Mass-loss rate estimates.]{Mean mass-loss rates $\overline{\dot{M}}$ derived with Eq.~\ref{eq:ourmdotformula} for all AGB, Post-AGB, supergiant, and hypergiant stars from the sample. Targets with line profiles that show clear deviations from the soft-parabola profile are marked with (!). $\sigma(\dot{M})$ indicates the spread of the \mdot-estimates if a minimum of three diagnostic lines is available. If variability is detected in the estimates, this is indicated in the fourth column. If no variability could be seen in the estimates, this is indicated with \textit{Constant}. The fifth column lists some general remarks. $N_{\mathrm{lines}}$ is the number of lines used. The telescopes with which the line data were obtained are listed in the seventh column. \router is given in the last column. }\label{tbl:mdotestimates}\\
\hline \hline \\[-2ex]
Target &$\overline{\dot{M}}$ &$|\log(\sigma(\dot{M})/\overline{\dot{M}})|$ &Mass loss &Remarks &$N_{\mathrm{lines}}$ &Telescopes &\router \\
 &(\msun/yr) & & & & & &($10^{16}$\,cm) \\
\hline \\[-2ex]
\endfirsthead
\multicolumn{3}{c}{\textbf{{\tablename} \thetable{}.} Continued} \\[0.5ex]
\hline \hline \\[-2ex]
Target &$\overline{\dot{M}}$ &$|\log(\sigma(\dot{M})/\overline{\dot{M}})|$ &Mass loss &Remarks &$N_{\mathrm{lines}}$ &Telescopes &\router \\
 &(\msun\,yr$^{-1}$) & & & & & &($10^{16}$\,cm) \\
\hline \\[-2ex]
\endhead
\hline\\[-2ex]
\multicolumn{3}{l}{{Continued on Next Page\ldots}} \\
\endfoot
\\[-1.8ex]
\endlastfoot
NML Cyg (!)&8.7$\times10^{-5}$& 0.28&Constant&Red supergiant&4&JCMT& 21.8\\
T Cet&8.8$\times10^{-8}$& 0.64&Constant&&4&JCMT&  0.9\\
WX Psc&1.9$\times10^{-5}$& 0.09&Possibly variable&&16&APEX, BLT, FCRAO, JCMT, OSO& 11.0\\
R Scl (!)&1.6$\times10^{-6}$& 0.16&Constant&Detached shell reported&5&APEX, SEST&  5.9\\
V669 Cas&5.5$\times10^{-5}$& 0.01&Possibly variable&&4&JCMT& 25.9\\
o Cet (!)&2.5$\times10^{-7}$& 0.37&Constant&&5&JCMT, SEST&  1.4\\
IK Tau&4.5$\times10^{-6}$& 0.11&Constant&&10&APEX, IRAM, JCMT&  5.0\\
TX Cam&6.5$\times10^{-6}$& 0.20&Constant&&3&JCMT&  5.9\\
NV Aur&1.8$\times10^{-5}$& 0.17&Constant&&4&JCMT& 10.7\\
Alpha Ori (!)&2.1$\times10^{-7}$& 0.69&Constant&Red supergiant&3&JCMT&  1.4\\
Red Rectangle&7.7$\times10^{-8}$&-&Constant&Post-AGB&1&APEX&  1.5\\
VY CMa (!)&2.8$\times10^{-4}$& 0.26&Constant&Red supergiant&6&JCMT, SEST& 38.6\\
Calabash Nebula&8.0$\times10^{-5}$&-&Constant&Post-AGB&1&APEX& 15.4\\
R Leo (!)&9.2$\times10^{-8}$& 0.45&Constant&&5&APEX, CSO, IRAM&  0.9\\
CW Leo&1.6$\times10^{-6}$& 0.12&Constant&&9&CSO, IRAM, JCMT, NRAO, SEST&  6.6\\
RW LMi&5.9$\times10^{-6}$& 0.02&Possibly variable&&5&CSO, IRAM& 12.1\\
GSC 08608-00509&8.8$\times10^{-5}$&-&Constant&Post-AGB&1&APEX& 23.8\\
U Hya&4.9$\times10^{-8}$& 0.39&Constant&&5&APEX, SEST&  1.2\\
V Hya (!)&6.1$\times10^{-5}$& 0.31&Constant&&4&APEX, SEST& 42.7\\
Y CVn&9.5$\times10^{-8}$& 0.11&Constant&&4&CSO, IRAM&  1.7\\
R Hya&1.6$\times10^{-7}$& 0.25&Constant&Detached shell reported&7&CSO, IRAM, JCMT, NRAO&  1.2\\
W Hya&7.8$\times10^{-8}$& 0.57&Constant&&3&APEX&  0.9\\
RX Boo&3.6$\times10^{-7}$& 0.03&Possibly variable&&7&CSO, IRAM, JCMT&  1.6\\
II Lup&3.9$\times10^{-6}$& 0.36&Constant&&6&APEX&  9.0\\
G Her&7.0$\times10^{-7}$& 0.18&Constant&&3&JCMT&  2.1\\
V438 Oph&4.1$\times10^{-8}$&-&Constant&&2&JCMT&  0.6\\
Cotton Candy Nebula&2.1$\times10^{-5}$&-&Constant&Post-AGB&1&APEX& 10.8\\
AFGL 5379&2.8$\times10^{-5}$& 0.34&Constant&&5&APEX, JCMT& 12.7\\
VX Sgr&6.1$\times10^{-5}$& 0.19&Constant&Red supergiant&3&JCMT& 20.2\\
NX Ser&6.2$\times10^{-5}$& 0.19&Constant&&4&JCMT& 23.7\\
OH 26.5+0.6&9.7$\times10^{-6}$&-&Constant&&2&JCMT&  7.9\\
IRC +20370&4.3$\times10^{-6}$& 0.06&Constant&&5&APEX, IRAM& 11.1\\
R Sct&2.1$\times10^{-7}$&-&Constant&&1&APEX&  1.3\\
AFGL 2343&1.4$\times10^{-3}$&-&Constant&Yellow hypergiant&1&APEX&115.3\\
W Aql&1.3$\times10^{-5}$& 0.17&Constant&&14&APEX, JCMT, NRAO, SEST& 15.1\\
OH 44.8-2.3&4.6$\times10^{-6}$& 0.30&Constant&&4&JCMT&  5.1\\
IRC +10420 (!)&3.6$\times10^{-3}$& 0.20&Constant&Yellow hypergiant&10&CSO, IRAM, JCMT&205.7\\
AFGL 2403&1.7$\times10^{-5}$& 0.30&Constant&&3&JCMT& 11.1\\
Chi Cyg&2.4$\times10^{-7}$& 0.04&Possibly variable&&9&CSO, IRAM, JCMT, NRAO&  2.1\\
X Pav&5.2$\times10^{-7}$&-&Constant&&1&APEX&  1.9\\
IRC -10529&4.5$\times10^{-6}$&-&Constant&&2&APEX&  5.2\\
RZ Sgr&5.8$\times10^{-7}$&-&Constant&&2&APEX&  3.4\\
V Cyg&4.0$\times10^{-7}$&-&Constant&&1&JCMT&  3.0\\
Mu Cep (!)&2.0$\times10^{-6}$&-&Constant&Red supergiant&2&JCMT&  3.5\\
EP Aqr (!)&3.1$\times10^{-7}$& 0.23&Constant&&4&JCMT&  1.5\\
GLMP 1048&1.5$\times10^{-5}$& 0.09&Constant&&4&JCMT&  9.7\\
OH 104.9+2.4&8.4$\times10^{-6}$& 0.21&Constant&&4&JCMT&  7.1\\
pi1 Gru (!)&8.5$\times10^{-7}$& 0.15&Constant&&3&APEX&  3.4\\
LL Peg&3.1$\times10^{-5}$& 0.27&Possibly variable&&9&APEX, CSO, IRAM, OSO& 38.6\\
LP And&4.6$\times10^{-6}$& 0.02&Possibly variable&&6&CSO, IRAM, JCMT, OSO& 12.4\\
R Cas (!)&4.0$\times10^{-7}$& 0.39&Constant&&4&JCMT&  1.7\\
\hline
\end{longtable}
\end{landscape}}\clearpage\normalsize

\begin{figure*}\begin{center}\includegraphics[width=.85\linewidth,angle=180]{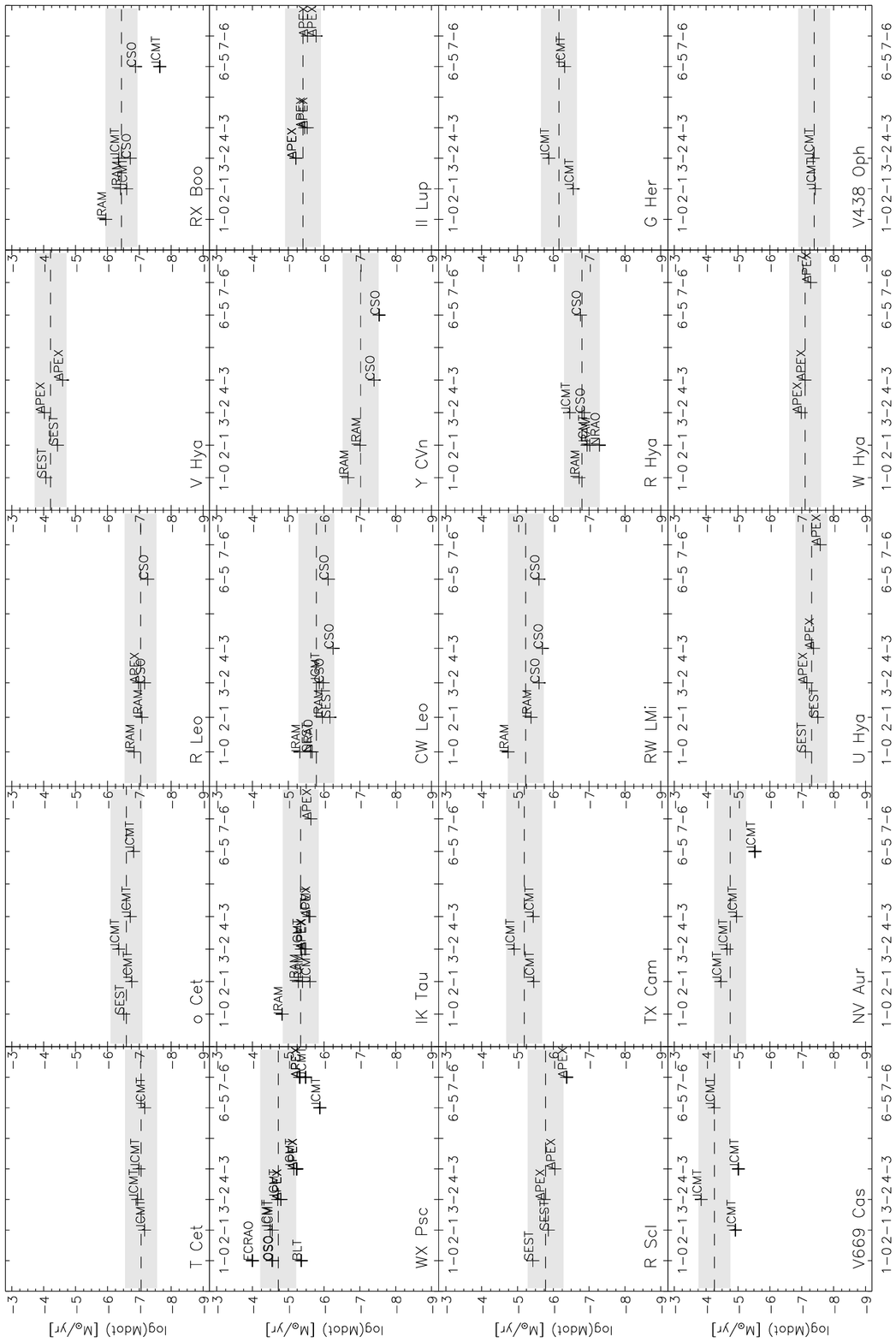}\\\vspace{.5in}\caption{\mdot-estimates from Eq.~\ref{eq:ourmdotformula} for the AGB stars in the sample. The dashed line indicates the mean value $\overline{\Mdot}$ listed in Table~\ref{tbl:mdotestimates}, the dashed line indicates the mean of the estimates. The shaded area indicates the typical uncertainty region around the mean value, which is a factor of three. \label{fig:mdotestimatesagb}}\end{center}\end{figure*}\clearpage
\begin{figure*}\begin{center}\includegraphics[width=.85\linewidth,angle=180]{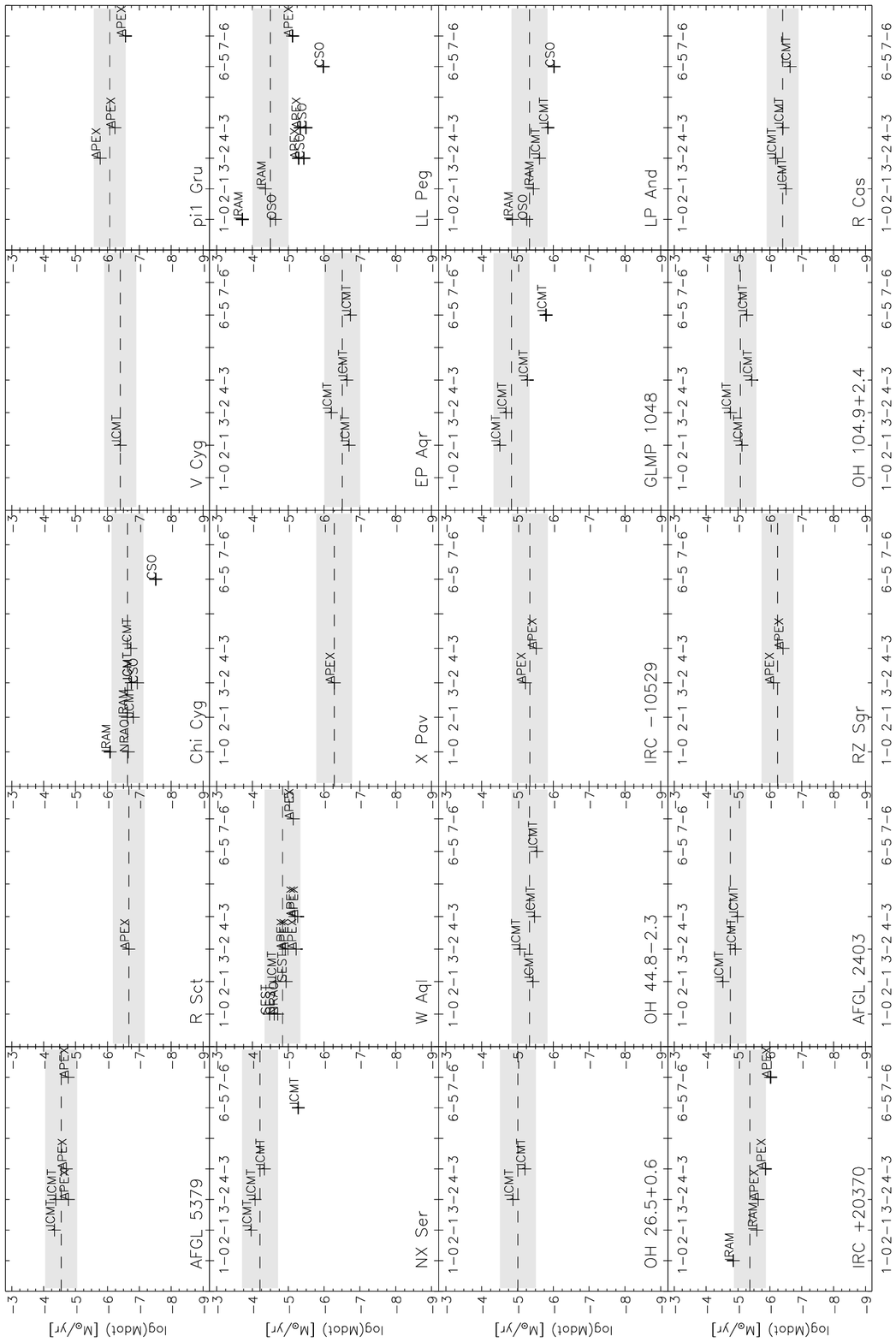}\\\vspace{.5in}\textbf{Fig.~\ref{fig:mdotestimatesagb}} (continued)\end{center}\end{figure*}\clearpage
\begin{figure*}\begin{center}\includegraphics[height=.9\linewidth,width=.5\linewidth,angle=90]{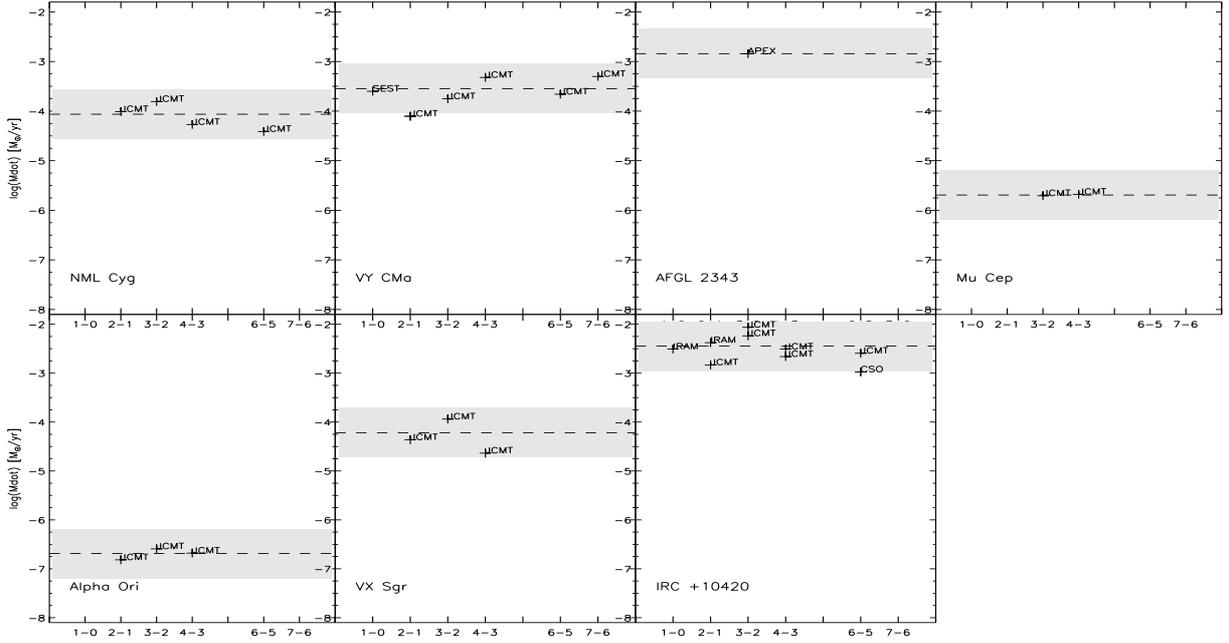}\caption{\mdot-estimates from Eq.~\ref{eq:ourmdotformula} for the supergiants and hypergiants in the sample. The dashed line indicates the mean value $\overline{\Mdot}$ listed in Table~\ref{tbl:mdotestimates}, the dashed line indicates the mean of the estimates. The shaded area indicates the typical uncertainty region around the mean value, which is a factor of three.\label{fig:mdotestimatesrsgyhg}}\end{center}\end{figure*}

\section{Results and discussion}\label{sect:resultsanddiscussion}
\subsection{Mass-loss rate \mdot}\label{subsect:samplemdotresults}
For all targets in the sample for which we could fit at least one rotational line profile with a soft parabola (Eq.~\ref{eq:softparabola}) we listed estimates for \mdot in Table~\ref{tbl:mdotestimates}. The estimates from all \twco lines for the sample AGB stars are shown in Fig.~\ref{fig:mdotestimatesagb}. For the supergiants and hypergiants they are shown in Fig.~\ref{fig:mdotestimatesrsgyhg}. In case the soft-parabola fit procedure yielded a negative $\beta$, the exponent $s_{10,J}$ in Eq.~\ref{eq:ourmdotformula} was assumed to be zero. 

The listed estimates are averages over all values obtained via Eq.~\ref{eq:ourmdotformula} for the fitted line transitions, and are written as $\overline{\Mdot}$. In Table~\ref{tbl:mdotestimates} we also listed the number of lines used and the spread of the estimates around the mean. 

We did not attempt to fit the line profiles for U\,Ant and S\,Sct, since these are composed of emission from the inner part of the CSE and from the detached shell at larger outflow velocities, respectively. As discussed in Sect.~\ref{subsubsect:individualtargets} some other targets show clear deviations from the fitted soft-parabola profile. The estimates for these targets, e.g. EP\,Aqr, are therefore subject to larger uncertainties than the spreads listed in Table~\ref{tbl:mdotestimates} and have to be interpreted with caution. 

If the mass-loss-rate estimates of all available lines are within or close to the typical factor three uncertainty region, it is reasonable to state that the mass-loss rate of the target has been constant throughout the regions of the envelope sampled by these transitions. For most AGB stars in the sample this is indeed the case, but there are some exceptions indicating a possible variability in mass loss in the part of the envelope that is traced.

\subsubsection{Variability}
\paragraph{Criterion:}
Since higher-$J$ transitions trace deeper, more recently formed envelope layers, a criterion to decide upon variability should consider the behaviour of the \mdot-estimates in view of the $J$-values. The presence of a \textit{picket-fence} structure in the estimates, i.e. with \mdot varying strongly between transitions could suggest variability of the mass-loss rate, but on rather short time scales of some hundreds of years. The presence of a strong overall gradient can also imply mass-loss-rate variability: e.g. decreasing \mdot for increasing $J$ can be indicative for a mass-loss rate that has decreased over time. When discussing the sensitivity of our results to the modelling of the cooling of the envelope in Sect.~\ref{subsubsect:sensitivityanalysis}, we pointed out that our estimates might intrinsically hold a shallow gradient. The differences associated with these model uncertainties between the highest and lowest estimates (for a constant \mdotinput) are, however, not expected to exceed a factor three for $J=1-0$ up to $J=7-6$. Therefore, if a much steeper gradient is found in the estimates for some target, with a total coverage exceeding an order of magnitude in \mdot, we are therefore inclined to state that the mass-loss rate of the respective target has varied over time.

\paragraph{Variability in the sample:}
The estimates for V669\,Cas, shown in Fig.~\ref{fig:mdotestimatesagb}, exhibit the picket-fence type of trend and suggest that variability in the mass-loss rate is traced for this star.

The estimates for NV\,Aur and GLMP\,1048 based on data of the $J=2-1,\,3-2,\,4-3,\,6-5$ transitions, the estimates for R\,Scl and IRC\,+20370, for $J=1-0,\,2-1,\,3-2,\,4-3,\,7-6$, and the estimates for CW\,Leo, for $J=1-0,\,2-1,\,3-2,\,4-3,\,6-5$ exhibit a spread of about an order of magnitude. This spread is within the uncertainties we expect around a \mdot-value and therefore points to a constant mass-loss rate throughout the traced regions of the envelope. The estimates for RW\,LMi, LP\,And and $\chi$\,Cyg with $J=1-0,\,2-1,\,3-2,\,4-3,\,6-5$, respectively cover factors about ten, twenty and thirty in \mdot. The estimates for RX\,Boo, for $J=1-0,\,2-1,\,3-2,\,6-5$, cover a factor about forty in \mdot. In all these cases, we find a strong downward gradient, which most likely indicates that the more inward regions, i.e. the excitation regions of the higher-$J$ lines, were produced by a lower mass-loss rate than the outer regions. A strong gradient is also found for both WX\,Psc and LL\,Peg for $J=1-0$ up to $J=6-5$. For both targets, however, the \mdot-estimate based on the $J=7-6$ transition is higher by about an order of magnitude than the one based on the $J=6-5$ transition, indicating a recent increase of the mass-loss rate, as was proposed by \cite{decin2007_wxpsc} for WX\,Psc. Qualitatively, this phenomenon is independent of the modelling of the \h2o-cooling, i.e. an increase would still be present if another cooling mechanism was considered in deriving the estimates. Obtaining a result corresponding to detailed modelling using an estimator as in Eq.~\ref{eq:ourmdotformula} can be used as a benchmark for the quality of the estimator.

It is likely that the \ohir stars and the more extreme Mira stars in the sample have already undergone mass-loss modulations, but to verify this more data are needed, especially of $J=1-0,\;6-5$ and $7-6$.

\paragraph{Data quality:}
The IRAM measurements\footnote{All IRAM measurements were previously presented in the literature.} in our sample are subject to somewhat larger uncertainties than our own APEX and JCMT data since some questions have arisen about the data reduction. \cite{decin2008_rhya} already reported large uncertainties on the IRAM data of R\,Hya and \cite{skinner1999} mentioned significant deviations between different measurements of the same lines towards CW\,Leo. The IRAM data \citeauthor{skinner1999} gathered from the literature and archives seemed to suffer from large uncertainties linked to pointing and calibration. They also discussed inconsistencies in the literature concerning the reported absolute line intensities. Very often confusion or unclarity arises about e.g. conversion from antenna temperatures to main-beam-brightness temperatures. For these reasons, we considered leaving the IRAM measurements out of the data set. In case of LL\,Peg this leads to a decrease of the estimated mass-loss rate with a factor 5.3, for IRC\,+20370 with a factor 3.9. For IRC\,+10420 and R\,Hya omitting the IRAM data has no effect on the \mdot-estimate. For the eight other objects with IRAM observations, the decrease is with a factor between 1.3 and 2.7. The \mdot-estimates ignoring IRAM data are always lower.

Omitting the IRAM data has an effect on the gradients seen in the estimates and discussed earlier. In many cases, the CO-sampling is now restricted to the transitions higher than $J=1-0$ and/or $J=2-1$, making possible trends less significant. In case of CW\,Leo, RX\,Boo, IRC\,+20370, RW\,LMi and LP\,And the supposed trends in the estimates are no longer present. For LL\,Peg the presence of an OSO $J=1-0$ measurement still gives the trend seen when the IRAM data are included.

For the \orich semi-regular variable RX\,Boo we find two estimates that could hint towards variability, i.e. those based on the IRAM $J=1-0$ and JCMT $J=6-5$ measurements. The latter is duplicated by a measurement carried out with CSO, and the estimate based on that observation is about an order of magnitude higher than the one derived from the JCMT. We are however inclined to attach more value to our JCMT observation, since the CSO calibration as described by \cite{teyssier2006} holds larger uncertainties, such as rather large pointing errors. We conclude from this that there is some, however somewhat uncertain, downward trend with lower \mdot-estimates for higher $J$, indicating the possibility of a decrease in mass-loss rate over time. 

\paragraph{Comparison to the literature:}
\cite{teyssier2006} note that a change in the mass-loss rate is needed to reproduce the data for three of their sample targets: R\,Hya, RW\,LMi and $\chi$\,Cyg. \cite{decin2008_rhya} constructed a mass-loss model for R\,Hya and showed that a detached shell is needed to explain the rotation-vibrational lines. The purely rotational lines could however be reproduced with a constant mass-loss rate. Our estimates do, indeed, not point to any significant variations in the mass-loss rate, as can be seen from Fig.~\ref{fig:mdotestimatesagb}. The \mdot-estimates for RW\,LMi cover one order of magnitude and show a (shallow) gradient with lower \mdot for higher $J$. This could point to a mass-loss rate that is presently lower than the one that produced the outer layers of the circumstellar envelope. \citeauthor{teyssier2006} specify a difference of only a factor 1.5 between the two mass-loss rates, so we can say that our estimates support this hypothesis of mass-loss-rate variability. In case of $\chi$\,Cyg \citeauthor{teyssier2006} mention a minimum change of a factor 4 between the former and the present mass-loss rate, with the present one being lowest. This result is again reflected in our estimates. 

The detached-shell structure of the \crich semi-regular variable R\,Scl was described by \cite{bergeat2005} as the consequence of a former mass-loss rate of $\sim$$5.5\times10^{-6}$\,\msun\,yr$^{-1}$ and a present mass-loss rate of $\sim$$3.5\times10^{-7}$\,\msun\,yr$^{-1}$. The range of mass-loss rates covered by our estimates agrees with these values. 

The mass-loss rates estimated from the individual CO emission lines for WX\,Psc --- shown in Fig.~\ref{fig:mdotestimatesagb} --- are in very good agreement with the detailed modelling of the star by \cite{decin2007_wxpsc}. They present a profile $\dot{M}(r)$ exhibiting a high \mdot-value, followed in time by a strong decrease and again an increase of the mass-loss rate. We do not reproduce the same large difference of three orders in magnitude as mentioned by \cite{decin2007_wxpsc}. A possible explanation is that the excitation regions of the individual observed transitions are not restricted to either the high-\mdot or the low-\mdot regions. The estimates then represent more than one of these regions and are a kind of 'mean' value of \mdot. However, the estimates we present in this paper for WX\,Psc reflect the detailed mass-loss modelling very well. 

\cite{winters2000} reported on possible time variability of the mass-loss rate of CW\,Leo. The present-day mass-loss rate is suggested to be lower than the mass-loss rate that produced the outer envelope layers by a factor of up to five. The decrease we see in our estimates, however, is by a factor $\sim$30. Also, the \mdot-values presented in the literature are about an order of magnitude higher than the $\overline{\dot{M}}$-value we derived. A possible explanation for these discrepancies is the \h2o-cooling uncertainty, leading to underestimates of \mdot. Another explanation is that the $J=7-6$ transition for the first time reveals a stronger variation in \mdot than could be traced with single-dish observations of CO transitions up to $J=4-3$ or $6-5$. The model presented by \cite{ramstedt2008} fitting lines $J=1-0$ up to $J=4-3$ shows clear deviations from the data. The discrepancies are systematic, i.e. the model underestimates the observed intensities more strongly for increasing $J$. This indicates the need for consideration of \mdot-variability for this target.

\cite{neri1998} suggested that LL\,Peg might have a bipolar detached shell and \cite{mauron2006} reported on a spiral structure in the envelope. The latter is indicative for variability in the outflow from the central star.

\paragraph{Supergiants and hypergiants:}
The estimates for all five RSGs in our sample suggest constant mass-loss rates; see Fig.~\ref{fig:mdotestimatesrsgyhg}. In case of the yellow hypergiant IRC\,+10420, \cite{humphreys1997,castro-carrizo2007} and \cite{dinh-v-trung2009}, among others, have put forward the possibility of a complex mass-loss history, with several episodes of high and low mass-loss rates. This resulted in multiple arc-like structures and shells around the central star.  For AFGL\,2343, the second hypergiant in our sample, we only used the $J=3-2$ line of \twco to estimate the mass-loss rate, so no conclusions on variability in mass-loss rate can be drawn from this. \cite{castro-carrizo2007} mention variations in the mass-loss rate for both hypergiants on time scales of $\sim$1000\,years.  In this study, where we only consider CO emission lines measured with single-dish telescopes, we lack sensitivity for the rapid density fluctuations. This is due to the overlap of the line forming regions of the different rotational transitions.

\begin{figure}
 \begin{center}
\subfigure[Histogram of the $\overline{\Mdot}$-values for all AGB stars, grouped per pulsational type.\label{subfig:histopuls}]{\includegraphics[angle=90,width=.85\linewidth]{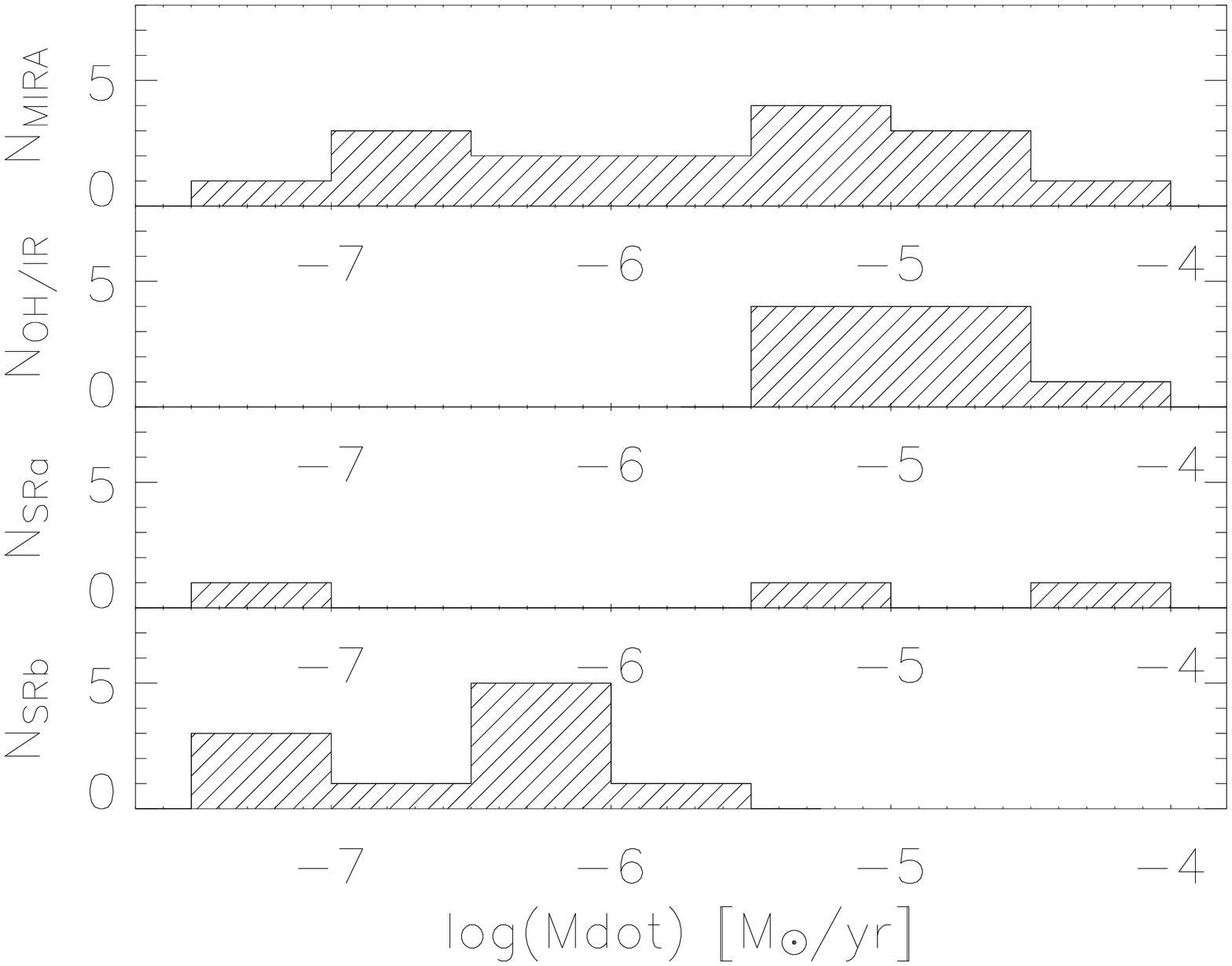}}
\subfigure[Histogram of the $\overline{\Mdot}$-values for all AGB stars, grouped per chemistry type. $N_{\mathrm{C}}$ is the number of \crich targets, $N_{\mathrm{O}}$ is the number of \orich targets, and $N_{\mathrm{S}}$ is the number of S-type targets. \label{subfig:histochem}]{\includegraphics[angle=90,width=.85\linewidth]{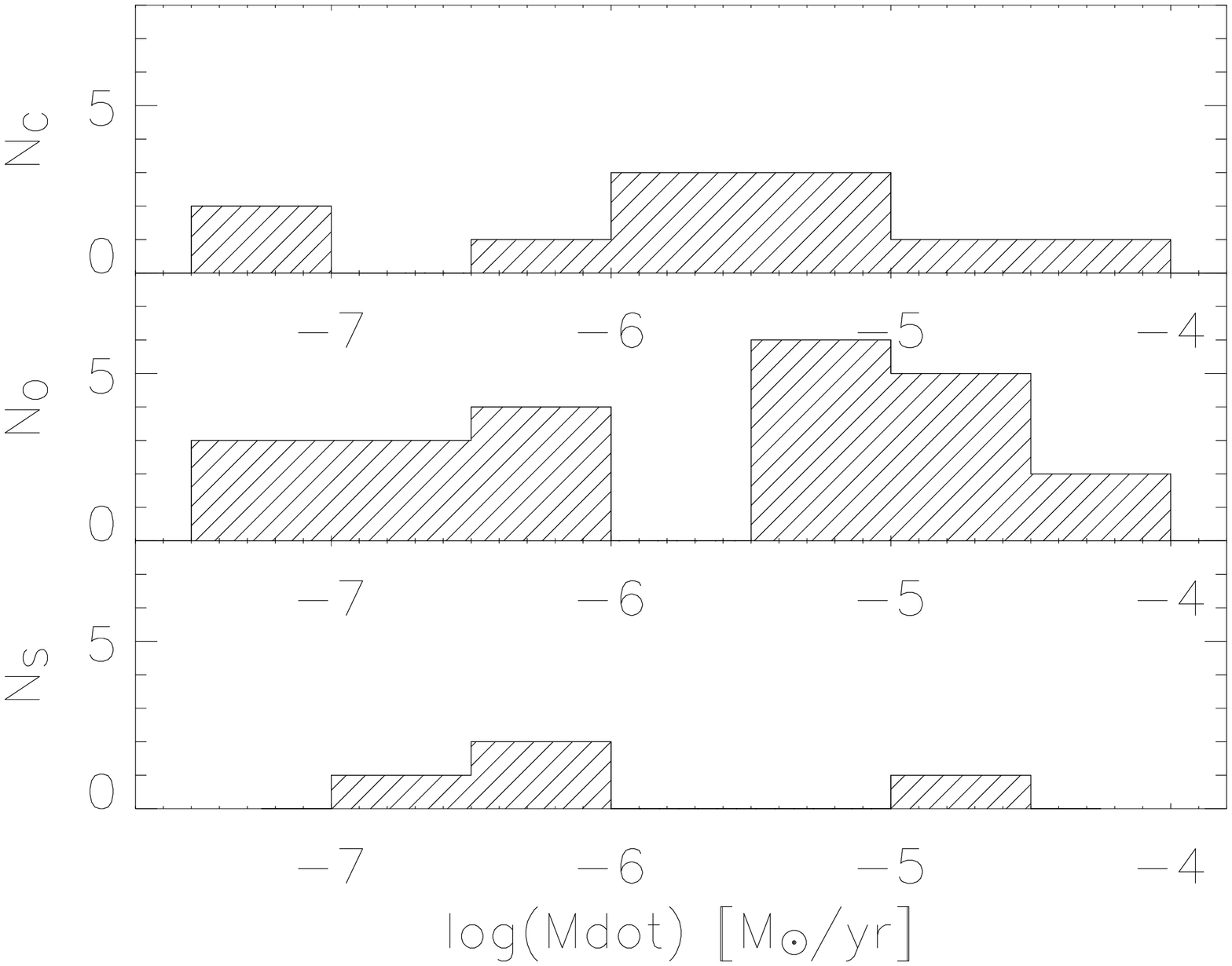}}
\subfigure[Histogram of the $\overline{\Mdot}$-values of all AGB stars.\label{subfig:histoagb}]{\includegraphics[angle=90,width=.85\linewidth]{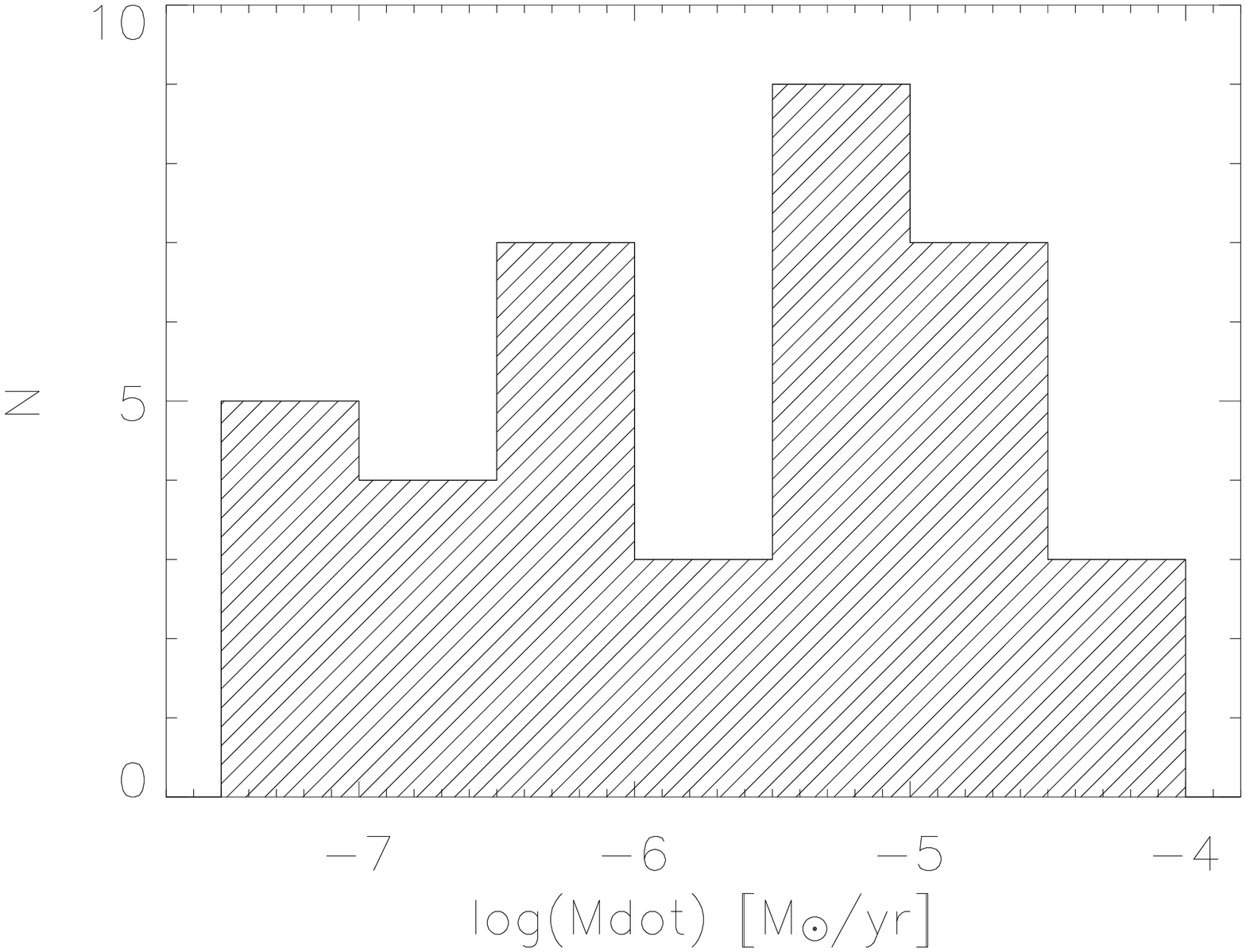}}
\caption{Histograms containing the $\overline{\Mdot}$-values estimated from Eq.~\ref{eq:ourmdotformula} and given in Table~\ref{tbl:mdotestimates}. The adopted bin size is $\Delta \log(\Mdot)=0.5$, which corresponds to about a factor three difference in mass-loss rate between consecutive bins. \label{fig:mdothistos}}
\end{center}
\end{figure}

\subsubsection{AGB mass loss in view of pulsational type and chemistry type}
The histograms in Fig.~\ref{fig:mdothistos} show that the spread on $\overline{\Mdot}$ for AGB stars is quite large, but there is a noticeable difference between the semi-regular variables on the one hand and the Miras and \ohir targets on the other --- see Fig.~\ref{subfig:histopuls}. The spread on \mdot for the 16 Miras is rather extended, with values ranging from $9\times10^{-8}$\,\msun\,yr$^{-1}$ (R\,Leo) up to $6\times10^{-5}$\,\msun\,yr$^{-1}$ (NX\,Ser) and a mean value of $10^{-5}$\,\msun\,yr$^{-1}$. 

The mass-loss rates for the 9 \ohir targets were all estimated in the range $4\times10^{-6}\leq\overline{\Mdot}\leq 6\times10^{-5}$\,\msun\,yr$^{-1}$, with an average of $2\times10^{-5}$\,\msun\,yr$^{-1}$. The estimates for these targets seem low, considering that \ohir stars have optically thick envelopes and can be heavily obscured. However, it has already been pointed out by e.g. \cite{delfosse1997} that the mass-loss rates derived from CO lines can be significantly lower than those derived from the infrared flux in the case of \ohir stars. The most likely explanation for the discrepancy is found in the recent ($\sim$1000\,yr ago) onset of a superwind phase. The latter would not be traced by the CO lines available in this data set, except possibly for WX\,Psc. 

The SRb targets have  lower mass-loss rates, with an average of $5\times10^{-7}$\,\msun\,yr$^{-1}$ and the highest value reaching $2\times10^{-6}$\,\msun\,yr$^{-1}$ (R\,Scl). These results are in accordance with the idea that the SR variables are likely progenitors of Miras \citep{whitelock2000,yesilyaprak2004}. 

When considering the histogram representing the AGB targets grouped per chemistry type (\orich, \crich or S-type) in Fig.~\ref{subfig:histochem}, no clear distinction between these groups can be made. Though this could be due to the bias of our sample towards \orich stars, there are no clear indications for significant discrepancies in \mdot between the different types. \cite{ramstedt2006} also reported no noticeable differences in \mdot-values for the different chemistries. Putting all AGB targets together for which \mdot-estimates could be made --- irrespective of pulsational or chemical type --- in Fig.~\ref{subfig:histoagb}, we get a minimum value of $4\times10^{-8}$\,\msun\,yr$^{-1}$ (SRb V438\,Oph), a maximum value of $6\times10^{-5}$\,\msun\,yr$^{-1}$ (O-rich Mira NX\,Ser) and a mean of $10^{-5}$\,\msun\,yr$^{-1}$.

\begin{figure} \includegraphics[angle=90,width=\linewidth]{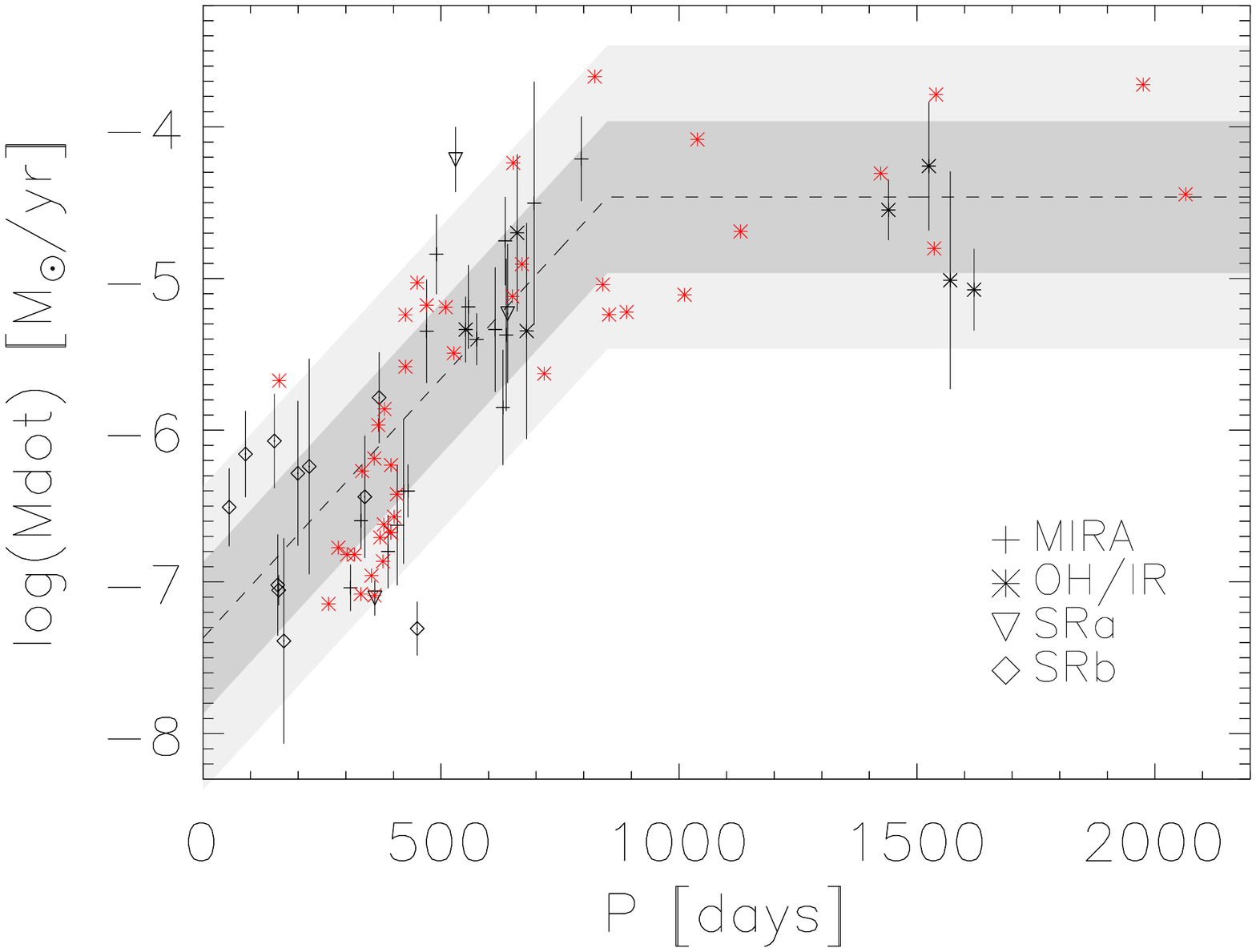}
\caption{Mean estimates for \mdot versus period of pulsation for the AGB stars in the sample are plotted in black. Different pulsational types are indicated with different plotting symbols: + for Miras, $\ast$ for \ohir stars, $\nabla$ for SRa and $\diamond$ for SRb type stars. The red $\ast$ \ohir data points were obtained from \citet[][ priv. comm.]{he2001}. A linear relation fits the Miras, \ohir stars and semi-regular variables in our sample with periods shorter than 850~days. \mdot levels off  to a constant value of $2.6\times10^{-5}$\,\msun\,yr$^{-1}$ for longer periods. The darker shaded region represents a factor three spread around this fit, the paler grey region represents a spread of a factor 10. See Sect.~\ref{subsubsect:mdotversusperiod} for further details.   \label{fig:mdotperiodfit}}
\end{figure}

\subsubsection{AGB mass loss as a function of pulsation period}\label{subsubsect:mdotversusperiod}
The $\overline{\Mdot}$-values for all AGB stars in the sample --- except for the RV\,Tauri star R\,Sct ---  are plotted versus pulsation period in Fig.~\ref{fig:mdotperiodfit} with error bars indicating the spread as listed in Table~\ref{tbl:mdotestimates}. In case only one or two estimates could be obtained, we assumed a spread of a factor three, which is appropriate considering the overall spread in the sample. The red data points in the graph were taken from \cite{he2001} and cover different types of \ohir stars with a large spread on the periods of pulsation. Five objects in the sample of \cite{he2001} are also present in our sample. Their pulsation periods and mass-loss rates are in accordance with our estimates, except in the case of V669\,Cas, where \citeauthor{he2001} use $P=1995$\,days \citep{slootmaker1985}, while we adopted the more recently determined $P=1525$\,days \citep{groenewegen1999}. In constructing a linear relation between the pulsation period and the logarithm of the mass-loss rate, we only took into account the data points in our sample --- i.e. we did not consider the data points of \cite{he2001} --- with periods shorter than 1000\,days. The positions in the graph of the objects with periods longer than 850\,days are represented by a constant \mdot of $3.4\times10^{-5}$\,\msun\,yr$^{-1}$. The linearly increasing part of the fit ($P\lesssim850$\,days) was constructed to represent the Miras, the short-period \ohir stars and the AGB semi-regulars. The constant fitting value represents a levelling-off or saturation of the AGB mass-loss rate for stars with long periods, as proposed by, e.g., \cite{vassiliadis1993}. The lack of data points for stars with longer periods complicates the determination of the cut-off period and the mean \mdot-value for the long-$P$ objects. Both were more or less arbitrarily chosen to yield a good fit. The fit in Fig.~\ref{fig:mdotperiodfit} is represented by 
\[\begin{array}{lcllr}
\log(\Mdot)&=&-7.37+3.42\times10^{-3}\times P & \qquad & P \lesssim 850\,\mathrm{days}
\\
&=&-4.46\, & \qquad & P \gtrsim 850\,\mathrm{days}\\
\end{array}\]
with \mdot in units of \msun\,yr$^{-1}$ and $P$ in days. Although \cite{he2001} note that the quality of their derived mass-loss rates may be questionable, we find that their data points of both short and long-period targets are well represented by our fit. The targets that deviate strongest from the fit are semi-regular variables, e.g. W\,Hya and U\,Hya, and the Miras NX\,Ser (O-rich) and W\,Aql (S-type). We note that in constructing the fit we did not take into account the effect of differences in stellar mass that are likely to cause extra spread on the mass-loss rate for fixed pulsation periods, as shown by \cite{vassiliadis1993}. We also point out the very good agreement between our fit and the results presented by \cite{schoeier2001}, who produced a similar diagram of \mdot versus pulsation period for a sample of 68 optically bright carbon stars (their Fig.\,12), and found a relation $\dot{M}\propto P^{2.3}$ for periods $P$ between 100 and 800 days.

\begin{table}\centering \setlength{\tabcolsep}{1.mm}
\caption{Line intensity ratios $I_{\mathrm{MB}}(^{12}\mathrm{CO})/I_{\mathrm{MB}}(^{13}\mathrm{CO})$ for the sample stars give first order estimates of their isotope abundance ratio \twthratio. An absolute flux error of $20\,\%$ for transitions $J=2-1,\;3-2$, and $4-3$ gives an error of $28\,\%$ on the respective line intensity ratios. For $J=6-5$, the absolute flux error is of the order of $30\,\%$, so the error on the line intensity ratio is about $42\,\%$. The \twthratio-value listed in the second to last column is the mean line intensity ratio, corrected with a factor 0.87 for the difference in line strengths (Eq.~\ref{eq:12c13c-corr}). The last column (M09) lists the values derived by \cite{milam2009} from CO data.}
\label{tbl:twthratio}
 \begin{tabular}{lrrrrrr}\hline\hline\\[-2ex]
		&\multicolumn{4}{c}{$I_{\mathrm{MB}}(^{12}\mathrm{CO})/I_{\mathrm{MB}}(^{13}\mathrm{CO})$}
						&		& M09	\\
Target  		&	$2-1$		&	$3-2$		&	$4-3$		&	$6-5$		&	\twthratio	&	\twthratio\\
\hline\\[-2ex]												
\textit{O-rich} 	&			&			&			&			&			&	\\
RX\,Boo			&	12.0	&	10.5	&	-	&	-	&	9.8	&	-\\
EP\,Aqr          	&	9.5	&	6.2	&	12.3	&	-	&	8.1	&	-\\
T\,Cet			&	11.5	&	14.4	&	-	&	-	&	11.2	&	-\\
o\,Cet           	&	12.2	&	8.0	&	-	&	3.3	&	6.8	&	-\\
IK\,Tau          	&	8.7	&	6.8	&	-	&	-	&	6.7	&	10\\
TX\,Cam			&	17.1	&	13.3	&	-	&	-	&	13.2	&	31\\
NV\,Aur          	&	5.4	&	-	&	-	&	-	&	4.7	&	-\\
R\,Hya			&	6.1	&	-	&	-	&	-	&	5.3	&	-\\
R\,Cas			&	20.8	&	12.4	&	-	&	-	&	14.4	&	-\\
WX\,Psc          	&	6.2	&	5.5	&	-	&	-	&	5.1	&	-\\
AFGL\,5379		&	8.3	&	-	&	-	&	-	&	7.2	&	-\\
OH\,44.8-2.3		&	3.8	&	7.6	&	-	&	-	&	4.9	&	-\\
NML\,Cyg		&	10.0	&	-	&	-	&	-	&	8.7	&	13\\
$\alpha$\,Ori		&	6.5	&	-	&	-	&	-	&	5.7	&	8\\
VY\,CMa          	&	19.7	&	14.3	&	-	&	-	&	14.8	&	25, 33, 46\\
VX\,Sgr			&	-	&	22.0	&	-	&	-	&	19.2	&	-\\
IRC\,+10420      	&	9.0	&	9.7	&	-	&	-	&	8.1	&	14\\
AFGL\,2343       	&	-	&	3.8	&	-	&	-	&	3.3	&	-\\
			&		&		&		&		&		&	\\
\textit{S-type} 	&		&		&		&		&		&	\\
$\pi^1$\,Gru		&	-	&	35.3	&	-	&	-	&	30.7	&	-\\  
W\,Aql           	&	16.1	&	12.8	&	-	&	-	&	12.6	&	-\\
$\chi$\,Cyg 		&	-	&	16.2	&	-	&	-	&	14.1	&	33\\
			&		&		&		&		&		&	\\
\textit{C-rich} 	&		&		&		&		&		&	\\
V\,Hya           	&	-	&	30.7	&	-	&	-	&	26.7	&	71, 72\\
R\,Scl           	&	-	&	10.6	&	-	&	-	&	9.2	&	-\\
II\,Lup          	&	-	&	4.6	&	-	&	-	&	4.0	&	-\\
IRC\,+20370      	&	-	&	8.1	&	-	&	-	&	7.1	&	-\\
V\,Cyg           	&	13.8	&	-	&	-	&	-	&	12.0	&	-\\  
LL\,Peg          	&	-	&	4.2	&	-	&	-	&	3.6	&	-\\   

\hline
\end{tabular}
\end{table}

\begin{figure}
 \includegraphics[angle=90,width=\linewidth]{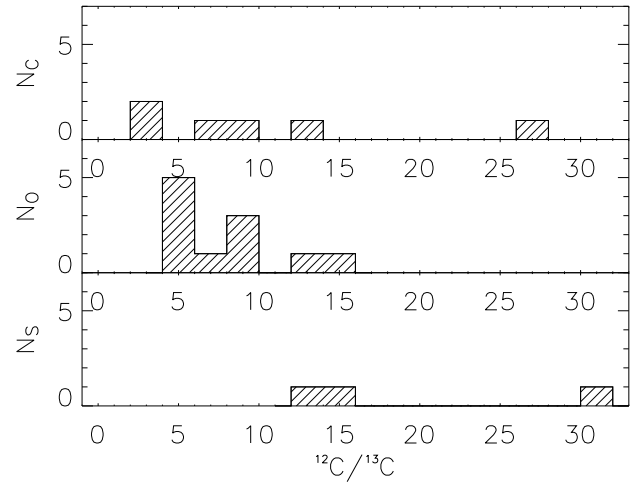}
\caption{Histogram with empirical estimates of the \twthratio isotope abundance ratio for the AGB stars in our sample with an adopted bin size of two units. The targets are grouped per characteristic chemistry: \crich objects (top panel), \orich targets (middle panel), and S-type targets (bottom panel). \label{fig:histo-12c13c}}
\end{figure}

\subsection{\twthratio isotope abundance ratio}\label{subsubsect:12C/13C}
For 29 sample stars we have observations of both \twco and \thco rotational lines, allowing us to estimate the \twthratio isotope abundance ratio. This ratio is a tracer of the star's chemical evolutionary state and nucleosynthesis \citep{sopka1989,schoeier2000,garciahernandez2007,milam2009}. \thc is produced during the interpulse phases when, after dredge-up, protons are partially mixed in the region between the hydrogen and helium burning shells. Reactions of the protons with \twc in this intershell region form \thc. 
In massive AGB stars ($3$\,\msun$\leq$ \mstar$\leq7$\,\msun), ``hot bottom burning'' (HBB) can be activated when the convective envelope penetrates the hydrogen-burning shell. During HBB, \twthratio is lowered, since the net production of \thc is higher than that of \twc in the re-activated CN-cycle. For HBB-stars, a ratio \twthratio  around 10 is expected, but could possibly range down to values about 3 or even 1 \citep{ohnaka1999}. HBB is expected to be active in the most massive and luminous oxygen-rich AGB stars. In this respect, \ohir stars --- typified by high luminosities, large infrared excesses and high mass-loss rates --- are the best candidates for having very low \twthratio isotope ratios \citep[][and references therein]{garciahernandez2007}. 

Observational \twthratio-values for AGB stars range from as low as 1 \citep{ohnaka1999} up to about 90 \citep{milam2009}. \citet[and references therein]{lebzelter2008} mention some clustering of empirical values of \twthratio between 50 and 70 for galactic C-rich stars, while their evolutionary models even predict values higher than 100. The solar value is about 89 \citep{anders1989}. \\

Line strength ratios of \twco and \thco transitions with identical $J$ provide a first order estimate of the isotope abundance ratio in case both lines are optically thin \citep{schoeier2000}. In case of optically thick lines, these estimates are only tentative, and lower than the actual isotopic ratio. \cite{schoeier2000} mention that the \twthratio values derived from their radiative transfer models are mostly higher than those estimated from intensity ratios. This effect is largest for their high \mdot-objects, where the difference amounts to a factor 6. 

We note here that the line intensity ratios for $J=2-1$ are consistently larger than those for $J=3-2$ for our sample stars. A linear least-squares fit for the targets with line intensity ratios available for both $J=2-1$ and $J=3-2$ shows that \[\frac{I(^{12}\mathrm{CO}(3-2))}{I(^{13}\mathrm{CO}(3-2))}=4.3+0.48\times\frac{I(^{12}\mathrm{CO}(2-1))}{ I(^{13}\mathrm{CO}(2-1))}.\]
The standard deviation from this linear fit is 2.1 units in $I(^{12}\mathrm{CO}(3-2))/I(^{13}\mathrm{CO}(3-2))$. The slope and offset in this relation clearly indicate that individual line intensity ratios will not lead to accurate values of the actual \twthratio abundance ratio and can only give rough estimates.

We have calculated \twthratio-values for our sample targets, correcting for line strength differences and --- for the sake of simplicity --- assuming the targets were unresolved by the beam, in the way this was done by \cite{schoeier2000} and as is given by Eq.~\ref{eq:12c13c-corr}. 
\begin{equation}
 \label{eq:12c13c-corr}
^{12}\mathrm{C}/^{13}\mathrm{C}
=\frac{I_{\mathrm{MB}}(^{12}\mathrm{CO}(J\rightarrow J-1))}{I_{\mathrm{MB}}(^{13}\mathrm{CO}(J\rightarrow J-1))}\times\left(\frac{\nu_{^{12}\mathrm{CO}(J\rightarrow J-1)}}{\nu_{^{13}\mathrm{CO}(J\rightarrow J-1)}}\right)^{-3}
\end{equation}
The frequency correction factor is equal to 0.87 for all values of $J$ considered. If multiple line intensity ratios were available for one $J$, these were averaged. The line intensity ratios and the resulting values for \twthratio are listed in Table~\ref{tbl:twthratio}. Fig.~\ref{fig:histo-12c13c} shows a histogram plot of the \twthratio-values for the AGB stars in the sample. It is clear from this graph that we derive very low values for \twthratio, but as mentioned before, these estimates are lower limits in case of large mass-loss rate objects due to optical depth effects, that are not taken into account using this approach. Since high-\mdot objects are more easily observed, the sample is biased towards the higher mass-loss rates. The values for the 12 Miras range from 3.6 (\crich LL\,Peg) up to 14.4 (\orich R\,Cas) and for the 3 \ohir stars they range from 4.9 (OH\,44.8-2.3) up to 7.2 (AFGL\,5379). The spread on \twthratio for the other stars in the sample is quite large and we do not see any clear differences in the first-order estimates of \twthratio when considering the different types of stars in the sample.

\cite{milam2009} found \twthratio$\sim$13 for NML\,Cyg and \twthratio$\sim$14 for IRC\,+10420, while for VY\,CMa they derived a much higher value between 25 and 46. We find  similar ratios with a first order estimate for VY\,CMa of 14.8, and values of 8.7 and 8.1 for NML\,Cyg and IRC\,+10420, respectively. We note that \cite{milam2009} deconvolved the line profiles for VY\,CMa and V\,Hya into separate velocity components, since the different structures probed could have discrepant chemical properties \citep[see also][for VY\,CMa]{ziurys2007}. They analysed the multiple components for \twthratio separately, resulting in three different values for VY\,CMa (46, 33 and 25) and 2 values for V\,Hya (71 and 72). \cite{milam2009} claim a $\pm30$\,\% uncertainty on their \twthratio-values.

\subsection{Wind driving efficiency}
\begin{figure}
 \includegraphics[angle=90,width=\linewidth]{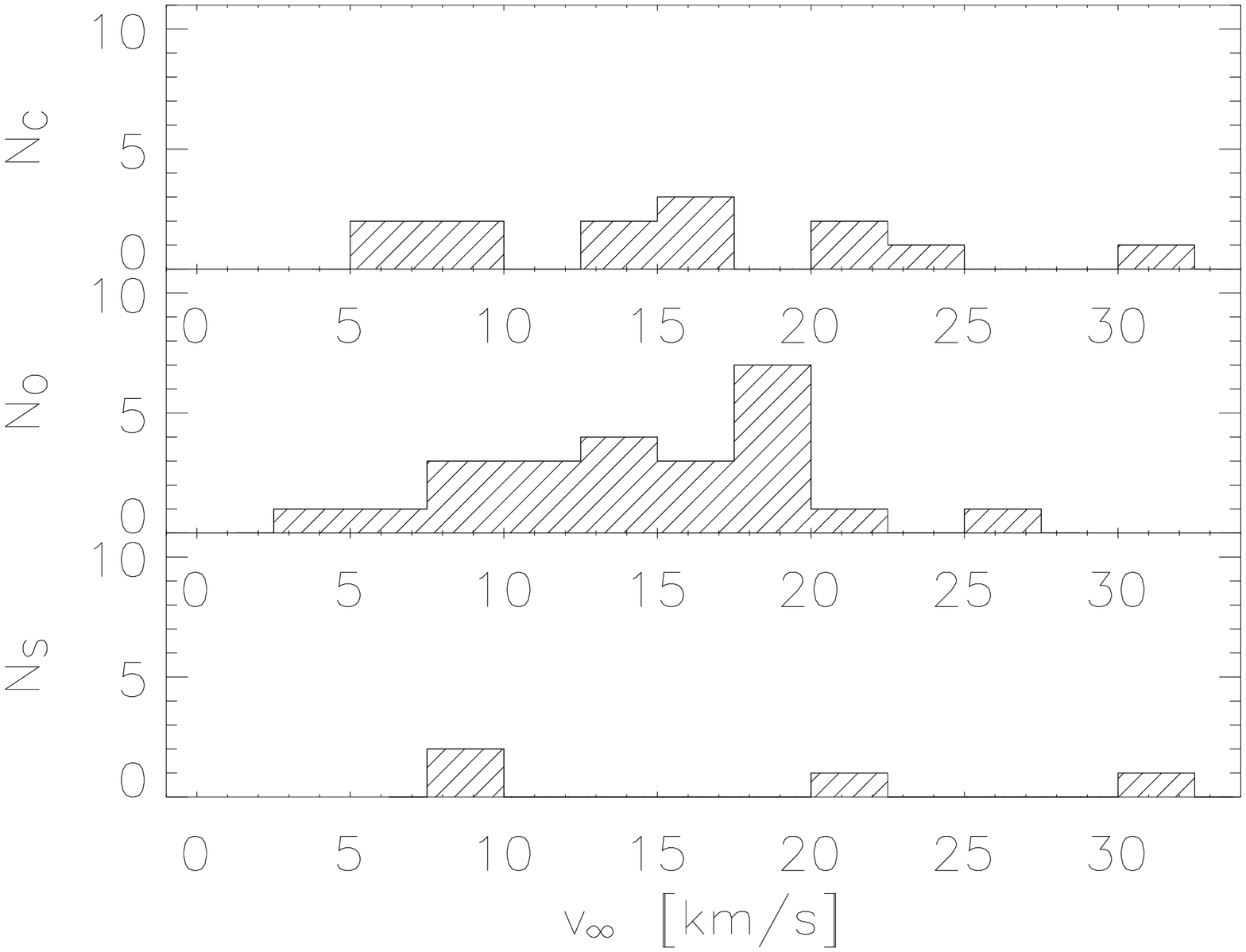} 
\caption{Comparative histogram plots of the expansion velocities (\vinfty) of the AGB stars in the sample. The bin size was set to $\Delta \Vinfty =2.5$\,km\,s$^{-1}$. The targets are grouped per characteristic chemistry: \crich objects (top panel), \orich targets (middle panel), and S-type targets (bottom panel). \label{fig:vinfty}}
\end{figure}

\cite{olofsson1993} and \cite{oudmaijer2008} suggest that the acceleration of winds around C-rich stars is more efficient than that of winds around O-rich stars because of differences in dust absorption properties. Comparing samples of 40 S-type, 77 \orich and 61 \crich stars, \cite{ramstedt2006} also find indications for a higher terminal velocity in \crich stars. 

Fig.~\ref{fig:vinfty} shows histograms of \vinfty in our sample for the three different chemical types of objects: \orich, \crich and S-type stars. The mean \vinfty-values for respectively 24 \orich and 13 \crich AGB targets in our sample are  14.5\,km\,s$^{-1}$ and 15.4\,km\,s$^{-1}$. These values are $\sim$4--10\,km\,s$^{-1}$ higher than those of \cite{ramstedt2006}. We note that the largest \vinfty-values for our \crich and S-type sample exceed those of \cite{ramstedt2006} by  about 10\,km\,s$^{-1}$. These extreme values are however the only ones in both samples --- V\,Hya in the C-rich and $\pi^1$\,Gru in the S-type sample --- so they are only weakly reflected in the samples' mean values. Our results do not lead to the same clear conclusion that the wind driving efficiency is higher for \crich than for \orich stars. They do, however, point in the same direction considering the sampling bias we have in favour of \orich objects and the occurrence of higher \vinfty-values for \crich than for \orich targets in our sample.


\section{Conclusions}\label{sect:conclusions}
The data set presented in this paper covers a very large number of high-quality measurements of CO rotational line transitions $J \rightarrow J-1$ with $J$ ranging from 1 up to 7. Observations of both \twco and \thco are included. It is the first time that a CO data set allowing one for such a detailed sampling of circumstellar envelopes of evolved objects is presented in the literature. 

A grid of radiative transfer models calculated with \gastronoom covering AGB stellar and wind parameters was used to derive a formalism allowing one to make estimates of mass-loss rates. The \mdot-estimator is valid for $J$ up to 16, while until now, only estimators valid for $J$ up to 4 have been presented in the literature. We also extended the validity range of the \mdot-estimator by correcting for saturation of the CO line intensities and therefore accounting for optically thick winds. The intrinsic spread of the estimator ensures good \mdot-estimates when all input parameters are well determined. We like to point out that, when only one transition is used to estimate a mass-loss rate, the $J=2-1$ transition will give the most reliable results.

For many of the sample AGB stars, the estimates based on the different rotational lines result in constant \mdot-values. The estimates for 7 targets do however hint at a variability of the mass-loss rate, by showing a large spread (more than an order of magnitude in \mdot) and clear trends. Uncertainties linked to the implemented \h2o-cooling could produce the effect of seemingly decreasing \mdot in time, but there are two convincing cases in which these uncertainties cannot account for the variations seen in the estimates. This is the case for  LL\,Peg and WX\,Psc, and we have shown that the \mdot-estimates for the latter target reflect the mass-loss history constructed through detailed radiative-transfer modelling by \cite{decin2007_wxpsc}. The fact that we \textit{(1)} do not have full sampling of all rotational $J$-levels\footnote{The rotational transitions CO($J\rightarrow J-1$) observable with ground-based telescopes are limited to $J=1-0,\, 2-1,\, 3-2,\, 4-3,\, 6-5$ and $7-6$. Many of the higher-frequency transitions will be observed with the \hifi and \pacs instruments on board the Herschel Space Observatory.} for all sample targets and that \textit{(2)} did not perform full radiative-transfer modelling for the sample, of course limit our ability to prove conclusively that a star has experienced a constant mass loss or mass-loss modulations. This kind of detailed modelling will be presented in a forthcoming paper by Lombaert et al. (in preparation). The examples discussed in this paper show, however, that the use of the estimator allows one not only to establish good mass-loss-rate estimates in case of constant mass-loss targets, but also to trace strong variability in the mass-loss rate of AGB stars, given a sufficiently large number of CO line transitions. The fact that our estimates bear quite close resemblance to the mass loss derived through detailed radiative transfer modelling, gives confidence that our estimator is of good quality. 

Further observations of mostly high-excitation rotational lines, such as $J=6-5$ and $J=7-6$ would make it possible to make a better distinction between constant or (possibly) variable mass loss. The $J=5-4$ transition will be measured by \hifi for many stars in our sample. These additional measurements will provide a more complete sampling of the circumstellar envelopes and the mass-loss history of these evolved stars.

\begin{acknowledgements}
	The authors wish to thank Sofie Dehaes and Sacha Hony for their contribution to the data reduction and Hans Olofsson and David Teyssier for their contributions to the data set presented in this paper. Carlos De Breuck from ESO is acknowledged for the helpful discussions on APEX data reduction. The authors thank Jan Cami for proposing and planning the observations of EP Aqr. We thank Joris Blommaert and Rens Waters for the many fruitful discussions. E.D. and L.D. acknowledge support from the Fund for Scientific Research of Flanders (FWO) under  grant number G.0470.07. The computations for this research have been done with the VIC HPC Cluster of the KULeuven. We are grateful to the LUDIT HPC team for their support. This publication makes use of data products from the Two Micron All Sky Survey, which is a joint project of the University of Massachusetts and the Infrared Processing and Analysis Center, funded by the National Aeronautics and Space Administration and the National Science Foundation. APEX data were taken  for programmes O-77.F-0605-2006, E-77.D-4004-2006, E-77.D-0781-2006, E-078.D-0534A-2007, and E-081.D-0207A-2008. The JCMT data were obtained in the framework of the science programmes M00AN09, M00BN19, M01AN04, M01BN10, M02BN05, M03AN02, M03BN06, M04AN20, M06BN03, M08AN01, and M08BN11. 
\end{acknowledgements}

\clearpage

\Online
\begin{appendix}
\section{CO data -- figures and tables}\label{sec:app_data}
\begin{figure*} 
\includegraphics[angle=180,width=0.9\linewidth]{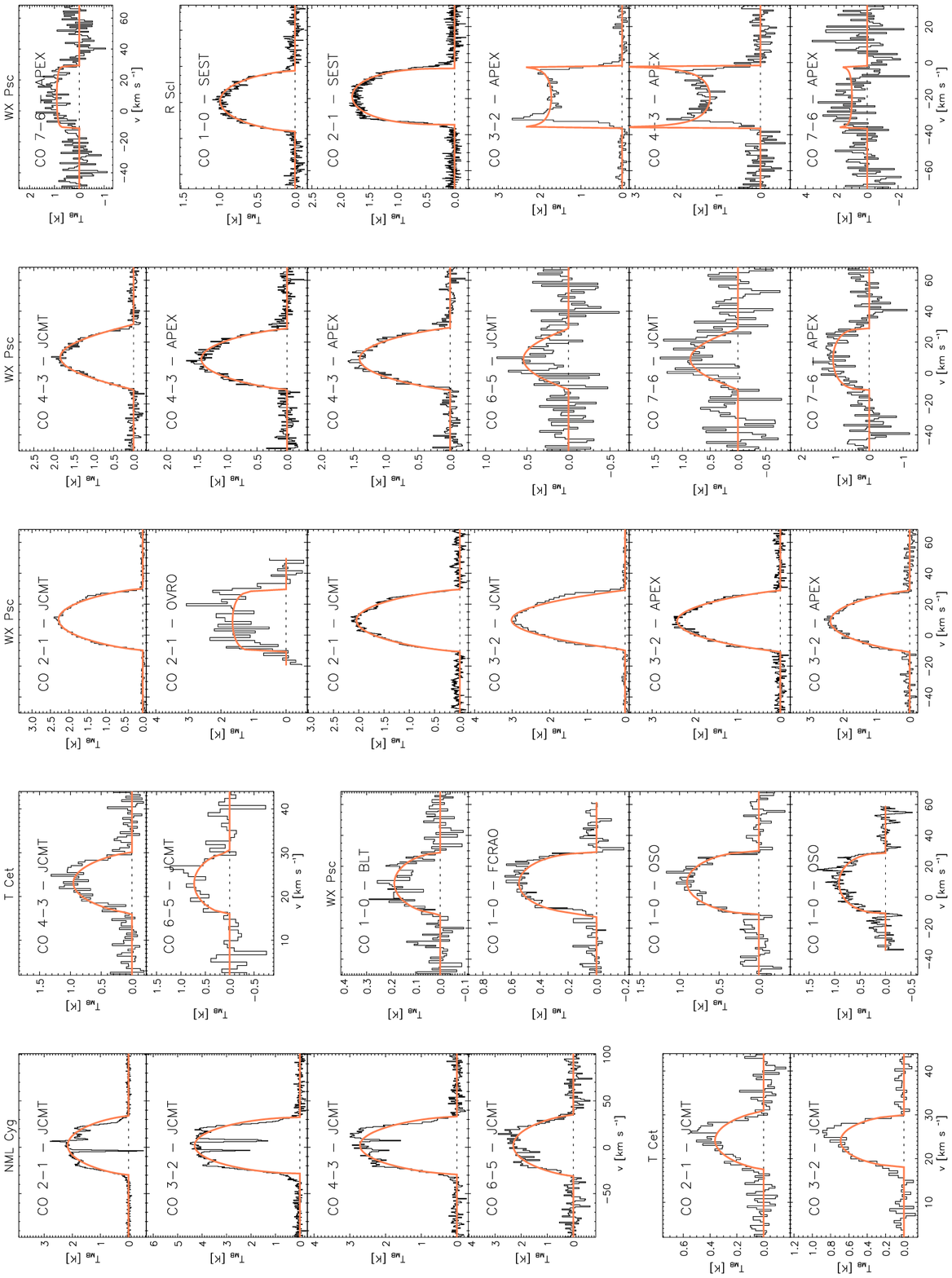}
\caption{Profiles of \twco with fitted soft-parabola profiles.}
\label{fig:12CO_parabola}
\end{figure*}

All \twco and \thco data of the sample  and the soft-parabola fits (Sect.~\ref{subsubsect:softparabola}) to these data are shown in Figs.~\ref{fig:12CO_parabola} and \ref{fig:13CO_parabola}. The most important line parameters --- main-beam temperature at the line centre and  integrated intensity --- and the $\beta$ parameter from the soft-parabola fitting procedure as described in Sect.~\ref{subsubsect:softparabola} are given in Tables~\ref{tbl:12CO} and \ref{tbl:13CO}. References for, or remarks on, the data are listed in Tables~\ref{tbl:12CO-REMARKS} and \ref{tbl:13CO-REMARKS} for the \twco and \thco data respectively. 

\longtabL{1}{\tiny\setlength{\tabcolsep}{0.5mm}
\begin{landscape}
\begin{longtable}{llrr|rrr|rrr|rrr|rrr|rrr|rrr}
 \caption[\twco-data presentation.]{Overview of all data presented in Fig.~\ref{fig:12CO_parabola}. The objects are ordered according to their IRAS numbers and are specified with their velocity in the frame of the Local Standard of Rest ($v_{\mathrm{LSR}}$) and the terminal velocity ($v_{\infty}$) of the CO envelope. For each measured transition, the main beam brightness temperature at the centre of the line ($T_{\mathrm{MB,c}}$) is given in K, together with the integrated intensity ($I_{\mathrm{MB}}$), given in K\,km\,s$^{-1}$. For the high-quality data, a soft parabola fit (see text) was performed and the according $\beta$-values are listed as the third entry per transition. In case of multiple observations of a specific transition, the data are given in the exact same order as in  Fig.~\ref{fig:12CO_parabola}.}\label{tbl:12CO}\\
\hline \hline \\[-2ex]
IRAS &Alternative name &$v_{\mathrm{LSR}}$ &$ v_{\infty}$ &\multicolumn{3}{c}{CO 1--0} &\multicolumn{3}{c}{CO 2--1} &\multicolumn{3}{c}{CO 3--2} &\multicolumn{3}{c}{CO 4--3} &\multicolumn{3}{c}{CO 6--5} &\multicolumn{3}{c}{CO 7--6} \\
  &  &\multicolumn{2}{c}{km/s} &$ T_{\mathrm{MB,c}} $ &$ I_{\mathrm{MB}} $ &$ \beta $ &$ T_{\mathrm{MB,c}} $ &$ I_{\mathrm{MB}} $ &$ \beta $ &$ T_{\mathrm{MB,c}} $ &$ I_{\mathrm{MB}} $ &$ \beta $ &$ T_{\mathrm{MB,c}} $ &$ I_{\mathrm{MB}} $ &$ \beta $ &$ T_{\mathrm{MB,c}} $ &$ I_{\mathrm{MB}} $ &$ \beta $ &$ T_{\mathrm{MB,c}} $ &$ I_{\mathrm{MB}} $ &$ \beta $ \\
\hline \\[-2ex]
\endfirsthead
\multicolumn{3}{c}{\textbf{{\tablename} \thetable{}.} Continued} \\[0.5ex]
\hline \hline \\[-2ex]
IRAS &Alternative name &$v_{\mathrm{LSR}}$ &$ v_{\infty}$ &\multicolumn{3}{c}{CO 1--0} &\multicolumn{3}{c}{CO 2--1} &\multicolumn{3}{c}{CO 3--2} &\multicolumn{3}{c}{CO 4--3} &\multicolumn{3}{c}{CO 6--5} &\multicolumn{3}{c}{CO 7--6} \\
  &  &\multicolumn{2}{c}{km/s} &$ T_{\mathrm{MB,c}} $ &$ I_{\mathrm{MB}} $ &$ \beta $ &$ T_{\mathrm{MB,c}} $ &$ I_{\mathrm{MB}} $ &$ \beta $ &$ T_{\mathrm{MB,c}} $ &$ I_{\mathrm{MB}} $ &$ \beta $ &$ T_{\mathrm{MB,c}} $ &$ I_{\mathrm{MB}} $ &$ \beta $ &$ T_{\mathrm{MB,c}} $ &$ I_{\mathrm{MB}} $ &$ \beta $ &$ T_{\mathrm{MB,c}} $ &$ I_{\mathrm{MB}} $ &$ \beta $ \\
\hline \\[-2ex]
\endhead
\hline\\[-2ex]
\multicolumn{3}{l}{{Continued on Next Page\ldots}} \\
\endfoot
\\[-1.8ex]
\endlastfoot
-&NML Cyg&    2.0&   33.0&- &- &- &   2.17 &  94.68 &   1.42 &   4.23 & 197.40 &   1.10 &   2.68 & 128.60 &   1.27 &   2.30 & 110.97 &   1.45 &- &- &- \\
00192-2020&T Cet&   23.0&    7.0&- &- &- &   0.35 &   3.91 &   1.45 &   0.69 &   7.18 &   1.12 &   0.95 &  10.12 &   1.45 &   0.73 &   8.26 &   1.14 &- &- &- \\
01037+1219&WX Psc&    9.0&   19.8&   0.19 &   3.61 &   1.10 &   2.29 &  65.64 &   1.53 &   2.98 &  80.79 &   1.95 &   1.88 &  49.81 &   2.16 &   0.54 &   8.59 &   2.15 &   0.86 &  22.78 &   2.27 \\
 " & " & " & " &   0.55 &  51.70 &   0.84 &   1.64 &  59.76 &   0.26 &   2.44 &  66.43 &   1.72 &   1.42 &  40.76 &   1.56 & - &- &- &   1.06 &  34.78 &   0.59 \\
 " & " & " & " &   0.92 &  25.45 &   0.99 &   2.11 &  59.32 &   1.47 &   2.43 &  67.15 &   1.60 &   1.44 &  40.21 &   1.66 & - &- &- &   0.90 &  36.45 &   0.15 \\
 " & " & " & " &   0.91 &  26.95 &   0.80 & - &- &- & - &- &- & - &- &- & - &- &- &- &- &- \\
01246-3248&R Scl&  -17.0&   17.0&   0.98 &  24.86 &   1.21 &   1.73 &  47.60 &   0.70 &   1.72 &  54.69 &  -0.22 &   1.21 &  45.70 &  -0.77 &- &- &- &   0.99 &  25.91 &  -0.33 \\
01304+6211&V669 Cas&  -55.0&   13.0&- &- &- &   0.28 &   1.72 &   2.80 &   0.68 &  10.77 &   1.99 &   0.14 &   2.12 &   0.08 &   0.35 &   7.12 &  -0.49 &- &- &- \\
02168-0312&o Cet&   46.4&    8.1&   1.25 &   8.09 &   6.89 &  13.79 &  63.74 &  17.16 &  21.76 & 107.17 &  15.33 &  20.01 & 101.63 &  12.43 &  16.25 &  86.23 &   9.81 &- &- &- \\
03507+1115&IK Tau&   33.8&   18.5&   1.68 &  63.35 &  -0.75 &   4.54 & 150.35 &   0.23 &   3.04 &  87.02 &   0.99 &   4.36 & 127.27 &   0.92 &- &- &- &   5.04 & 122.61 &   1.91 \\
 " & " & " & " & - &- &- &   2.93 &  89.62 &   0.54 &   4.04 & 115.71 &   1.01 &   4.25 & 119.43 &   1.09 & - &- &- &- &- &- \\
 " & " & " & " & - &- &- &   5.97 & 200.84 &   0.12 &   3.23 &  94.22 &   1.01 & - &- &- & - &- &- &- &- &- \\
04566+5606&TX Cam&   10.8&   21.2&- &- &- &   2.34 &  63.58 &   2.10 &   5.41 & 144.84 &   2.22 &   2.77 &  76.30 &   2.24 &- &- &- &- &- &- \\
05073+5248&NV Aur&    3.0&   19.2&- &- &- &   1.38 &  36.31 &   1.75 &   1.51 &  38.76 &   1.91 &   1.35 &  35.43 &   2.21 &   0.37 &   8.88 &   1.10 &- &- &- \\
05524+0723&Alpha Ori&    3.5&   14.0&- &- &- &   0.02 &  -0.29 &- &   1.98 &  49.06 &   0.25 &   2.60 &  60.01 &   0.65 &- &- &- &- &- &- \\
 " & " & " & " & - &- &- &   0.66 &  16.00 &   0.06 & - &- &- & - &- &- & - &- &- &- &- &- \\
06176-1036&Red Rectangle&    0.3&    8.3&- &- &- &- &- &- &   0.84 &   4.60 &  16.94 &- &- &- &- &- &- &- &- &- \\
07209-2540&VY CMa&   21.0&   46.5&   0.07 &  44.20 &  10.48 &   0.98 &  65.40 &   1.59 &   2.85 & 171.76 &   2.00 &   6.48 & 405.72 &   1.84 &   4.84 & 247.31 &   3.69 &   7.65 & 423.46 &   3.12 \\
07399-1435&Calabash Nebula&   40.0&  110.0&- &- &- &- &- &- &   1.73 & 111.03 &  19.67 &- &- &- &- &- &- &- &- &- \\
09448+1139&R Leo&   -1.0&    9.0&   0.37 &   4.07 &   2.01 &   2.19 &  28.43 &   1.81 &   2.99 &  33.54 &   2.93 &- &- &- &   4.46 &  38.10 &   4.49 &- &- &- \\
 " & " & " & " & - &- &- & - &- &- &   1.89 &  22.63 &   1.47 & - &- &- & - &- &- &- &- &- \\
09452+1330&CW Leo&  -25.0&   14.5&  17.44 & 411.94 &   0.53 &  38.17 & 830.58 &   1.04 &  40.92 & 852.24 &   1.36 &  30.08 & 639.17 &   1.40 &  58.25 &1130.61 &   1.92 &- &- &- \\
 " & " & " & " &   9.68 & 222.79 &   0.99 &  23.27 & 491.83 &   2.13 &  31.98 & 672.09 &   1.41 & - &- &- & - &- &- &- &- &- \\
 " & " & " & " &   9.49 & 230.77 &   0.38 & - &- &- & - &- &- & - &- &- & - &- &- &- &- &- \\
10131+3049&RW LMi&   -1.8&   20.8&   4.10 & 132.65 &   0.77 &   7.75 & 215.61 &   1.94 &   3.02 &  74.82 &   3.04 &   4.21 & 126.23 &   2.04 &   7.03 & 182.66 &   2.77 &- &- &- \\
10197-5750&GSC 08608-00509&   -5.0&   28.0&- &- &- &- &- &- &   0.75 &  24.76 &   3.09 &- &- &- &- &- &- &- &- &- \\
10329-3918&U Ant&   24.0&   20.5&   0.22 &   9.96 &- &   0.32 &  11.53 &- &   0.50 &   9.59 &- &   0.33 &   2.76 &- &- &- &- &   0.05 &  -4.69 &- \\
10350-1307&U Hya&  -31.0&    8.5&   0.42 &   4.79 &   1.72 &   0.91 &  10.76 &   4.28 &   2.64 &  29.28 &   2.51 &   2.94 &  30.64 &   2.81 &- &- &- &   3.29 &  25.13 &   5.11 \\
10491-2059&V Hya&  -17.0&   30.0&   0.81 &  22.93 &   7.12 &   1.47 &  37.99 &   8.87 &   3.47 & 105.43 &   5.42 &   1.53 &  53.47 &   3.75 &- &- &- &   1.31 &  16.81 &- \\
12427+4542&Y CVn&   21.0&    7.5&   0.93 &   9.07 &   0.87 &   1.88 &  24.62 &   0.43 &- &- &- &   1.36 &  17.00 &   0.37 &   1.25 &  16.85 &   0.48 &- &- &- \\
13269-2301&R Hya&  -10.0&   12.5&   0.13 &   1.58 &   5.23 &   1.11 &  13.00 &   5.51 &   4.01 &  41.49 &   7.76 &- &- &- &   6.44 &  51.68 &  13.55 &- &- &- \\
 " & " & " & " & - &- &- &   1.42 &  17.02 &   5.82 &   1.64 &  14.82 &   9.03 & - &- &- & - &- &- &- &- &- \\
 " & " & " & " & - &- &- &   0.69 &   6.35 &   6.31 & - &- &- & - &- &- & - &- &- &- &- &- \\
13428-6232&GLMP 363&-&-&- &- &- &- &- &- &  -0.00 &   2.41 &- &- &- &- &- &- &- &- &- &- \\
13462-2807&W Hya&   41.0&    8.5&- &- &- &- &- &- &   2.61 &  36.50 &   0.44 &   3.06 &  41.08 &   0.95 &- &- &- &   3.06 &  32.06 &   2.21 \\
14219+2555&RX Boo&    2.0&    9.0&   1.10 &  16.47 &   0.53 &   3.19 &  50.12 &   0.36 &   3.93 &  58.09 &   0.60 &- &- &- &   0.43 &   4.13 &   4.60 &- &- &- \\
 " & " & " & " & - &- &- &   2.06 &  30.97 &   0.70 &   1.54 &  23.30 &   0.75 & - &- &- &   2.57 &  30.02 &   3.05 &- &- &- \\
15194-5115&II Lup&  -15.0&   23.0&- &- &- &- &- &- &   4.66 & 144.41 &   1.64 &   4.93 & 155.90 &   1.61 &- &- &- &   4.89 & 151.43 &   1.77 \\
 " & " & " & " & - &- &- & - &- &- &   4.55 & 144.38 &   1.52 &   3.96 & 128.39 &   1.40 & - &- &- &   3.18 &  82.09 &   3.28 \\
16262-2619&Alpha Sco&-&-&- &- &- &- &- &- &   0.02 &   0.46 &- &- &- &- &- &- &- &- &- &- \\
16269+4159&G Her&   20.4&   13.0&- &- &- &   0.41 &   6.15 &   4.29 &   1.33 &  23.21 &   3.30 &- &- &- &   1.18 &  17.02 &   2.88 &- &- &- \\
17123+1107&V438 Oph&    9.8&    4.2&- &- &- &   0.20 &   1.19 &   1.99 &   0.31 &   1.60 &   1.75 &- &- &- &- &- &- &- &- &- \\
17150-3224&Cotton Candy Nebula&   15.0&   24.0&- &- &- &- &- &- &   0.78 &  18.84 &   5.15 &- &- &- &- &- &- &- &- &- \\
17411-3154&AFGL 5379&  -21.2&   25.0&- &- &- &   1.59 &  49.95 &   2.65 &   2.81 &  80.36 &   3.41 &   2.36 &  64.80 &   3.38 &- &- &- &   1.83 &  54.44 &   3.00 \\
 " & " & " & " & - &- &- & - &- &- &   1.76 &  50.45 &   3.48 & - &- &- & - &- &- &- &- &- \\
17443-2949&PN RPZM 39&-&-&- &- &- &- &- &- &   0.24 &   5.26 &- &- &- &- &- &- &- &- &- &- \\
17501-2656&V4201 Sgr&-&-&- &- &- &   0.52 &   1.00 &- &   0.30 &   0.11 &- &- &- &- &- &- &- &- &- &- \\
18050-2213&VX Sgr&    6.5&   24.3&- &- &- &   0.74 &  32.47 &   0.30 &   2.41 &  94.47 &   0.73 &   1.18 &  42.12 &   0.95 &- &- &- &- &- &- \\
18059-3211&Gomez Nebula&    2.7&    6.0&- &- &- &   0.47 &   2.79 &   3.86 &   0.91 &   4.37 &   8.95 &- &- &- &- &- &- &- &- &- \\
18100-1915&OH 11.52 -0.58&-&-&- &- &- &- &- &- &  -0.60 &  -3.27 &- &   0.02 &   1.91 &- &- &- &- &  -0.66 &   0.07 &- \\
18257-1000&V441 Sct&  115.0&-&- &- &- &- &- &- &   0.23 &   2.94 &- &   0.13 &   4.76 &- &- &- &- &  -0.06 &  -7.52 &- \\
18308-0503&AFGL 5502&   43.0&   14.7&- &- &- &  43.96 & 253.44 &  40.33 &  45.72 & 289.44 &  33.42 &  52.00 & 301.50 &  38.30 &  16.27 &  94.75 &  34.52 &- &- &- \\
18327-0715&OH 24.69 +0.24&   40.0&-&- &- &- &- &- &- &  -0.09 &  -7.40 &- &  -0.10 &  -2.95 &- &- &- &- &   0.91 &  -1.46 &- \\
18333+0533&NX Ser&   33.5&   17.8&- &- &- &   1.20 &  26.08 &   2.72 &   1.29 &  28.15 &   3.16 &   0.97 &  22.02 &   1.96 &   0.20 &   4.09 &   2.49 &- &- &- \\
18348-0526&OH 26.5+0.6&   27.4&   17.0&- &- &- &  -0.18 &   7.02 &- &   1.09 &  23.57 &   2.19 &   0.85 &  18.98 &   1.85 &- &- &- &- &- &- \\
18361-0647&OH 25.50 -0.29&   27.5&-&- &- &- &- &- &- &   1.76 &   3.29 &- &   2.31 &   6.62 &- &- &- &- &   0.05 &   1.57 &- \\
18397+1738&IRC +20370&   -0.5&   16.0&   3.65 &  84.14 &   1.12 &   4.52 &  93.39 &   2.07 &   2.93 &  60.34 &   2.21 &   3.45 &  66.09 &   2.85 &- &- &- &   3.10 &  49.77 &   3.51 \\
18432-0149&V1360 Aql&   40.0&-&- &- &- &- &- &- &  -0.41 &   6.38 &- &  -1.23 &  -2.43 &- &- &- &- &   1.47 &  -5.88 &- \\
18448-0545&R Sct&   57.0&    6.0&- &- &- &- &- &- &   0.77 &   3.85 &   8.17 &- &- &- &- &- &- &- &- &- \\
18460-0254&V1362 Aql&  100.0&-&- &- &- &   0.32 &  18.17 &- &  -0.35 &   6.81 &- &   0.02 &  -5.16 &- &- &- &- &   0.16 &   1.21 &- \\
 " & " & " & " & - &- &- & - &- &- &  -0.30 &  -7.24 &- & - &- &- & - &- &- &- &- &- \\
18476-0758&S Sct&   15.0&    8.0&   0.13 &   1.27 &- &   0.12 &   1.23 &- &   0.20 &   1.54 &- &- &- &- &- &- &- &- &- &- \\
18488-0107&V1363 Aql&   75.0&-&- &- &- &- &- &- &   2.21 &  18.75 &- &   0.64 &   9.98 &- &- &- &- &   0.01 &  -0.29 &- \\
18498-0017&V1365 Aql&   60.0&-&- &- &- &   0.35 &  10.72 &- &   0.36 &  13.07 &- &   0.34 &   8.50 &- &- &- &- &  -0.12 &   1.11 &- \\
 " & " & " & " & - &- &- & - &- &- &   0.07 &   3.99 &- & - &- &- & - &- &- &- &- &- \\
19067+0811&V1368 Aql&   60.0&-&- &- &- &- &- &- &  -0.62 &   1.50 &- &  -0.51 &   0.44 &- &- &- &- &   0.01 &   5.61 &- \\
19110+1045&KJK G45.07&   60.0&   25.0&- &- &- &- &- &- &- &- &- &  32.42 & 729.07 &   6.42 &- &- &- &- &- &- \\
19114+0002&V1427 Aql&   99.0&   40.0&- &- &- &- &- &- &   1.69 &  84.28 &   2.47 &- &- &- &- &- &- &- &- &- \\
19126-0708&W Aql&  -25.0&   20.0&   0.97 &  27.03 &   0.93 &   2.76 &  80.88 &   1.16 &   3.89 &  96.71 &   2.51 &   4.70 & 119.33 &   2.32 &- &- &- &   4.97 & 134.55 &   1.49 \\
 " & " & " & " &   0.99 &  26.93 &   1.71 &   1.88 &  51.58 &   2.13 &   4.75 & 118.44 &   2.51 &   4.06 &  97.28 &   2.21 & - &- &- &   1.73 &  29.21 &   0.31 \\
 " & " & " & " &   1.27 &  34.06 &   1.77 & - &- &- &   5.02 & 127.23 &   2.26 &   4.08 &  96.31 &   1.07 & - &- &- &- &- &- \\
 " & " & " & " & - &- &- & - &- &- &   2.16 &  52.98 &   2.49 & - &- &- & - &- &- &- &- &- \\
19192+0922&OH 44.8-2.3&  -70.5&   17.7&- &- &- &   0.38 &   9.67 &   1.42 &   0.63 &  15.40 &   1.99 &   0.41 &   9.99 &   1.50 &   0.53 &   9.59 &   2.57 &- &- &- \\
19244+1115&IRC +10420&   75.5&   42.3&   0.90 &  49.94 &   2.10 &   1.61 &  94.76 &   1.44 &   3.23 & 178.76 &   1.93 &   2.94 & 149.88 &   2.72 &   2.77 & 139.62 &   2.46 &- &- &- \\
 " & " & " & " & - &- &- &   2.46 & 158.99 &   0.87 &   3.75 & 227.95 &   1.73 &   2.10 & 118.49 &   1.81 &   1.63 &  80.74 &   2.08 &- &- &- \\
19283+1944&AFGL 2403&   29.0&   17.0&- &- &- &   0.54 &  10.48 &   3.01 &   0.24 &   5.53 &   2.30 &   0.40 &   8.83 &   2.43 &- &- &- &- &- &- \\
19486+3247&Chi Cyg&   10.5&    8.5&   3.20 &  35.55 &   0.38 &   6.32 &  90.86 &   0.26 &   3.49 &  47.70 &   0.75 &  10.25 & 125.99 &   1.30 &   1.45 &  23.08 &   0.14 &- &- &- \\
 " & " & " & " &   0.80 &  10.01 &   1.06 &   4.11 &  57.77 &   0.60 &   5.32 &  75.11 &   0.61 & - &- &- & - &- &- &- &- &- \\
 " & " & " & " & - &- &- & - &- &- &   5.46 &  75.53 &   0.62 & - &- &- & - &- &- &- &- &- \\
20075-6005&X Pav&  -18.5&   11.0&- &- &- &- &- &- &   1.24 &  20.18 &   1.14 &- &- &- &- &- &- &- &- &- \\
20077-0625&IRC -10529&  -18.0&   16.5&- &- &- &- &- &- &   1.91 &  36.97 &   3.27 &   1.56 &  33.48 &   2.13 &- &- &- &   0.49 &  19.74 &- \\
20120-4433&RZ Sgr&  -30.0&    9.0&- &- &- &- &- &- &   1.13 &  12.76 &   1.92 &   0.80 &   9.78 &   1.99 &- &- &- &   0.03 &  -1.88 &- \\
20396+4757&V Cyg&   15.0&   15.0&- &- &- &   2.65 &  43.61 &   2.62 &- &- &- &- &- &- &- &- &- &- &- &- \\
21419+5832&Mu Cep&   23.0&   35.0&- &- &- &   0.04 &   2.63 &- &   0.23 &  13.46 &   0.85 &   0.44 &  28.31 &  -0.20 &- &- &- &- &- &- \\
21439-0226&EP Aqr&  -33.5&   11.5&- &- &- &   5.12 &  30.07 &  75.47 &   7.88 &  42.15 &  72.53 &   7.94 &  42.19 &  68.50 &   7.33 &  31.35 &  99.32 &- &- &- \\
21554+6204&GLMP 1048&  -18.7&   18.6&- &- &- &   0.53 &  12.23 &   2.74 &   0.54 &  10.22 &   3.82 &   0.26 &   4.42 &   5.65 &   0.07 &   1.20 &   3.17 &- &- &- \\
22177+5936&OH 104.9+2.4&  -26.0&   18.3&- &- &- &   0.20 &   5.09 &   1.46 &   0.46 &  10.56 &   2.39 &   0.11 &   3.37 &   0.61 &   0.22 &   6.50 &   0.44 &  -0.54 & -10.81 &- \\
22196-4612&pi1 Gru&  -11.0&   30.0&- &- &- &- &- &- &   3.84 &  99.14 &   9.01 &   3.32 &  79.75 &  10.77 &- &- &- &   1.81 &  39.85 &   8.23 \\
23166+1655&LL Peg&  -31.0&   16.0&   4.75 & 101.41 &   2.02 &   6.45 & 118.99 &   3.23 &   4.50 &  76.16 &   4.00 &   4.85 &  87.58 &   3.43 &   1.02 &  22.40 &   1.95 &   7.79 & 152.69 &   2.73 \\
 " & " & " & " &   1.56 &  26.97 &   2.61 & - &- &- &   2.40 &  40.23 &   3.53 &   4.03 &  72.19 &   3.28 & - &- &- &- &- &- \\
23320+4316&LP And&  -17.0&   14.0&   4.28 &  83.92 &   1.16 &   7.90 & 153.59 &   1.72 &   3.81 &  67.60 &   2.25 &   4.14 &  74.17 &   2.34 &   3.09 &  54.90 &   1.85 &- &- &- \\
 " & " & " & " &   1.63 &  33.70 &   0.94 & - &- &- & - &- &- & - &- &- & - &- &- &- &- &- \\
23558+5106&R Cas&   25.5&   13.5&- &- &- &   2.68 &  56.45 &   1.14 &   5.20 &  99.03 &   2.08 &   6.05 & 109.97 &   2.21 &   5.01 &  76.20 &   3.42 &- &- &- \\
\hline
\end{longtable}
\end{landscape}}\clearpage\normalsize

\longtabL{2}{\tiny\setlength{\tabcolsep}{0.7mm}
\begin{landscape}
\begin{longtable}{ll|cc|cc|cc|cc|cc|cc}
 \caption[\twco-data origin.]{References for the presented \twco data. In case of multiple observations of a specific transition, the data are given in the exact same order as in  Fig.~\ref{fig:12CO_parabola}. References or remarks are (1) \cite{knapp1985}, (2) \cite{margulis1990}, (3) \cite{nyman1992}, (4) \cite{sopka1989}, (5) \cite{knapp1982}, (6) \cite{olofsson1993}, (7) \cite{neri1998}, (8) archive data, (9) \cite{teyssier2006}, (10) \cite{groenewegen1998}, (11) \cite{huggins1988}, (12) \cite{wang1994}, (13) \cite{wannier1986}, (14) \cite{bieging1994}, (-) this study.}\label{tbl:12CO-REMARKS}\\
\hline \hline \\[-2ex]
IRAS &Alternative name &\multicolumn{2}{c}{CO 1--0} &\multicolumn{2}{c}{CO 2--1} &\multicolumn{2}{c}{CO 3--2} &\multicolumn{2}{c}{CO 4--3} &\multicolumn{2}{c}{CO 6--5} &\multicolumn{2}{c}{CO 7--6} \\
  &  &Telescope &Reference &Telescope &Reference &Telescope &Reference &Telescope &Reference &Telescope &Reference &Telescope &Reference \\
\hline \\[-2ex]
\endfirsthead
\multicolumn{3}{c}{\textbf{{\tablename} \thetable{}.} Continued} \\[0.5ex]
\hline \hline \\[-2ex]
IRAS &Alternative name &\multicolumn{2}{c}{CO 1--0} &\multicolumn{2}{c}{CO 2--1} &\multicolumn{2}{c}{CO 3--2} &\multicolumn{2}{c}{CO 4--3} &\multicolumn{2}{c}{CO 6--5} &\multicolumn{2}{c}{CO 7--6} \\
  &  &Telescope &Reference &Telescope &Reference &Telescope &Reference &Telescope &Reference &Telescope &Reference &Telescope &Reference \\
\hline \\[-2ex]
\endhead
\hline\\[-2ex]
\multicolumn{3}{l}{{Continued on Next Page\ldots}} \\
\endfoot
\\[-1.8ex]
\endlastfoot
-&NML Cyg&- &- &JCMT &- &JCMT &- &JCMT &- &JCMT &- &- &- \\
00192-2020&T Cet&- &- &JCMT &- &JCMT &- &JCMT &- &JCMT &- &- &- \\
01037+1219&WX Psc&BLT &1 &JCMT &- &JCMT &- &JCMT &- &JCMT &- &JCMT &- \\
 " & " &FCRAO &2 &OVRO &5 &APEX &- &APEX &- &- &- &APEX &- \\
 " & " &OSO &3 &JCMT &- &APEX &- &APEX &- &- &- &APEX &- \\
 " & " &OSO &4 &- &- &- &- &- &- &- &- &- &- \\
01246-3248&R Scl&SEST &6 &SEST &6 &APEX &- &APEX &- &- &- &APEX &- \\
01304+6211&V669 Cas&- &- &JCMT &- &JCMT &- &JCMT &- &JCMT &- &- &- \\
02168-0312&o Cet&SEST &3 &JCMT &- &JCMT &- &JCMT &- &JCMT &- &- &- \\
03507+1115&IK Tau&IRAM &7 &IRAM &7 &APEX &- &JCMT &8 &- &- &APEX &- \\
 " & " &- &- &JCMT &8 &JCMT &8 &APEX &- &- &- &- &- \\
 " & " &- &- &IRAM &- &APEX &- &- &- &- &- &- &- \\
04566+5606&TX Cam&- &- &JCMT &- &JCMT &- &JCMT &- &- &- &- &- \\
05073+5248&NV Aur&- &- &JCMT &- &JCMT &- &JCMT &- &JCMT &- &- &- \\
05524+0723&Alpha Ori&- &- &JCMT &- &JCMT &- &JCMT &- &- &- &- &- \\
 " & " &- &- &JCMT &8 &- &- &- &- &- &- &- &- \\
06176-1036&Red Rectangle&- &- &- &- &APEX &- &- &- &- &- &- &- \\
07209-2540&VY CMa&SEST &3 &JCMT &- &JCMT &- &JCMT &- &JCMT &- &JCMT &- \\
07399-1435&Calabash Nebula&- &- &- &- &APEX &- &- &- &- &- &- &- \\
09448+1139&R Leo&IRAM &9 &IRAM &9 &APEX &- &- &- &CSO &9 &- &- \\
 " & " &- &- &- &- &CSO &9 &- &- &- &- &- &- \\
09452+1330&CW Leo&IRAM &10 &IRAM &10 &JCMT &10 &CSO &9 &CSO &9 &- &- \\
 " & " &NRAO &11 &SEST &6 &CSO &12 &- &- &- &- &- &- \\
 " & " &SEST &6 &- &- &- &- &- &- &- &- &- &- \\
10131+3049&RW LMi&IRAM &9 &IRAM &9 &CSO &9 &CSO &9 &CSO &9 &- &- \\
10197-5750&GSC 08608-00509&- &- &- &- &APEX &- &- &- &- &- &- &- \\
10329-3918&U Ant&SEST &6 &SEST &6 &APEX &- &APEX &- &- &- &APEX &- \\
10350-1307&U Hya&SEST &6 &SEST &6 &APEX &- &APEX &- &- &- &APEX &- \\
10491-2059&V Hya&SEST &6 &SEST &6 &APEX &- &APEX &- &- &- &APEX &- \\
12427+4542&Y CVn&IRAM &7 &IRAM &7 &- &- &CSO &9 &CSO &9 &- &- \\
13269-2301&R Hya&IRAM &9 &IRAM &9 &JCMT &- &- &- &CSO &9 &- &- \\
 " & " &- &- &JCMT &- &CSO &9 &- &- &- &- &- &- \\
 " & " &- &- &NRAO &13 &- &- &- &- &- &- &- &- \\
13428-6232&GLMP 363&- &- &- &- &APEX &- &- &- &- &- &- &- \\
13462-2807&W Hya&- &- &- &- &APEX &- &APEX &- &- &- &APEX &- \\
14219+2555&RX Boo&IRAM &9 &IRAM &9 &JCMT &- &- &- &JCMT &- &- &- \\
 " & " &- &- &JCMT &- &CSO &9 &- &- &CSO &9 &- &- \\
15194-5115&II Lup&- &- &- &- &APEX &- &APEX &- &- &- &APEX &- \\
 " & " &- &- &- &- &APEX &- &APEX &- &- &- &APEX &- \\
16262-2619&Alpha Sco&- &- &- &- &JCMT &- &- &- &- &- &- &- \\
16269+4159&G Her&- &- &JCMT &- &JCMT &- &- &- &JCMT &- &- &- \\
17123+1107&V438 Oph&- &- &JCMT &- &JCMT &- &- &- &- &- &- &- \\
17150-3224&Cotton Candy Nebula&- &- &- &- &APEX &- &- &- &- &- &- &- \\
17411-3154&AFGL 5379&- &- &JCMT &- &JCMT &- &APEX &- &- &- &APEX &- \\
 " & " &- &- &- &- &APEX &- &- &- &- &- &- &- \\
17443-2949&PN RPZM 39&- &- &- &- &JCMT &- &- &- &- &- &- &- \\
17501-2656&V4201 Sgr&- &- &JCMT &- &JCMT &- &- &- &- &- &- &- \\
18050-2213&VX Sgr&- &- &JCMT &- &JCMT &- &JCMT &- &- &- &- &- \\
18059-3211&Gomez Nebula&- &- &JCMT &- &JCMT &- &- &- &- &- &- &- \\
18100-1915&OH 11.52 -0.58&- &- &- &- &APEX &- &APEX &- &- &- &APEX &- \\
18257-1000&V441 Sct&- &- &- &- &APEX &- &APEX &- &- &- &APEX &- \\
18308-0503&AFGL 5502&- &- &JCMT &- &JCMT &- &JCMT &- &JCMT &- &- &- \\
18327-0715&OH 24.69 +0.24&- &- &- &- &APEX &- &APEX &- &- &- &APEX &- \\
18333+0533&NX Ser&- &- &JCMT &- &JCMT &- &JCMT &- &JCMT &- &- &- \\
18348-0526&OH 26.5+0.6&- &- &JCMT &- &JCMT &- &JCMT &- &- &- &- &- \\
18361-0647&OH 25.50 -0.29&- &- &- &- &APEX &- &APEX &- &- &- &APEX &- \\
18397+1738&IRC +20370&IRAM &7 &IRAM &7 &APEX &- &APEX &- &- &- &APEX &- \\
18432-0149&V1360 Aql&- &- &- &- &APEX &- &APEX &- &- &- &APEX &- \\
18448-0545&R Sct&- &- &- &- &APEX &- &- &- &- &- &- &- \\
18460-0254&V1362 Aql&- &- &JCMT &- &JCMT &- &APEX &- &- &- &APEX &- \\
 " & " &- &- &- &- &APEX &- &- &- &- &- &- &- \\
18476-0758&S Sct&SEST &6 &SEST &6 &APEX &- &- &- &- &- &- &- \\
18488-0107&V1363 Aql&- &- &- &- &APEX &- &APEX &- &- &- &APEX &- \\
18498-0017&V1365 Aql&- &- &JCMT &- &JCMT &- &APEX &- &- &- &APEX &- \\
 " & " &- &- &- &- &APEX &- &- &- &- &- &- &- \\
19067+0811&V1368 Aql&- &- &- &- &APEX &- &APEX &- &- &- &APEX &- \\
19110+1045&KJK G45.07&- &- &- &- &- &- &JCMT &- &- &- &- &- \\
19114+0002&V1427 Aql&- &- &- &- &APEX &- &- &- &- &- &- &- \\
19126-0708&W Aql&NRAO &14 &JCMT &- &APEX &- &APEX &- &- &- &APEX &- \\
 " & " &SEST &3 &SEST &- &APEX &- &APEX &- &- &- &APEX &- \\
 " & " &SEST &- &- &- &APEX &- &APEX &- &- &- &- &- \\
 " & " &- &- &- &- &APEX &- &- &- &- &- &- &- \\
19192+0922&OH 44.8-2.3&- &- &JCMT &- &JCMT &- &JCMT &- &JCMT &- &- &- \\
19244+1115&IRC +10420&IRAM &7 &JCMT &- &JCMT &- &JCMT &- &JCMT &- &- &- \\
 " & " &- &- &IRAM &7 &JCMT &8 &JCMT &8 &CSO &9 &- &- \\
19283+1944&AFGL 2403&- &- &JCMT &- &JCMT &- &JCMT &- &- &- &- &- \\
19486+3247&Chi Cyg&IRAM &7 &IRAM &7 &CSO &9 &JCMT &8 &CSO &9 &- &- \\
 " & " &NRAO &14 &JCMT &8 &JCMT &8 &- &- &- &- &- &- \\
 " & " &- &- &- &- &JCMT &8 &- &- &- &- &- &- \\
20075-6005&X Pav&- &- &- &- &APEX &- &- &- &- &- &- &- \\
20077-0625&IRC -10529&- &- &- &- &APEX &- &APEX &- &- &- &APEX &- \\
20120-4433&RZ Sgr&- &- &- &- &APEX &- &APEX &- &- &- &APEX &- \\
20396+4757&V Cyg&- &- &JCMT &- &- &- &- &- &- &- &- &- \\
21419+5832&Mu Cep&- &- &JCMT &- &JCMT &- &JCMT &- &- &- &- &- \\
21439-0226&EP Aqr&- &- &JCMT &- &JCMT &- &JCMT &- &JCMT &- &- &- \\
21554+6204&GLMP 1048&- &- &JCMT &- &JCMT &- &JCMT &- &JCMT &- &- &- \\
22177+5936&OH 104.9+2.4&- &- &JCMT &- &JCMT &- &JCMT &- &JCMT &- &JCMT &- \\
22196-4612&pi1 Gru&- &- &- &- &APEX &- &APEX &- &- &- &APEX &- \\
23166+1655&LL Peg&IRAM &9 &IRAM &9 &APEX &- &APEX &- &CSO &9 &APEX &- \\
 " & " &OSO &3 &- &- &CSO &9 &CSO &9 &- &- &- &- \\
23320+4316&LP And&IRAM &7 &IRAM &7 &JCMT &8 &JCMT &8 &CSO &9 &- &- \\
 " & " &OSO &3 &- &- &- &- &- &- &- &- &- &- \\
23558+5106&R Cas&- &- &JCMT &- &JCMT &- &JCMT &- &JCMT &- &- &- \\
\hline
\end{longtable}
\end{landscape}}\clearpage\normalsize

\begin{figure*} 
\includegraphics[angle=180,width=0.9\linewidth]{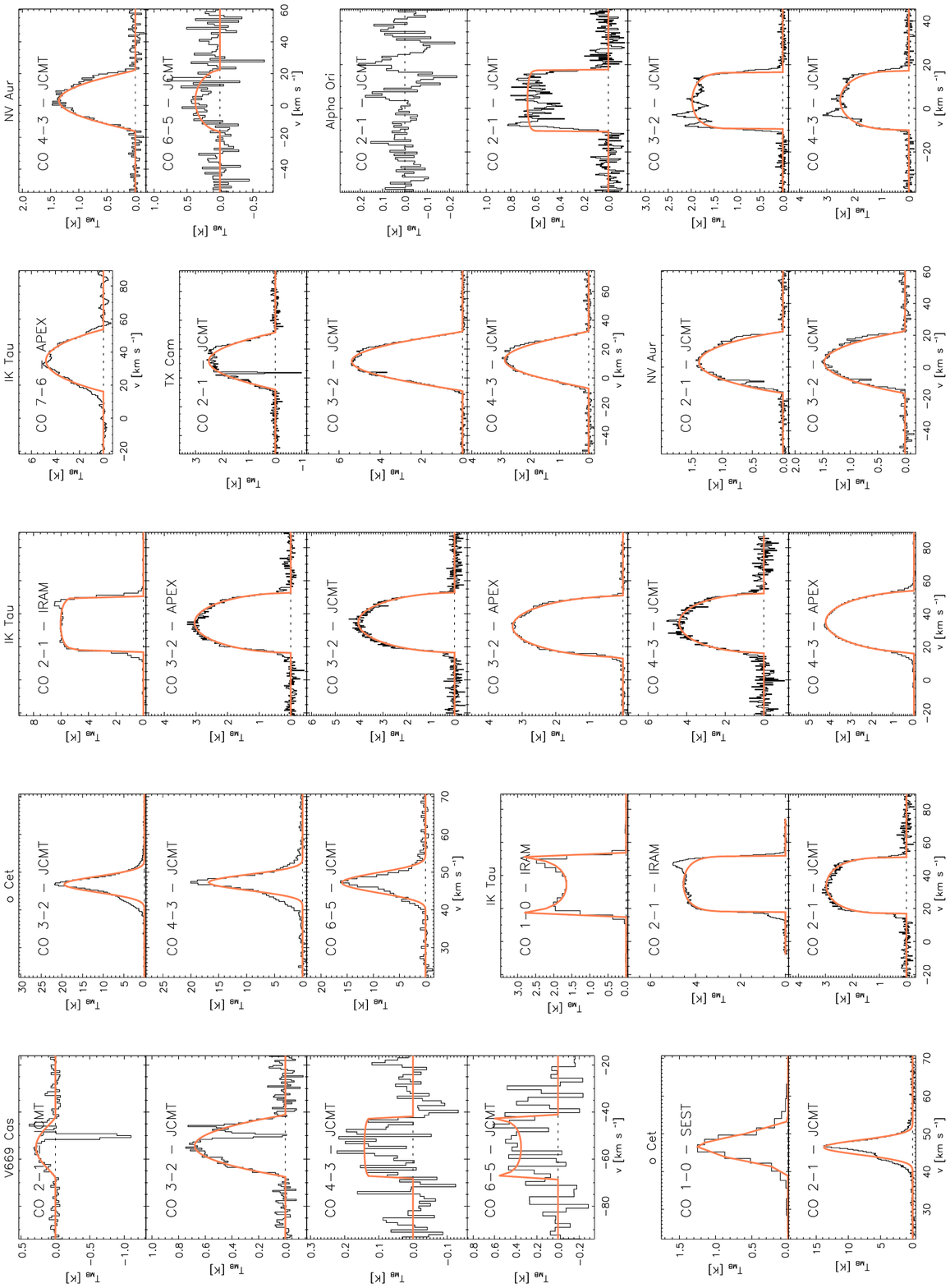}
\begin{center} \textbf{Fig.\,\ref{fig:12CO_parabola}.} (continued) \end{center}
\end{figure*}
\begin{figure*} 
\includegraphics[angle=180,width=0.9\linewidth]{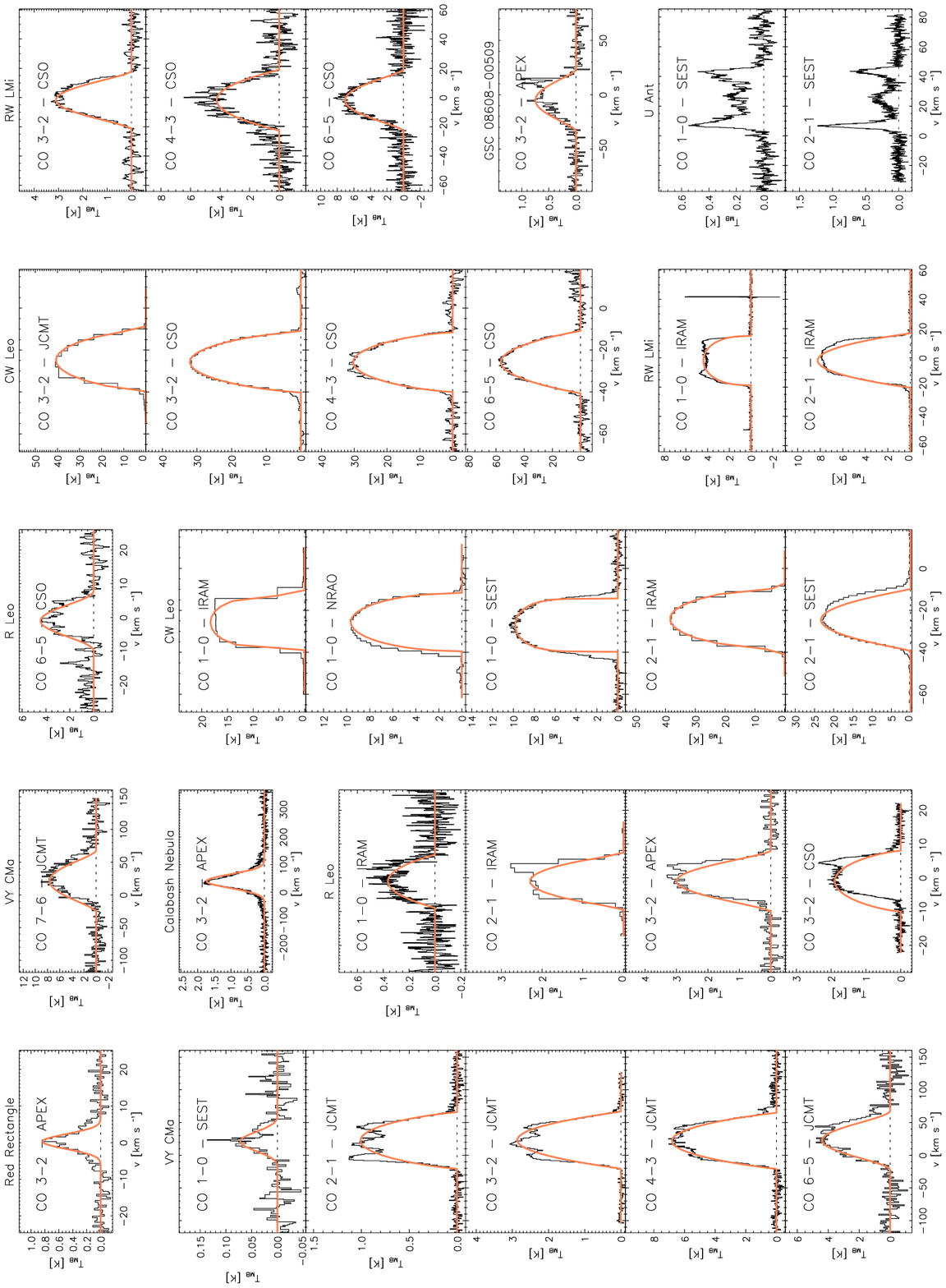}
\begin{center} \textbf{Fig.\,\ref{fig:12CO_parabola}.} (continued) \end{center}
\end{figure*}
\begin{figure*} 
\includegraphics[angle=180,width=0.9\linewidth]{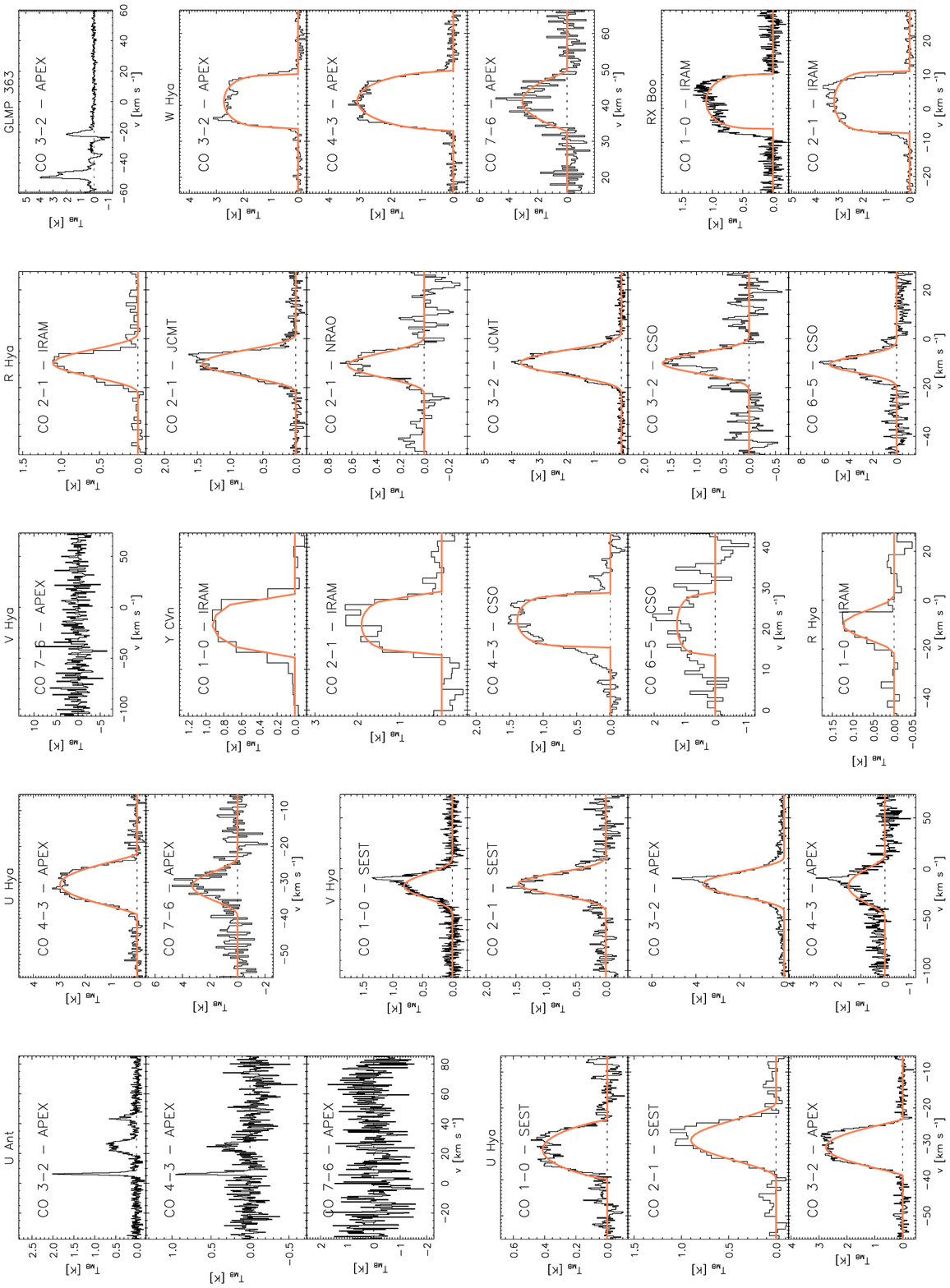}
\begin{center} \textbf{Fig.\,\ref{fig:12CO_parabola}.} (continued) \end{center}
\end{figure*}
\begin{figure*} 
\includegraphics[angle=180,width=0.9\linewidth]{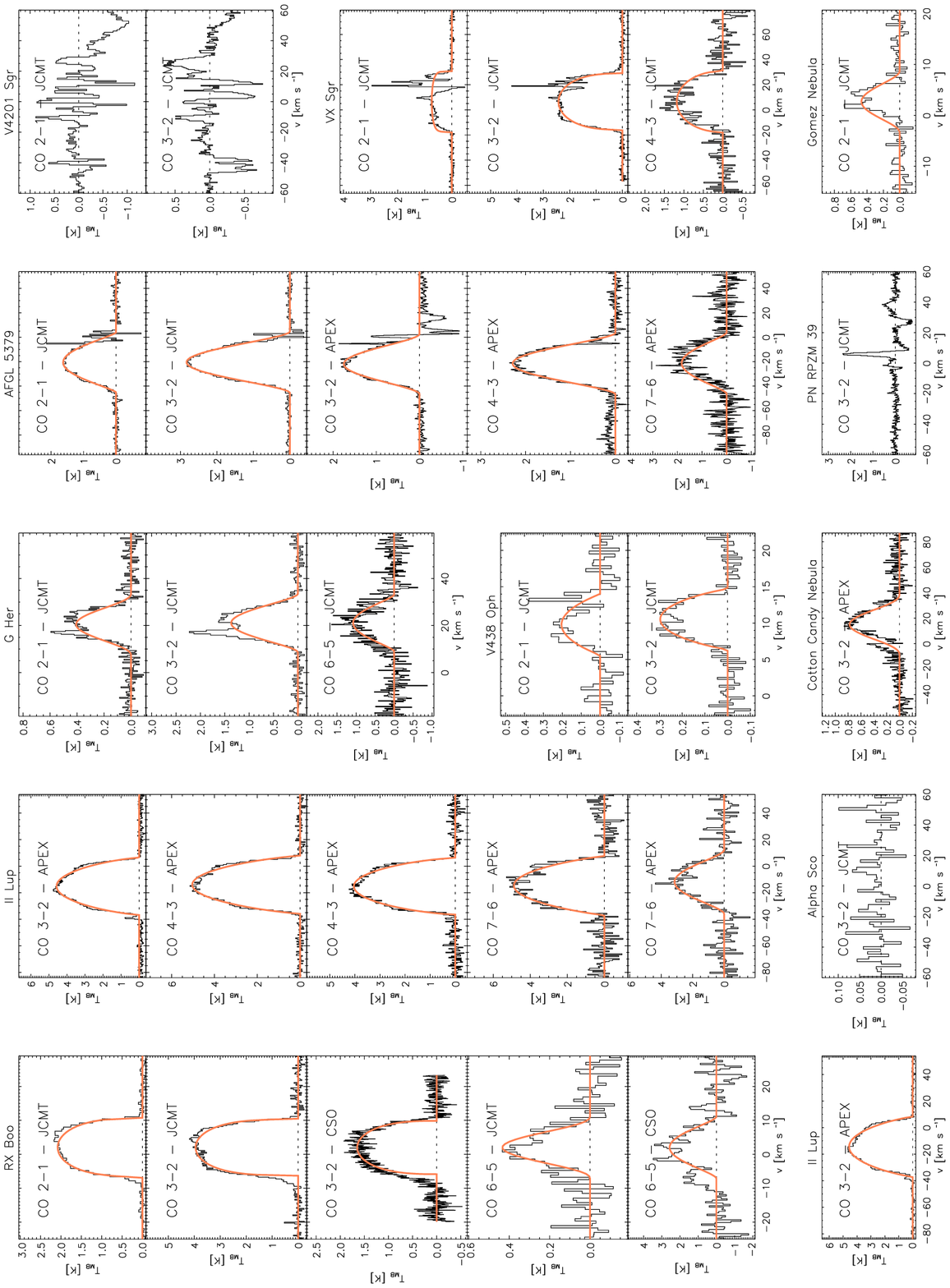}
\begin{center} \textbf{Fig.\,\ref{fig:12CO_parabola}.} (continued) \end{center}
\end{figure*}
\begin{figure*} 
\includegraphics[angle=180,width=0.9\linewidth]{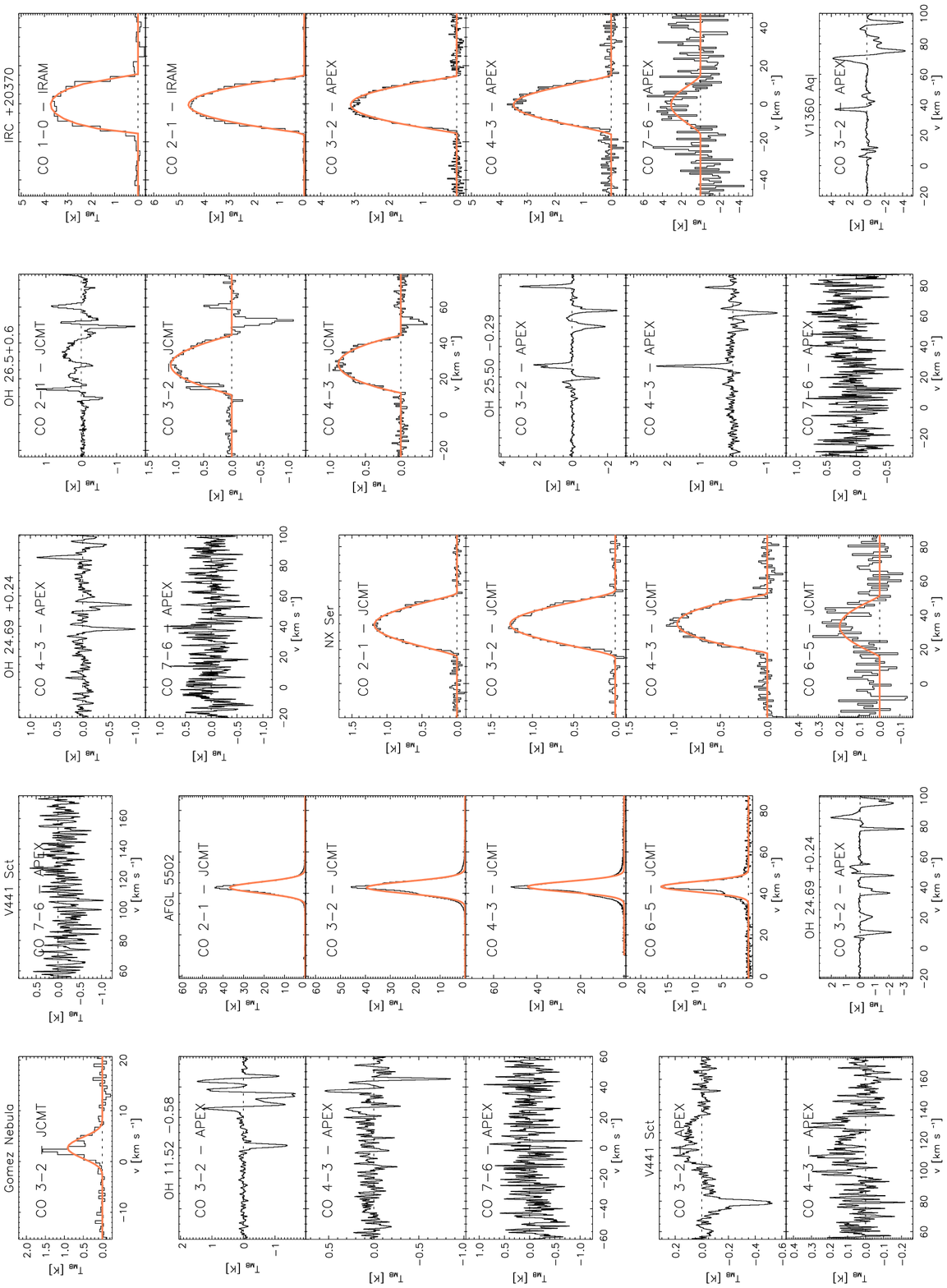}
\begin{center} \textbf{Fig.\,\ref{fig:12CO_parabola}.} (continued) \end{center}
\end{figure*}
\begin{figure*} 
\includegraphics[angle=180,width=0.9\linewidth]{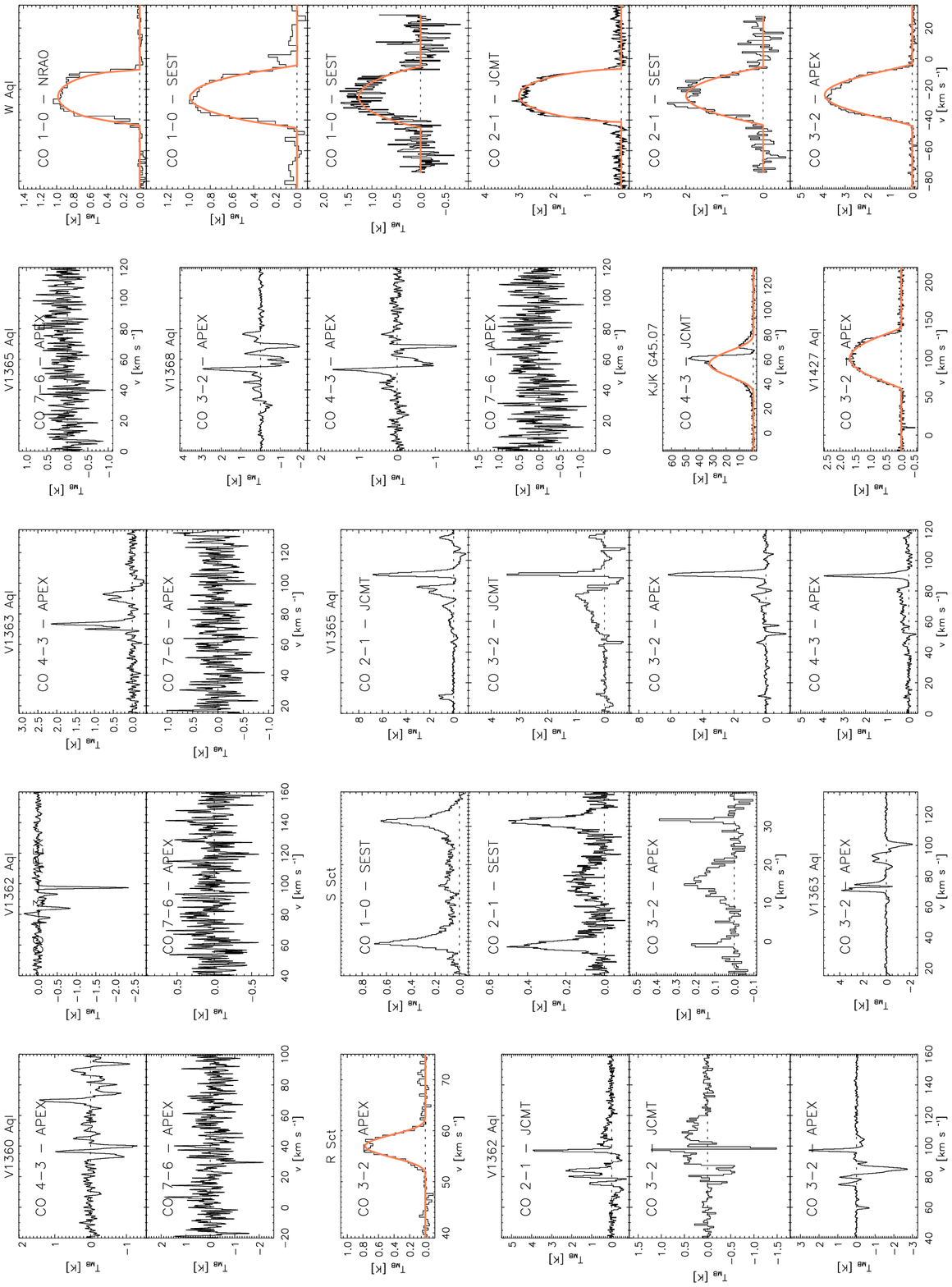}
\begin{center} \textbf{Fig.\,\ref{fig:12CO_parabola}.} (continued) \end{center}
\end{figure*}
\begin{figure*} 
\includegraphics[angle=180,width=0.9\linewidth]{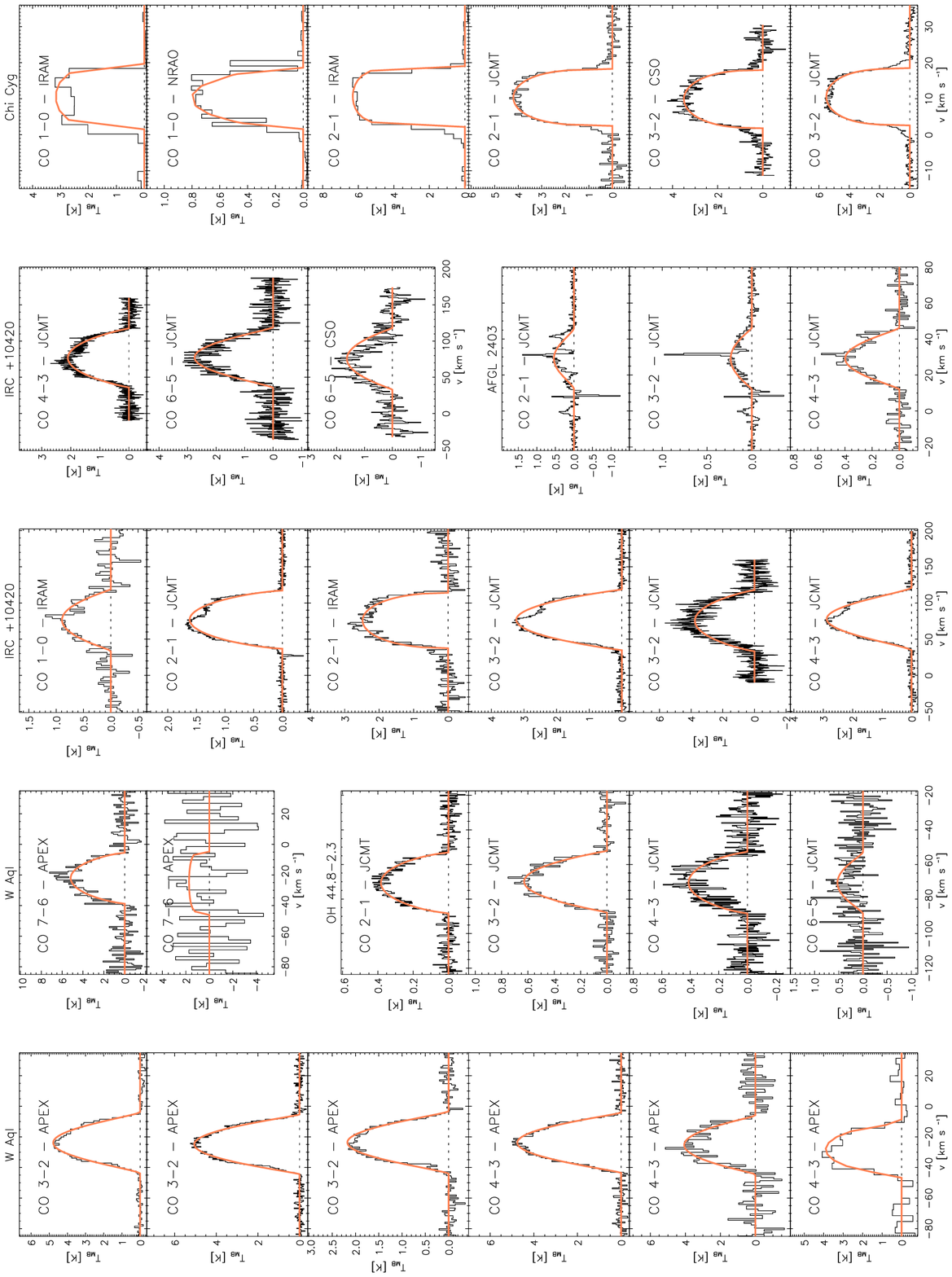}
\begin{center} \textbf{Fig.\,\ref{fig:12CO_parabola}.} (continued) \end{center}
\end{figure*}
\begin{figure*} 
\includegraphics[angle=180,width=0.9\linewidth]{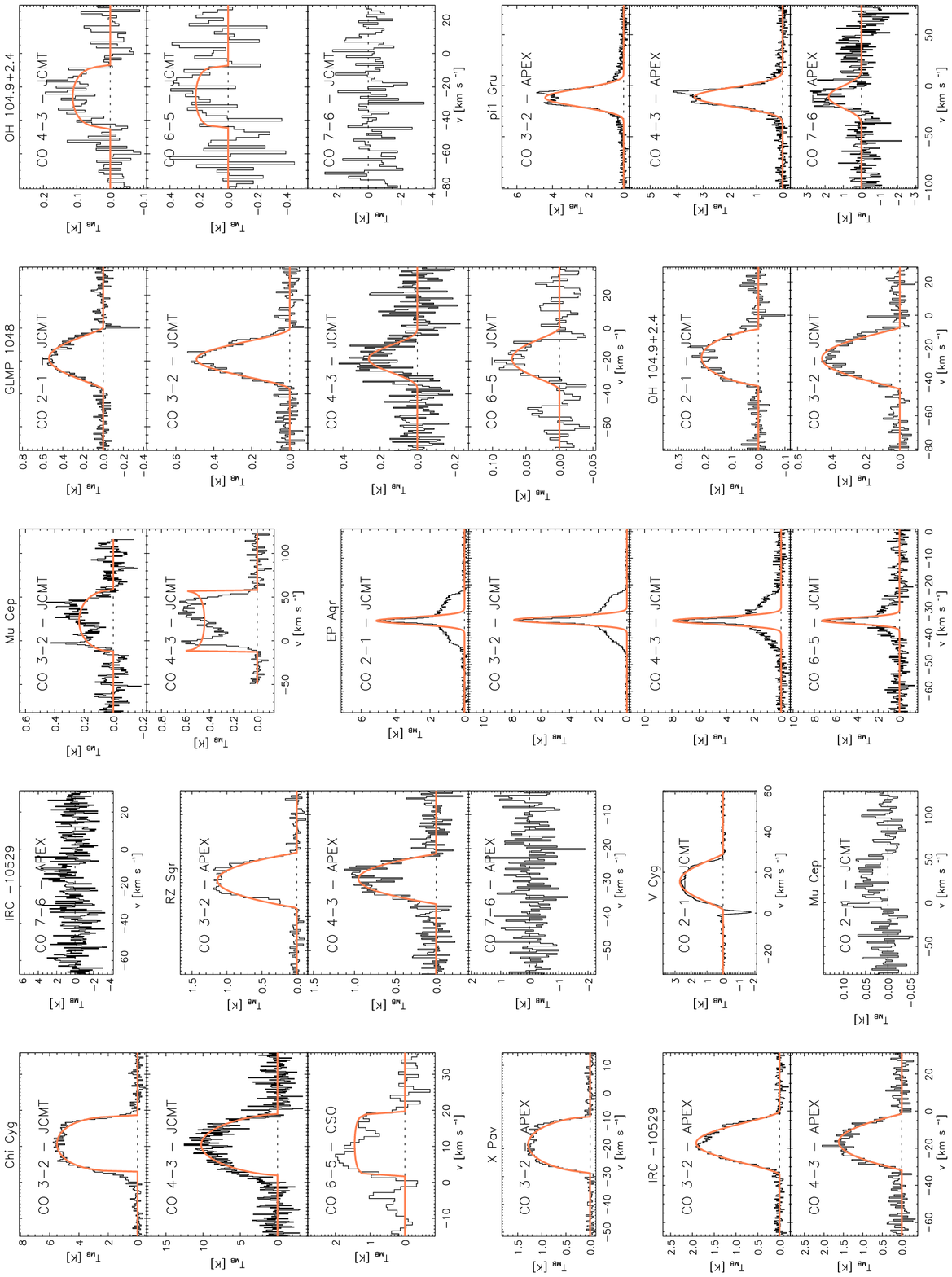}
\begin{center} \textbf{Fig.\,\ref{fig:12CO_parabola}.} (continued) \end{center}
\end{figure*}
\begin{figure*} 
\includegraphics[angle=180,width=0.9\linewidth]{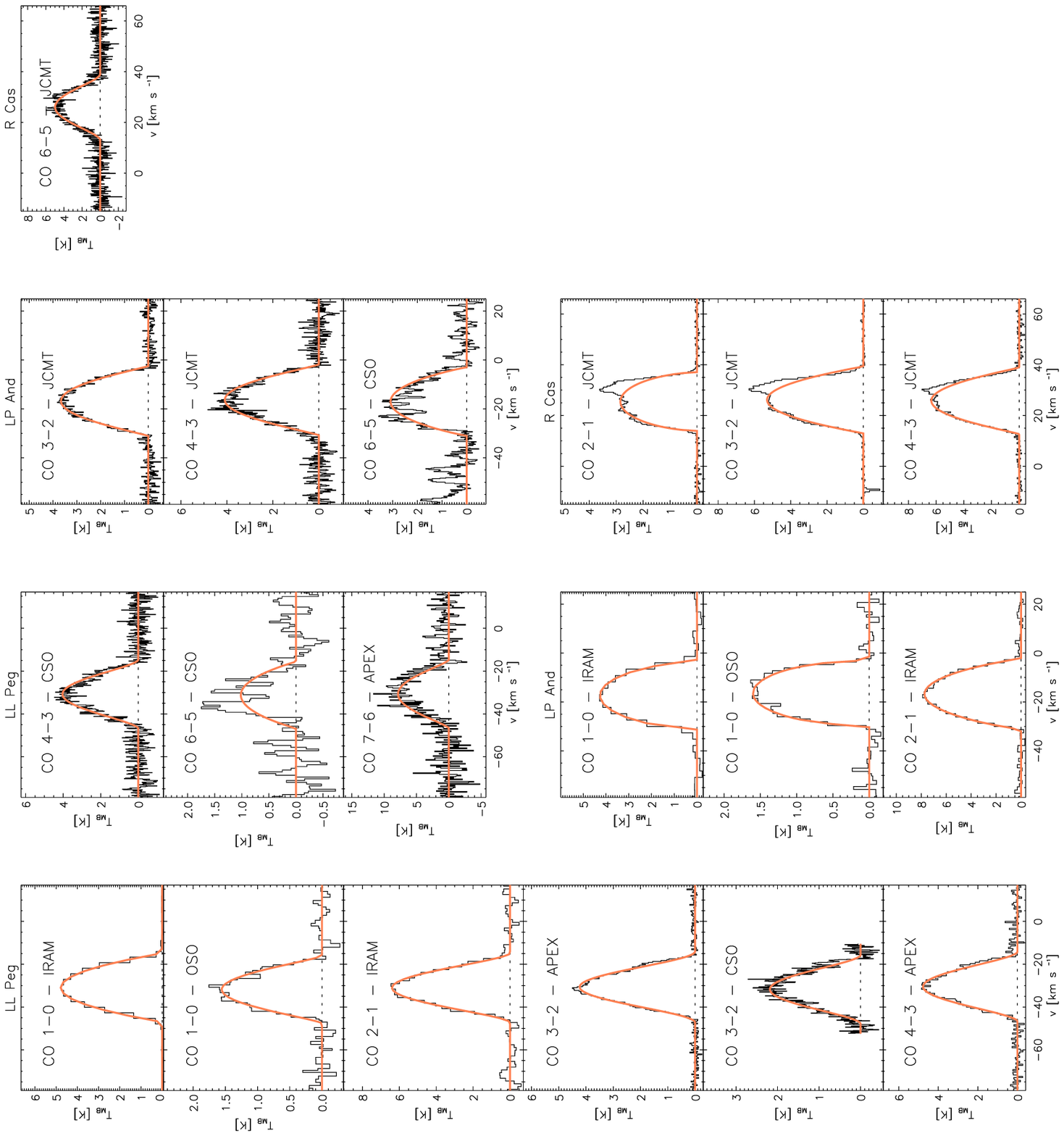}
\begin{center} \textbf{Fig.\,\ref{fig:12CO_parabola}.} (continued) \end{center}
\end{figure*}

\longtabL{3}{\normalsize\setlength{\tabcolsep}{0.7mm}
\begin{landscape}
\begin{longtable}{llrr|rrr|rrr|rrr|rrr}
 \caption[\thco-data presentation.]{Same as Table~\ref{tbl:12CO}, for \thco-data presented in Fig.~\ref{fig:13CO_parabola}.}\label{tbl:13CO}\\
\hline \hline \\[-2ex]
IRAS &Alternative name &$v_{\mathrm{LSR}}$ &$ v_{\infty}$ &\multicolumn{3}{c}{$^{13}$ CO 2--1} &\multicolumn{3}{c}{$^{13}$ CO 3--2} &\multicolumn{3}{c}{$^{13}$ CO 4--3} &\multicolumn{3}{c}{$^{13}$ CO 6--5} \\
  &  &\multicolumn{2}{c}{km/s} &$ T_{\mathrm{MB,c}} $ &$ I_{\mathrm{MB}} $ &$ \beta $ &$ T_{\mathrm{MB,c}} $ &$ I_{\mathrm{MB}} $ &$ \beta $ &$ T_{\mathrm{MB,c}} $ &$ I_{\mathrm{MB}} $ &$ \beta $ &$ T_{\mathrm{MB,c}} $ &$ I_{\mathrm{MB}} $ &$ \beta $ \\
\hline \\[-2ex]
\endfirsthead
\multicolumn{3}{c}{\textbf{{\tablename} \thetable{}.} Continued} \\[0.5ex]
\hline \hline \\[-2ex]
IRAS &Alternative name &$v_{\mathrm{LSR}}$ &$ v_{\infty}$ &\multicolumn{3}{c}{$^{13}$ CO 2--1} &\multicolumn{3}{c}{$^{13}$ CO 3--2} &\multicolumn{3}{c}{$^{13}$ CO 4--3} &\multicolumn{3}{c}{$^{13}$ CO 6--5} \\
  &  &\multicolumn{2}{c}{km/s} &$ T_{\mathrm{MB,c}} $ &$ I_{\mathrm{MB}} $ &$ \beta $ &$ T_{\mathrm{MB,c}} $ &$ I_{\mathrm{MB}} $ &$ \beta $ &$ T_{\mathrm{MB,c}} $ &$ I_{\mathrm{MB}} $ &$ \beta $ &$ T_{\mathrm{MB,c}} $ &$ I_{\mathrm{MB}} $ &$ \beta $ \\
\hline \\[-2ex]
\endhead
\hline\\[-2ex]
\multicolumn{3}{l}{{Continued on Next Page\ldots}} \\
\endfoot
\\[-1.8ex]
\endlastfoot
-&NML Cyg&    2.0&   33.0&   0.21 &   9.49 &   1.52 &- &- &- &- &- &- &- &- &- \\
00192-2020&T Cet&   23.0&    7.0&  -0.05 &   0.34 &- &   0.03 &   0.50 &  -1.47 &- &- &- &- &- &- \\
01037+1219&WX Psc&    9.0&   19.8&   0.36 &  12.93 &   0.07 &   0.48 &  16.01 &   0.48 &- &- &- &- &- &- \\
 " & " & " & " &   0.23 &   8.09 &   0.11 &   0.34 &  11.39 &   0.21 & - &- &- &- &- &- \\
 " & " & " & " & - &- &- &   0.36 &  11.44 &   0.50 & - &- &- &- &- &- \\
01246-3248&R Scl&  -17.0&   17.0&- &- &- &   0.11 &   5.17 &  -1.55 &- &- &- &- &- &- \\
02168-0312&o Cet&   46.4&    8.1&   1.39 &   5.22 &  19.70 &   3.13 &  13.35 &  16.91 &- &- &- &   5.17 &  26.54 &  11.62 \\
03507+1115&IK Tau&   33.8&   18.5&   0.28 &  10.34 &  -0.48 &   0.34 &  13.79 &  -0.68 &- &- &- &- &- &- \\
04566+5606&TX Cam&   10.8&   21.2&   0.10 &   3.72 &  -0.17 &   0.29 &  10.88 &   0.06 &- &- &- &- &- &- \\
05073+5248&NV Aur&    3.0&   19.2&   0.22 &   6.75 &   0.58 &- &- &- &- &- &- &- &- &- \\
05524+0723&Alpha Ori&    3.5&   14.0&   0.09 &   2.46 &   0.03 &- &- &- &- &- &- &- &- &- \\
07209-2540&VY CMa&   21.0&   46.5&   0.04 &   3.32 &   0.57 &   0.18 &  12.01 &   1.07 &- &- &- &- &- &- \\
10491-2059&V Hya&  -17.0&   30.0&- &- &- &   0.14 &   3.43 &   8.49 &- &- &- &- &- &- \\
13269-2301&R Hya&  -10.0&   12.5&   0.47 &   2.78 &   5.26 &- &- &- &- &- &- &- &- &- \\
14219+2555&RX Boo&    2.0&    9.0&   0.14 &   2.58 &  -1.32 &   0.57 &   5.51 &   3.04 &- &- &- &- &- &- \\
15194-5115&II Lup&  -15.0&   23.0&- &- &- &   0.76 &  31.06 &  -0.06 &- &- &- &- &- &- \\
 " & " & " & " & - &- &- &   0.92 &  32.48 &   0.54 & - &- &- &- &- &- \\
17411-3154&AFGL 5379&  -21.2&   25.0&   0.16 &   6.05 &   0.84 &   0.05 &   2.70 &- &- &- &- &- &- &- \\
18050-2213&VX Sgr&    6.5&   24.3&- &- &- &   0.11 &   4.29 &   0.67 &- &- &- &- &- &- \\
18059-3211&Gomez Nebula&    2.7&    6.0&- &- &- &   0.23 &   1.81 &   3.25 &- &- &- &- &- &- \\
18308-0503&AFGL 5502&   43.0&   14.7&  22.97 &  77.97 & 102.80 &  24.38 &  95.37 &  87.94 &- &- &- &- &- &- \\
18397+1738&IRC +20370&   -0.5&   16.0&- &- &- &   0.28 &   7.41 &   0.76 &- &- &- &- &- &- \\
19114+0002&V1427 Aql&   99.0&   40.0&- &- &- &   0.43 &  22.20 &   2.12 &- &- &- &- &- &- \\
19126-0708&W Aql&  -25.0&   20.0&   0.14 &   5.03 &  -0.26 &   0.23 &   8.10 &  -0.28 &- &- &- &- &- &- \\
 " & " & " & " & - &- &- &   0.25 &   8.68 &   0.49 & - &- &- &- &- &- \\
 " & " & " & " & - &- &- &   0.23 &   7.68 &   0.48 & - &- &- &- &- &- \\
19192+0922&OH 44.8-2.3&  -70.5&   17.7&   0.10 &   2.56 &   0.13 &   0.08 &   2.04 &   0.49 &- &- &- &- &- &- \\
19244+1115&IRC +10420&   75.5&   42.3&   0.17 &  10.57 &   1.17 &   0.37 &  23.49 &   0.93 &- &- &- &- &- &- \\
19486+3247&Chi Cyg&   10.5&    8.5&- &- &- &   0.24 &   4.65 &  -0.89 &- &- &- &- &- &- \\
20396+4757&V Cyg&   15.0&   15.0&   0.16 &   3.16 &   0.20 &- &- &- &- &- &- &- &- &- \\
21439-0226&EP Aqr&  -33.5&   11.5&   0.67 &   3.18 & 208.64 &   1.42 &   6.76 & 136.35 &   0.87 &   3.44 &  16.11 &- &- &- \\
22196-4612&pi1 Gru&  -11.0&   30.0&- &- &- &   0.10 &   2.81 &   5.25 &- &- &- &- &- &- \\
23166+1655&LL Peg&  -31.0&   16.0&- &- &- &   0.86 &  18.36 &   1.62 &- &- &- &- &- &- \\
23558+5106&R Cas&   25.5&   13.5&   0.11 &   2.71 &   0.50 &   0.37 &   8.01 &   1.03 &- &- &- &- &- &- \\
\hline
\end{longtable}
\end{landscape}}\clearpage\normalsize

\longtabL{4}{\normalsize\setlength{\tabcolsep}{0.7mm}
\begin{landscape}
\begin{longtable}{ll|cc|cc|cc|cc}
 \caption[\thco-data origin.]{References for the presented \thco data. In case of multiple observations of a specific transition, the data are given in the exact same order as in  Fig.~\ref{fig:13CO_parabola}. References or remarks are (1) archive data, (-) this study.}\label{tbl:13CO-REMARKS}\\
\hline \hline \\[-2ex]
IRAS &Alternative name &\multicolumn{2}{c}{$^{13}$ CO 2--1} &\multicolumn{2}{c}{$^{13}$ CO 3--2} &\multicolumn{2}{c}{$^{13}$ CO 4--3} &\multicolumn{2}{c}{$^{13}$ CO 6--5} \\
  &  &Telescope &Reference &Telescope &Reference &Telescope &Reference &Telescope &Reference \\
\hline \\[-2ex]
\endfirsthead
\multicolumn{3}{c}{\textbf{{\tablename} \thetable{}.} Continued} \\[0.5ex]
\hline \hline \\[-2ex]
IRAS &Alternative name &\multicolumn{2}{c}{$^{13}$ CO 2--1} &\multicolumn{2}{c}{$^{13}$ CO 3--2} &\multicolumn{2}{c}{$^{13}$ CO 4--3} &\multicolumn{2}{c}{$^{13}$ CO 6--5} \\
  &  &Telescope &Reference &Telescope &Reference &Telescope &Reference &Telescope &Reference \\
\hline \\[-2ex]
\endhead
\hline\\[-2ex]
\multicolumn{3}{l}{{Continued on Next Page\ldots}} \\
\endfoot
\\[-1.8ex]
\endlastfoot
-&NML Cyg&JCMT &- &- &- &- &- &- &- \\
00192-2020&T Cet&JCMT &- &JCMT &- &- &- &- &- \\
01037+1219&WX Psc&JCMT &- &JCMT &- &- &- &- &- \\
 " & " &JCMT &- &APEX &- &- &- &- &- \\
 " & " &- &- &APEX &- &- &- &- &- \\
01246-3248&R Scl&- &- &APEX &- &- &- &- &- \\
02168-0312&o Cet&JCMT &- &JCMT &- &- &- &JCMT &- \\
03507+1115&IK Tau&JCMT &1 &APEX &- &- &- &- &- \\
04566+5606&TX Cam&JCMT &- &JCMT &- &- &- &- &- \\
05073+5248&NV Aur&JCMT &- &- &- &- &- &- &- \\
05524+0723&Alpha Ori&JCMT &- &- &- &- &- &- &- \\
07209-2540&VY CMa&JCMT &- &JCMT &- &- &- &- &- \\
10491-2059&V Hya&- &- &APEX &- &- &- &- &- \\
13269-2301&R Hya&JCMT &- &- &- &- &- &- &- \\
14219+2555&RX Boo&JCMT &- &JCMT &- &- &- &- &- \\
15194-5115&II Lup&- &- &APEX &- &- &- &- &- \\
 " & " &- &- &APEX &- &- &- &- &- \\
17411-3154&AFGL 5379&JCMT &- &JCMT &- &- &- &- &- \\
18050-2213&VX Sgr&- &- &JCMT &- &- &- &- &- \\
18059-3211&Gomez Nebula&- &- &JCMT &- &- &- &- &- \\
18308-0503&AFGL 5502&JCMT &- &JCMT &- &- &- &- &- \\
18397+1738&IRC +20370&- &- &APEX &- &- &- &- &- \\
19114+0002&V1427 Aql&- &- &APEX &- &- &- &- &- \\
19126-0708&W Aql&JCMT &- &APEX &- &- &- &- &- \\
 " & " &- &- &APEX &- &- &- &- &- \\
 " & " &- &- &APEX &- &- &- &- &- \\
19192+0922&OH 44.8-2.3&JCMT &- &JCMT &- &- &- &- &- \\
19244+1115&IRC +10420&JCMT &- &JCMT &- &- &- &- &- \\
19486+3247&Chi Cyg&- &- &JCMT &1 &- &- &- &- \\
20396+4757&V Cyg&JCMT &- &- &- &- &- &- &- \\
21439-0226&EP Aqr&JCMT &- &JCMT &- &JCMT &- &- &- \\
22196-4612&pi1 Gru&- &- &APEX &- &- &- &- &- \\
23166+1655&LL Peg&- &- &APEX &- &- &- &- &- \\
23558+5106&R Cas&JCMT &- &JCMT &- &- &- &- &- \\
\hline
\end{longtable}
\end{landscape}}\clearpage\normalsize

\begin{figure*} 
\includegraphics[angle=180,width=0.9\linewidth]{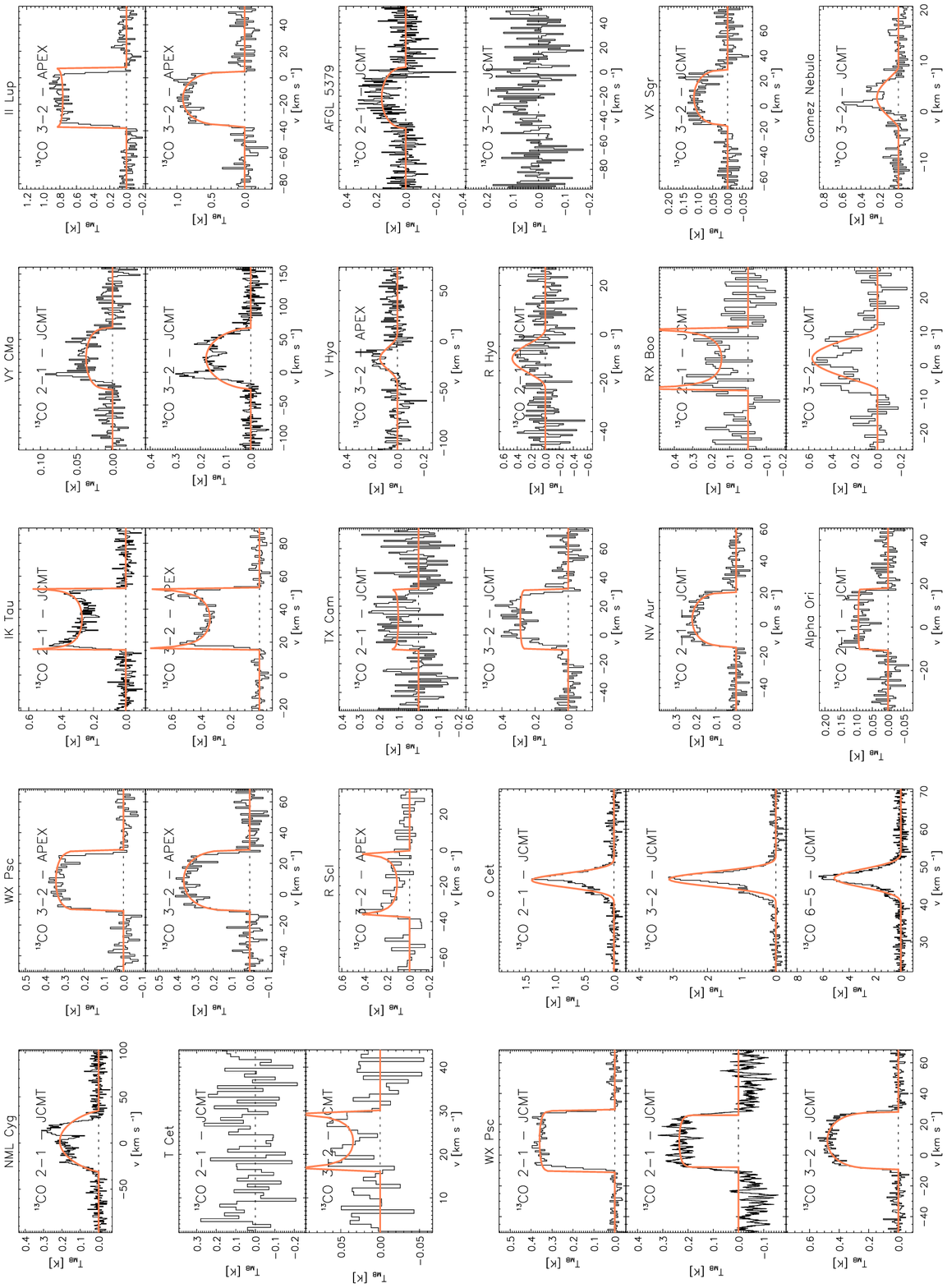}
\caption{Profiles of \thco with fitted soft-parabola profiles.}
\label{fig:13CO_parabola}
\end{figure*}
\begin{figure*} 
\includegraphics[angle=180,width=0.9\linewidth]{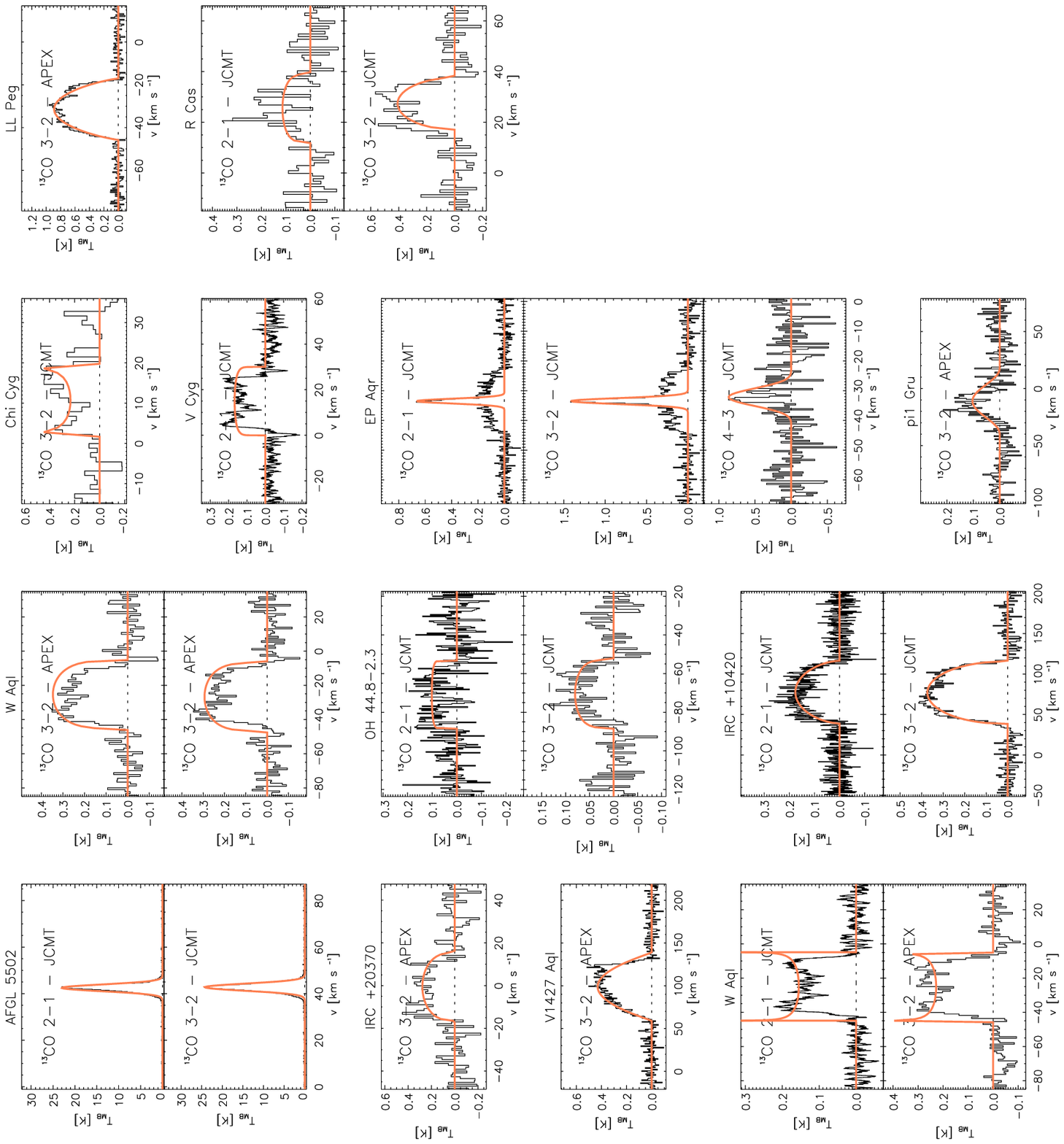}
\begin{center} \textbf{Fig.\,\ref{fig:13CO_parabola}.} (continued) \end{center}
\end{figure*}

\clearpage 

\section{Dereddening}\label{subsect:dereddening}
Since the effective temperatures of late-type stars are well-determined by the intrinsic \VmK colour (Sect.~\ref{sect:teff}), we need a way to derive this colour accurately. The magnitudes of both the $V$ and $K$ bands --- at 0.550\,\um and 2.2\,\um, respectively --- are reddened due to interstellar and circumstellar contributions. The data were corrected for interstellar extinction using the dust maps of \cite{schlegel1998}, which give the interstellar reddening at infinity, $E_{B-V,\infty}$, for a given set of galactic latitudes $b$ and longitudes $l$. This value was transformed into interstellar reddening at a given distance $d$ using
\begin{equation}
 E_{B-V} = E_{B-V,\infty} \left[1-\exp{(-10\; \frac{d}{\mathrm{kpc}} \sin|b|)}\right],
\end{equation}
following \cite{feast1990}.
The resulting value for $E_{B-V}$ was combined with $R_V=3.1$ and $A_K=0.112\;A_V$ \citep{schlegel1998} to give interstellar extinction coefficients for the $V$ and $K$ bands.

Circumstellar extinction was not accounted for, even though it is assumed to be significant for high mass-loss rate LPV targets. 
\cite{knapp1998} use the mass-loss rate \mdot to determine the circumstellar extinction in the $K$ band. Since the aim of the paper is to establish empirical relations between stellar parameters and \mdot, this method can not be applied to the data for the sake of consistency.
\cite{knapp2003} suggest a formalism to calculate the circumstellar extinction in the $K$ band, $A_{K\mathrm{,CS}}$, as a function of $S_{12}/S_{2.2}$, the ratio of flux densities at 12\,\um and 2.2\,\um, respectively. This method is only valid for low optical depths \citep{knapp2003}. When applied to the sample, unrealistic values for 
$A_{K\mathrm{,CS}}$ as high as several tens (LP\,And) or hundreds (CW\,Leo) of magnitudes are indeed obtained for the objects with presumed high \mdot values. To decide, however, whether or not this correction should be applied for a given target seems arbitrary. For these reasons we did not correct for circumstellar reddening.

\section{Basic stellar parameters}\label{sect:basicstellarparameters}
In this appendix we describe the methods that we used to obtain the effective temperature, stellar luminosity and distance towards objects. The results are given in Table~\ref{tbl:fundamentalparameters} in the paper.

\subsection{Effective temperature}\label{sect:teff}
\begin{figure}
 \includegraphics[width=\linewidth]{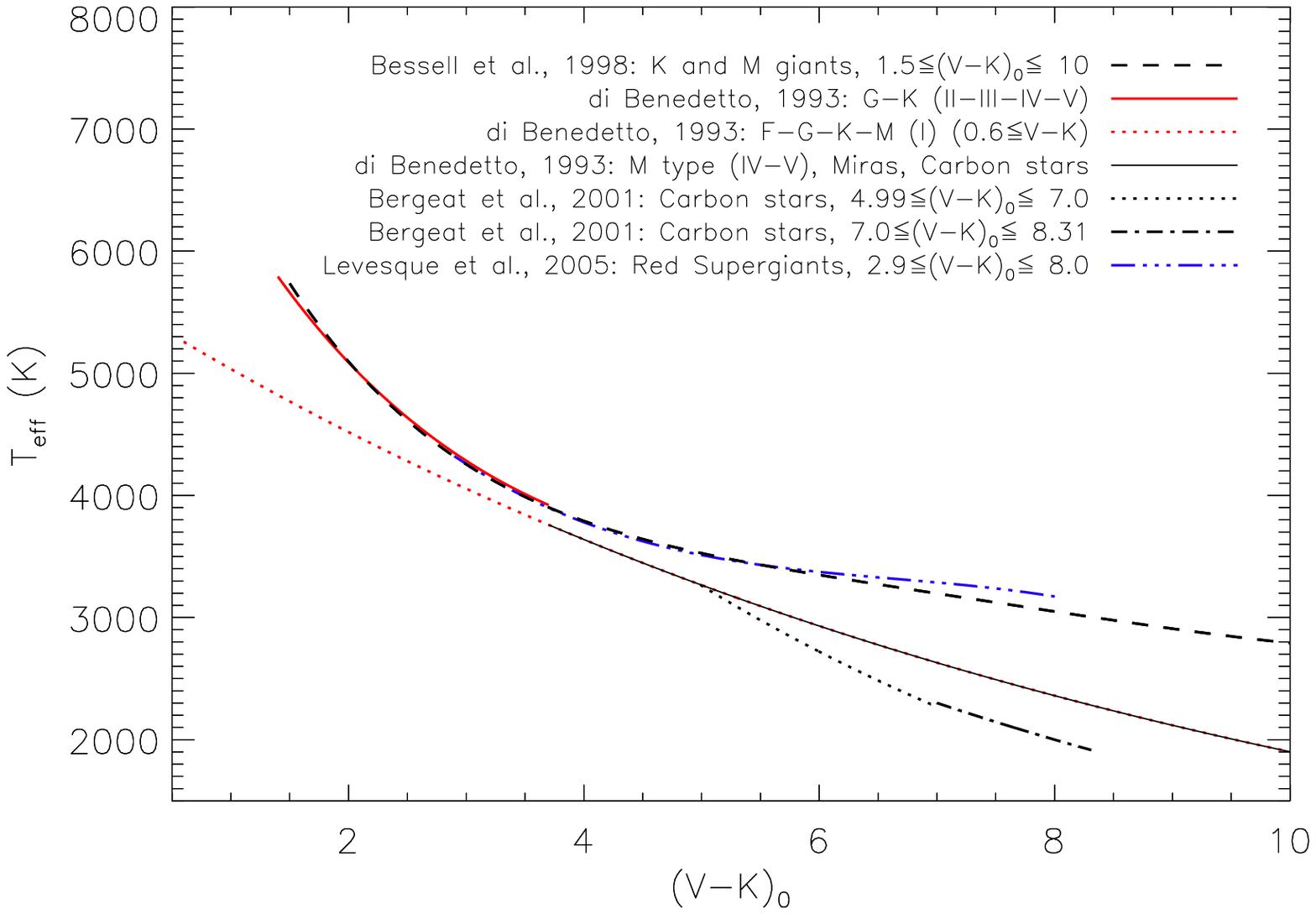}
\caption{Different calibration relations linking the (dereddened) broad-band colour $(V{\rm-}K)_0$ and \teff. Relations were taken from
\cite{dibenedetto1993,bessell1998,bergeat2001,levesque2005}. \label{fig:vkteffcalibration}}
\end{figure}

Since broad-band near-infrared colours are very good temperature criteria, being relatively independent of stellar gravities and abundances, they are often used to determine effective temperatures. The \VmK colour is a good \teff-indicator for cool, late-type stars, e.g. GKM giants and supergiants \citep{dibenedetto1993,bessell1998,bergeat2001,levesque2005}. For hotter A through K stars the $V-I$ colour is more suitable \citep{bessell1998}.

In all \teff versus \VmK relations the Johnson-Cousins-Glass photometric system is used. $K$ magnitudes for the sample stars that were obtained in the systems of 2MASS ($K_{\mathrm{s}}$) and SAAO ($K_{\mathrm{SAAO}}$) were transformed via 
\begin{equation}\label{eq:ktransform}
K=K_{\mathrm{s}}+0.044=K_{\mathrm{SAAO}}-0.0275.
\end{equation}
following \citep{bessell2005}. Since the accuracy of the $K_{\mathrm{SAAO}}$-measurements \citep[$\sim$0.03\,mag;][]{olivier2001} is almost a factor 10 better than that of $K_{\mathrm{s}}$-data ($\sim$0.2\,mag), preference was given to the SAAO-measurements when both were available. The errors on the magnitudes are expected to be small relative to the amplitude in the $K$ band caused by pulsations of the stars. The error bars on the \teff values determined here, were calculated assuming an error on $(V{\rm-}K)_0$ of 1\,mag, leading to rather conservative temperature estimates.

In all equations listed $(V{\rm-}K)_0$ is the intrinsic \VmK broad-band colour with \textit{(1)} both the $V$ and $K$ magnitudes at maximum light, or \textit{(2)} both at their mean value if maximum light was not available, and \textit{(3)} corrected for interstellar reddening in the way described in Sect.~\ref{subsect:dereddening}. 

\subsubsection{Long-period variables: Miras, semi-regulars and irregular variables}
\cite{bessell1998} defined a \teff versus \VmK calibration for KM giants (luminosity class III), given by 
\begin{eqnarray}\label{eq:vkteffbessell}
 T_{\mathrm{eff}}&=&9037.597-3101.282\times(V{\rm-}K)_0 \\
	&&+717.7044\times(V{\rm-}K)_0^2\nonumber\\&&-85.83809\times(V{\rm-}K)_0^3\nonumber \\
	&&+5.021194\times(V{\rm-}K)_0^4\nonumber\\&&-0.1137841\times(V{\rm-}K)_0^5.\nonumber
\end{eqnarray}
This relation is valid for 1.5\,mag$\,\leq(V{\rm-}K)_0\leq\,$10\,mag (see Fig.~\ref{fig:vkteffcalibration}). However, \cite{bessell1998} mention that their static model atmospheres cannot adequately represent the M giants cooler than 3000\,K, since most of these are long-period variables (LPVs) with shock waves traversing and greatly extending their atmospheres.

\cite{dibenedetto1993} presented a temperature calibration in which \teff is derived from angular-diameter measurements based on Michelson interferometry. In case of variable stars, measurements at maximum light were used. The effective temperature of GK stars of luminosity classes II-III-IV-V is described by 
\begin{eqnarray}
\label{eq:dibenedetto1}
\log T_{\mathrm{eff}}&=&a-(b+0.0118)(V{\rm-}K)_0\\
&&+0.0118(V{\rm-}K)_0^2 \nonumber
\end{eqnarray}
for 1.4\,mag$\,\leq(V{\rm-}K)_0\leq\,$3.7\,mag, for FGKM supergiants by 
\begin{equation}
\label{eq:dibenedetto2}
\log T_{\mathrm{eff}}=(a-0.205)-(b-0.086)(V{\rm-}K)_0
\end{equation}
for $(V{\rm-}K)_0\geq0.6$\,mag and for M-type giants of type IV, Miras and carbon stars by Eq.~\ref{eq:dibenedetto2} for 3.7\,mag$\,\leq(V{\rm-}K)_0\leq\,$10\,mag (see Fig.~\ref{fig:vkteffcalibration}). 
The constants $a$ and $b$ in Eqs.~\ref{eq:dibenedetto1} and \ref{eq:dibenedetto2}, and the validity ranges in terms of $(V{\rm-}K)_0$ are listed in Table~\ref{tbl:dibenedetto}.

For \orich LPVs, except for \ohir stars, the effective temperature was derived via Eq.~\ref{eq:dibenedetto2} \citep{dibenedetto1993}. The uncertainty on the temperature would then be of the order of 1\,\% according to \cite{dibenedetto1993}.

\cite{bergeat2001} focussed on \crich objects and derived a relation for \teff versus \VmK given by Eqs.~\ref{eq:bergeat1}, \ref{eq:bergeat2} and \ref{eq:bergeat3}. The notation $T_{\mathrm{eff,d}}$ denotes that the relation is based on angular-diameter measurements, \teff is the final temperature adopted for their sample stars. 
\begin{eqnarray}
\label{eq:bergeat1}
\log T_{\mathrm{eff,d}}&=&-0.079_{(\pm0.008)} (V{\rm-}K)_0+3.91_{(\pm0.02)}\\&&\qquad\qquad \qquad\qquad {\rm for }\;4.99\leq(V{\rm-}K)_0\leq7 \nonumber\\
\label{eq:bergeat2}
\log T_{\mathrm{eff,d}}&=&-0.061_{(\pm0.008)} (V{\rm-}K)_0+3.79_{(\pm0.02)}\\&&\qquad\qquad \qquad\qquad {\rm for }\;7\leq(V{\rm-}K)_0\leq8.31\nonumber\\
\label{eq:bergeat3}
 \log T_{\mathrm{eff,d}}&=&1.003\log T_{\mathrm{eff}}-0.009
\end{eqnarray}
These \teff values are estimated to be accurate to within an error margin of $\pm 140$\,K \citep{bergeat2001}. Because this \teff-calibration is based on a sample of 54 \crich stars, preference was given to this calibration, rather than to the one by \cite{dibenedetto1993}, in case of carbon stars. The validity range was adopted to be $4.99\,\leq (V{\rm-}K)_0\leq\,$8.31\,mag, consistent with the minimum and maximum in Table~5 of \cite{bergeat2001}.

It is clearly visible from Fig.~\ref{fig:vkteffcalibration} that the carbon stars have lower predicted effective temperatures than the \orich stars. This is in agreement with the results of \cite{marigo2002}, showing a clear dichotomy in \teff between both chemical types of AGB stars. This dichotomy is caused by different molecular opacities in stellar atmospheres with differing C/O abundance ratios. 
\\

When using these \teff estimators, one should be aware of the effect of circumstellar reddening on the $V$ and $K$ band data. This reddening is stronger for objects with higher values of \mdot and depends on the chemistry of the central star \citep{knapp1998,knapp2003}. As mentioned in Appendix~\ref{subsect:dereddening}, the data were not corrected for circumstellar reddening. 
\\

For \orich targets with $(V{\rm-}K)_0$ outside the valid ranges for the above mentioned relations, the temperature corresponding to the spectral type listed in Table~\ref{tbl:fundamentalparameters} was used \citep[Table~\ref{tbl:marigo_teffvspt};][]{marigo2008}. An error bar of 500\,K was assumed for these values. For the \crich targets with $(V{\rm-}K)_0$ outside the valid ranges the effective temperatures found in literature were used. References are given in Table~\ref{tbl:fundamentalparameters}.

\begin{table}
\centering
\caption{Effective temperature \teff versus spectral type for \orich targets as given by \cite{marigo2008}. They are used in this paper in case of \orich LPVs which have their dereddened $(V{\rm-}K)_0$ broad-band colour outside the valid ranges for the listed \teff versus $(V{\rm-}K)_0$ relations.}
\label{tbl:marigo_teffvspt}
 \begin{tabular}{ccccccc}
\hline\hline\\[-2ex]
 Spectral Type& M0 & M1 & M2 & M3 & M4 & M5 \\
\teff (K)&3850&3750&3650&3550&3490&3397\\
\hline\\[-2ex]
Spectral Type& M6 & M7 & M8 & M9 & M10 & \\
\teff (K)&3297&3129&2890&2667&2500&\\
\hline
\end{tabular}
\end{table}

\begin{table}\centering
\caption{Coefficients $a$ and $b$ for the \teff calibration from \VmK colours via Eqs.~\ref{eq:dibenedetto1} and \ref{eq:dibenedetto2}. Taken from \cite{dibenedetto1993}.}
\label{tbl:dibenedetto}
 \begin{tabular}{llccc}
\hline\hline\\[-2ex]
Spectral type & Luminosity class & $(V{\rm-}K)_0$ & $a$ & $b$ \\\hline\\[-2ex]
G-K&	II-III-IV-V&	$1.4-3.7$&	3.927& 0.122\\
M&	III&		$3.7-10$&		3.833&0.101\\
M&	II&		$3.7-10$&		3.859&0.108\\
F-G-K-M&	I&	$0.6-10$&		3.954&0.133\\
M&	IV-V&		$3.7-10$&		3.954&0.133\\
M&	Mira&		$3.7-10$&	3.954&0.133\\
C&	Carbon star&	$3.7-10$&	3.954&0.133\\	\hline
\end{tabular}
\end{table}

\subsubsection{\ohir stars}
In the case of \ohir stars the photosphere is not easily probed since the CSE is optically thick and consequently the infrared sources often do not have optical counterparts. Therefore, the \VmK colour is no longer a meaningful diagnostic of \teff. For these objects an effective temperature of $2750\pm750$\,K was assumed.

\subsubsection{Red supergiants}
The \teff-calibration of \cite{bessell1998} is not valid for red supergiants (RSGs) because of their low $\log g$-values. \cite{levesque2005} derived a \teff-scale for RSGs in the Milky Way given by Eq.~\ref{eq:levesqueRSG}. This relation  between the broad-band colour \VmK and the effective temperature is consistent with the results from their grid of {\sc marcs} models and is valid for $2.9{\rm \,mag}\leq(V{\rm-}K)_0\leq8.0{\rm \,mag}$. Adopting this new \teff-calibration for RSGs yields a better agreement between evolutionary tracks and observationally determined locations of RSGs in the HR-diagram \citep[see][Fig.~3]{massey2008}. This \teff-scale for RSGs closely follows the relation of \cite{bessell1998} defined for KM \textit{giants} (see Fig.~\ref{fig:vkteffcalibration}), but since Eq.~\ref{eq:levesqueRSG} was derived especially for RSGs, preference was given to this relation over Eq.~\ref{eq:vkteffbessell}. It is given by
\begin{eqnarray}
 \label{eq:levesqueRSG}
T_{\mathrm{eff}}&=&7741.9-1831.83(V{\rm-}K)_0\\
&&+263.135(V{\rm-}K)_0^2 -13.1943(V{\rm-}K)_0^3. \nonumber 
\end{eqnarray}
\cite{levesque2005} expect that the effective temperatures for M-type supergiants were obtained with a precision of 50\,K.

For RSGs with $(V{\rm-}K)_0<2.9$ or $(V{\rm-}K)_0>8.0$, the temperature was determined using the scale of \teff versus spectral type --- in case of galactic targets --- presented by \cite{levesque2005}.

\subsubsection{Yellow hypergiants}
The adopted \teff-calibration of \cite{dibenedetto1993} for FGKM targets of luminosity class I seems in accordance with the results presented by \cite{dejager1998} and \cite{oudmaijer2008} for the yellow hypergiants (YHGs) IRC\,+10420 and AFGL\,2343. Since both objects are known to exhibit large \teff-variations on time scales of a few decades \citep{oudmaijer2008}, this calibration is somewhat uncertain.

\subsubsection{Alternative methods}
\cite{haniff1995} derive effective temperatures from angular diameters and bolometric fluxes obtained through fits to optical and infrared photometry. The infrared flux method (IRFM), presented by \cite{blackwell1977}, consists of determining the ratio of the bolometric flux to the flux in a selected photometric band and comparing these results to predictions from model atmospheres. Since the goal of this paper is not to individually model all sample targets, we did not pursue these methods, nor did we perform full fits of the spectral energy distributions (SEDs). The latter would require detailed radiative transfer analysis for each star.

\subsection{Luminosity}\label{sect:luminosity}
The luminosities of Miras, \ohir stars, semi-regular pulsators of types a and b and red supergiants were determined through empirical period-bolometric magnitude relations presented in the literature \citep{feast1989,groenewegen1996,whitelock1991,barthes1999,whitelock2000,yesilyaprak2004}. These relations are then used to calculate luminosities adopting the solar bolometric magnitude \mbolsun$=4.75$\,mag.

\begin{figure}
 \includegraphics[height=\linewidth,angle=90]{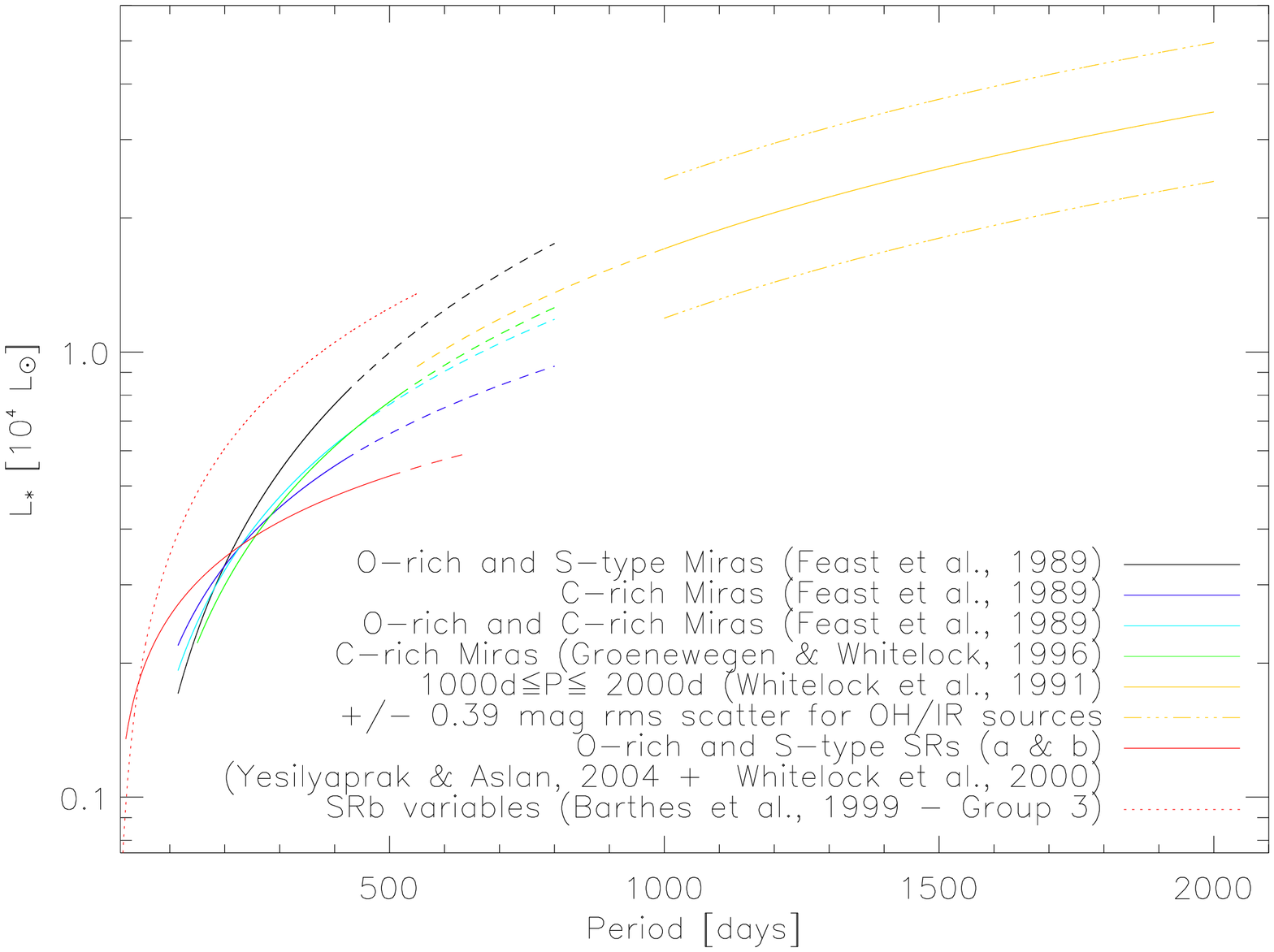}
\caption{$P-$\lstar relations taken from \cite{groenewegen1996,feast1989,whitelock1991,yesilyaprak2004,barthes1999}, and \cite{whitelock2000_bck}. Dashed lines indicate extrapolations of the $P-$\lstar relations from the literature to shorter or longer periods of pulsation. See text for details. \label{fig:plcalibration}} 
\end{figure}

\subsubsection{Miras and \ohir stars}
\cite{feast1989} derived period-luminosity, $P-$\lstar, relations for \orich and \crich Miras in the LMC with periods ranging from 115 up to 420~days. \cite{groenewegen1996} revised the $P-$\lstar relation for \crich Miras and extended the period range up to periods of 520~days. \cite{whitelock1991} developed a similar relation for \ohir targets with periods of pulsation exceeding 1000~days. These relations - and their extrapolations to shorter or longer pulsation periods - are shown in Fig.~\ref{fig:plcalibration}.

The relation derived by \cite{feast1989} for \orich Miras was assumed to be valid also for galactic \orich Miras and was adjusted with a distance modulus of 18.5\,mag. It is given by 
\begin{equation}
 \label{eq:orichPL}
M_{\mathrm{bol,O-MIRA,OH/IR}}=-3.00_{\pm0.24}\log P+2.85_{\pm0.57},
\end{equation}
with $P$ in days. 

\cite{whitelock1991} mention a rather large scatter of 0.39\,mag on their derived $P-$\lstar relation for \ohir variables  with periods ranging up to 2000~days. The large range in predicted luminosities associated with this scatter has been made visible in the figure. Note that the extrapolated relation for \orich Miras derived by \cite{feast1989} matches the higher luminosity edge derived for \ohir targets very well. \cite{whitelock1991} also point out that there are suggestions that the luminosities of \ohir sources with $P>1000$~days probably follow this extrapolated $P-$\lstar relation that was established for \orich Miras by \cite{feast1989}. Because of these arguments and the strong evolutionary connection between \orich Miras and \ohir stars (see Sect.~\ref{subsect:sampledescription}) we adopt the $P-$\lstar relation of \cite{feast1989} for \orich Miras, given in Eq.~\ref{eq:orichPL}, for the determination of luminosities for all \orich Miras and \ohir targets, regardless of their period of pulsation. Relation~\ref{eq:orichPL} was also used in determining the luminosity of S-type Miras in the sample.

Luminosities of \crich Miras were determined via
\begin{equation}
 \label{eq:crichPL}
M_{\mathrm{bol,C-MIRA}}=-2.59\log P+2.02_{(\pm0.26)}
\end{equation}
following \cite{groenewegen1996}. This relation predicts luminosities substantially higher than does the relation derived by \cite{feast1989}. Preference was given to the former since the sample of \cite{groenewegen1996} is almost three times as large as the one used by \cite{feast1989} (54 versus 20 objects). For longer-period \crich Miras, this $P-$\lstar relation was extrapolated, similar to what was done in the \orich case.

\subsubsection{Semi-regular pulsators: SRa \& SRb}
Many semi-regular type a (SRa) objects differ from Miras only by their smaller amplitudes in the $V$ band ($\Delta V < 2.5$\,mag). Objects classified as semi-regular~b (SRb) variables have rather poorly defined periodicity, show a superposition of multiple pulsation periods, or have alternating intervals of periodic and slow irregular changes. 

Semi-regulars and Miras are linked in terms of evolution, since SRs of types a and b are thought to be the progenitors of Miras \citep{whitelock2000,yesilyaprak2004}. Therefore some type of $P-$\lstar relation, analogous to those for Miras is probably justified for these LPVs. 

\cite{yesilyaprak2004} presented period-magnitude relations for all magnitudes in the Johnson system and the IRAS system for SRa and SRb stars. To determine the bolometric magnitude, and hence the luminosity, only the $J$ and $K$ period -- magnitude relations were used. 
\begin{eqnarray}
 \label{eq:srj}
J&=&-1.03_{(\pm0.56)}\log P-3.47_{(\pm0.98)}\\
\label{eq:srk}
K&=&-1.38_{(\pm0.58)}\log P -3.99_{(\pm1.00)}
\end{eqnarray}
\citeauthor{yesilyaprak2004} mention a scatter of 0.63\,mag and 0.85\,mag for the $J$ and $K$ relations, respectively. The bolometric correction on the $K$ magnitude was calculated via 
\begin{eqnarray}
\label{eq:srbck}
BC_K&=&10.86-38.10 (J-K)_0 \\
&&+64.16(J-K)_0^2-50.72(J-K)_0^3 \nonumber\\
&&+19.48(J-K)_0^4-2.94(J-K)_0^5 \nonumber
\end{eqnarray}
taken from \cite{whitelock2000_bck}, where
\begin{eqnarray}
\label{eq:bck}
M_{\mathrm{bol}}&=&K+BC_K. 
\end{eqnarray}

Relations~\ref{eq:srj} and \ref{eq:srk} were however only adopted in estimating the luminosities for SRa variables, and not for the SRb variables in the sample. \cite{barthes1999} presented $P-(J-K)$ relations for long-period variables, including a large number of SRb type stars. The sample of \cite{barthes1999} is divided in four groups, and the SRb type stars are largely split up in two of those  --- \textit{Groups 2} and \textit{3}, corresponding to younger and older kinematic properties respectively. Their \textit{Group 2} is far less luminous than their \textit{Group 3} and preference is given to the relations for their \textit{Group 3} - Eqs.~\ref{eq:mkbarthesgroup3} and \ref{eq:jkbarthesgroup3} - since these estimates are in best accordance with luminosities derived in the literature for the SRb objects in the sample. Bolometric corrections were again derived via Eq.~\ref{eq:srbck}.
\begin{eqnarray}
 \label{eq:mkbarthesgroup3}
K_{\mathrm{SRb}}&=&-2.11_{(\pm0.13)} (\log P-1.75)-6.51_{(\pm0.05)}\\
\label{eq:jkbarthesgroup3}
(J-K)_0&=&0.17_{(\pm0.04)} \log P+0.85_{(\pm0.08)}
\end{eqnarray}

The uncertainties on calculated luminosities only take into account the uncertainties on the coefficients in Eqs.~\ref{eq:orichPL} through \ref{eq:jkbarthesgroup3}. The error bars on the pulsation periods are assumed to be small compared to the errors on the coefficients in the above relations.

\subsubsection{Red supergiants}
The luminosity of the RSGs in the sample was determined as a function of period of pulsation and effective temperature. Eq.~\ref{eq:dambispkRSG} gives the relation between the $K$ magnitude and the pulsation period for RSGs in the milky way galaxy as presented by \cite{dambis1993}. This relation was completed with the bolometric correction on the $K$ band magnitudes given by \cite{levesque2005} (Eq.~\ref{eq:levesquebckRSG}). This method leads to luminosities ranging from $\sim$$10^4$\,\lsun up to $\sim$$5\;10^5$\,\lsun. It is given by
\begin{eqnarray}
\label{eq:dambispkRSG}
K&=&-10.10_{(\pm0.08)}-3.44(\log P-2.7)\\ 
\label{eq:levesquebckRSG}
BC_K&=&5.574-0.7589\times\frac{T_{\mathrm{eff}}}{1000\;\mathrm{K}}
\end{eqnarray}

\subsubsection{Yellow hypergiants}
No general period-luminosity relations can be established for these targets because of their very strong variability. The latter is reflected in the multiple motions from blue to red and vice versa in the HR diagram \citep{dejager1998,oudmaijer2008}. For both AFGL\,2343 and IRC\,+10420, the two YHGs in the sample, luminosities were taken from \cite{dejager1998}, yielding values for $\log L_{\star}/L_{\odot}$ of 5.30 and 5.80, respectively. 

\subsubsection{Alternative methods}
More recently, \cite{barthes1999} and \cite{knapp2003} derived $P-K$ relations for \orich LPVs, including Miras and semi-regulars of types a and b. Using these relations in determining the luminosities of Miras would introduce extra uncertainties, since bolometric corrections for $K$ band magnitudes would have to be derived. This is the main reason why the above mentioned $P-$\mbol relations of \cite{feast1989} and \cite{groenewegen1996} were used for Miras.

\subsection{Distance}\label{sect:distance}
Distances towards the sample stars have been determined in various ways. 
\textit{(1)} OH-maser phase lag distances were used for most \ohir stars, since these provide a very accurate distance estimate \citep{vanlangevelde1990}.
\textit{(2)} Hipparcos-data were used when available with a relative error lower than 50\,\%. 
\textit{(3)} values in the literature derived from SED fitting were used and assumed to have uncertainties of 50\,\%, unless stated otherwise in the original papers.

\subsubsection{Alternative method}
An alternative method to derive distances is to compare observed magnitudes with absolute magnitudes predicted through formalisms as presented in Sect.~\ref{sect:luminosity}. The relations that could be used for distance determination from magnitudes for Miras are
\begin{eqnarray}
\label{eq:kOMira}
K_{\mathrm{O-MIRA}}&=&-3.47_{(\pm0.19)}\log P+0.98_{(\pm0.45)}\\
\label{eq:kCMira}
K_{\mathrm{C-MIRA}}&=&-3.56\log P+1.14
\end{eqnarray}
Since the $K$ magnitude is less sensitive to metallicity than the bolometric magnitude $m_{\rm bol}$, it is a better distance indicator \citep[][and references therein]{whitelock2000}. The strong variability in the $K$ band of most sample stars, however, gives rise to large uncertainties on absolute magnitudes. An additional constraint to the application of this method to determine distances is that the circumstellar material can redden the $K$ band magnitudes significantly. 
\end{appendix}


\begin{thebibliography}{104}
\expandafter\ifx\csname natexlab\endcsname\relax\defNatureexlab#1{#1}\fi

\bibitem[{{Anders} \& {Grevesse}(1989)}]{anders1989}
{Anders}, E. \& {Grevesse}, N. 1989, \gca, 53, 197

\bibitem[{{Barth{\`e}s} {et~al.}(1999){Barth{\`e}s}, {Luri}, {Alvarez}, \&
  {Mennessier}}]{barthes1999}
{Barth{\`e}s}, D., {Luri}, X., {Alvarez}, R., \& {Mennessier}, M.~O. 1999,
  A\&AS, 140, 55

\bibitem[{{Bergeat} \& {Chevallier}(2005)}]{bergeat2005}
{Bergeat}, J. \& {Chevallier}, L. 2005, A\&A, 429, 235

\bibitem[{{Bergeat} {et~al.}(2001){Bergeat}, {Knapik}, \&
  {Rutily}}]{bergeat2001}
{Bergeat}, J., {Knapik}, A., \& {Rutily}, B. 2001, A\&A, 369, 178

\bibitem[{{Bessell}(2005)}]{bessell2005}
{Bessell}, M.~S. 2005, \araa, 43, 293

\bibitem[{{Bessell} {et~al.}(1998){Bessell}, {Castelli}, \&
  {Plez}}]{bessell1998}
{Bessell}, M.~S., {Castelli}, F., \& {Plez}, B. 1998, A\&A, 333, 231

\bibitem[{{Bieging} \& {Latter}(1994)}]{bieging1994}
{Bieging}, J.~H. \& {Latter}, W.~B. 1994, ApJ, 422, 765

\bibitem[{{Blackwell} \& {Shallis}(1977)}]{blackwell1977}
{Blackwell}, D.~E. \& {Shallis}, M.~J. 1977, \mnras, 180, 177

\bibitem[{{Bujarrabal} {et~al.}(2008){Bujarrabal}, {Young}, \&
  {Fong}}]{bujarrabal2008}
{Bujarrabal}, V., {Young}, K., \& {Fong}, D. 2008, A\&A, 483, 839

\bibitem[{{Castro-Carrizo} {et~al.}(2007){Castro-Carrizo}, {Quintana-Lacaci},
  {Bujarrabal}, {Neri}, \& {Alcolea}}]{castro-carrizo2007}
{Castro-Carrizo}, A., {Quintana-Lacaci}, G., {Bujarrabal}, V., {Neri}, R., \&
  {Alcolea}, J. 2007, A\&A, 465, 457

\bibitem[{{Dambis}(1993)}]{dambis1993}
{Dambis}, A.~K. 1993, Astronomy Letters, 19, 172

\bibitem[{{de Jager}(1998)}]{dejager1998}
{de Jager}, C. 1998, A\&Ar, 8, 145

\bibitem[{{Decin} {et~al.}(2008{Natureexlab{a}}){Decin}, {Blomme}, {Reyniers},
  {Ryde}, {Hinkle}, \& {de Koter}}]{decin2008_rhya}
{Decin}, L., {Blomme}, L., {Reyniers}, M., {et~al.} 2008{Natureexlab{a}}, A\&A,
  484, 401

\bibitem[{{Decin} {et~al.}(2008{Natureexlab{b}}){Decin}, {Cherchneff}, {Hony},
  {Dehaes}, {De Breuck}, \& {Menten}}]{decin2008_parent}
{Decin}, L., {Cherchneff}, I., {Hony}, S., {et~al.} 2008{Natureexlab{b}}, A\&A,
  480, 431

\bibitem[{{Decin} {et~al.}(2006){Decin}, {Hony}, {de Koter}, {Justtanont},
  {Tielens}, \& {Waters}}]{decin2006_vycma}
{Decin}, L., {Hony}, S., {de Koter}, A., {et~al.} 2006, A\&A, 456, 549

\bibitem[{{Decin} {et~al.}(2007){Decin}, {Hony}, {de Koter}, {Molenberghs},
  {Dehaes}, \& {Markwick-Kemper}}]{decin2007_wxpsc}
---. 2007, A\&A, 475, 233

\bibitem[{{Delfosse} {et~al.}(1997){Delfosse}, {Kahane}, \&
  {Forveille}}]{delfosse1997}
{Delfosse}, X., {Kahane}, C., \& {Forveille}, T. 1997, A\&A, 320, 249

\bibitem[{{di Benedetto}(1993)}]{dibenedetto1993}
{di Benedetto}, G.~P. 1993, A\&A, 270, 315

\bibitem[{{Dinh-V.-Trung} {et~al.}(2009){Dinh-V.-Trung}, {Muller}, {Lim},
  {Kwok}, \& {Muthu}}]{dinh-v-trung2009}
{Dinh-V.-Trung}, {Muller}, S., {Lim}, J., {Kwok}, S., \& {Muthu}, C. 2009,
  ApJ, 697, 409

\bibitem[{{Fa{\'u}ndez} {et~al.}(2004){Fa{\'u}ndez}, {Bronfman}, {Garay},
  {Chini}, {Nyman}, \& {May}}]{faundez2004}
{Fa{\'u}ndez}, S., {Bronfman}, L., {Garay}, G., {et~al.} 2004, A\&A, 426, 97

\bibitem[{{Faure} \& {Josselin}(2008)}]{faure2008}
{Faure}, A. \& {Josselin}, E. 2008, A\&A, 492, 257

\bibitem[{{Feast} {et~al.}(1989){Feast}, {Glass}, {Whitelock}, \&
  {Catchpole}}]{feast1989}
{Feast}, M.~W., {Glass}, I.~S., {Whitelock}, P.~A., \& {Catchpole}, R.~M. 1989,
  \mnras, 241, 375

\bibitem[{{Feast} \& {Whitelock}(2000)}]{feast2000}
{Feast}, M.~W. \& {Whitelock}, P.~A. 2000, \mnras, 317, 460

\bibitem[{{Feast} {et~al.}(1990){Feast}, {Whitelock}, \& {Carter}}]{feast1990}
{Feast}, M.~W., {Whitelock}, P.~A., \& {Carter}, B.~S. 1990, \mnras, 247, 227

\bibitem[{{Garc{\'{\i}}a-Hern{\'a}ndez}
  {et~al.}(2007){Garc{\'{\i}}a-Hern{\'a}ndez}, {Garc{\'{\i}}a-Lario}, {Plez},
  {Manchado}, {D'Antona}, {Lub}, \& {Habing}}]{garciahernandez2007}
{Garc{\'{\i}}a-Hern{\'a}ndez}, D.~A., {Garc{\'{\i}}a-Lario}, P., {Plez}, B.,
  {et~al.} 2007, A\&A, 462, 711

\bibitem[{{Green} {et~al.}(1993){Green}, {Maluendes}, \& {McLean}}]{green1993}
{Green}, S., {Maluendes}, S., \& {McLean}, A.~D. 1993, ApJS, 85, 181

\bibitem[{{Groenewegen} {et~al.}(1999){Groenewegen}, {Baas}, {Blommaert},
  {Stehle}, {Josselin}, \& {Tilanus}}]{groenewegen1999}
{Groenewegen}, M.~A.~T., {Baas}, F., {Blommaert}, J.~A.~D.~L., {et~al.} 1999,
  A\&AS, 140, 197

\bibitem[{{Groenewegen} {et~al.}(1998){Groenewegen}, {van der Veen}, \&
  {Matthews}}]{groenewegen1998}
{Groenewegen}, M.~A.~T., {van der Veen}, W.~E.~C.~J., \& {Matthews}, H.~E.
  1998, A\&A, 338, 491

\bibitem[{{Groenewegen} \& {Whitelock}(1996)}]{groenewegen1996}
{Groenewegen}, M.~A.~T. \& {Whitelock}, P.~A. 1996, \mnras, 281, 1347

\bibitem[{{Habing} \& {Olofsson}(2003)}]{habing2003}
{Habing}, H.~J. \& {Olofsson}, H., eds. 2003, {Asymptotic giant branch stars}

\bibitem[{{Haniff} {et~al.}(1995){Haniff}, {Scholz}, \& {Tuthill}}]{haniff1995}
{Haniff}, C.~A., {Scholz}, M., \& {Tuthill}, P.~G. 1995, \mnras, 276, 640

\bibitem[{{He} \& {Chen}(2001)}]{he2001}
{He}, J.~H. \& {Chen}, P.~S. 2001, \aj, 121, 2752

\bibitem[{{Huggins} {et~al.}(1988){Huggins}, {Olofsson}, \&
  {Johansson}}]{huggins1988}
{Huggins}, P.~J., {Olofsson}, H., \& {Johansson}, L.~E.~B. 1988, ApJ, 332,
  1009

\bibitem[{{Humphreys} {et~al.}(1997){Humphreys}, {Smith}, {Davidson}, {Jones},
  {Gehrz}, {Mason}, {Hayward}, {Houck}, \& {Krautter}}]{humphreys1997}
{Humphreys}, R.~M., {Smith}, N., {Davidson}, K., {et~al.} 1997, \aj, 114, 2778

\bibitem[{{Iben} \& {Renzini}(1983)}]{iben1983}
{Iben}, Jr., I. \& {Renzini}, A. 1983, \araa, 21, 271

\bibitem[{{Josselin} \& {Plez}(2007)}]{josselin2007}
{Josselin}, E. \& {Plez}, B. 2007, A\&A, 469, 671

\bibitem[{{Justtanont} {et~al.}(1996){Justtanont}, {de Jong}, {Helmich},
  {Waters}, {de Graauw}, {Loup}, {Izumiura}, {Yamamura}, {Beintema}, {Lahuis},
  {Roelfsema}, \& {Valentijn}}]{justtanont1996}
{Justtanont}, K., {de Jong}, T., {Helmich}, F.~P., {et~al.} 1996, A\&A, 315,
  L217

\bibitem[{{Justtanont} {et~al.}(1994){Justtanont}, {Skinner}, \&
  {Tielens}}]{justtanont1994}
{Justtanont}, K., {Skinner}, C.~J., \& {Tielens}, A.~G.~G.~M. 1994, ApJ, 435,
  852

\bibitem[{{Kahane} \& {Jura}(1994)}]{kahane1994}
{Kahane}, C. \& {Jura}, M. 1994, A\&A, 290, 183

\bibitem[{{Kemper} {et~al.}(2003){Kemper}, {Stark}, {Justtanont}, {de Koter},
  {Tielens}, {Waters}, {Cami}, \& {Dijkstra}}]{kemper2003}
{Kemper}, F., {Stark}, R., {Justtanont}, K., {et~al.} 2003, A\&A, 407, 609

\bibitem[{{Kerschbaum} \& {Olofsson}(1999)}]{kerschbaum1999}
{Kerschbaum}, F. \& {Olofsson}, H. 1999, A\&AS, 138, 299

\bibitem[{{Knapp} \& {Morris}(1985)}]{knapp1985}
{Knapp}, G.~R. \& {Morris}, M. 1985, ApJ, 292, 640

\bibitem[{{Knapp} {et~al.}(1982){Knapp}, {Phillips}, {Leighton}, {Lo},
  {Wannier}, {Wootten}, \& {Huggins}}]{knapp1982}
{Knapp}, G.~R., {Phillips}, T.~G., {Leighton}, R.~B., {et~al.} 1982, ApJ, 252,
  616

\bibitem[{{Knapp} {et~al.}(2003){Knapp}, {Pourbaix}, {Platais}, \&
  {Jorissen}}]{knapp2003}
{Knapp}, G.~R., {Pourbaix}, D., {Platais}, I., \& {Jorissen}, A. 2003, A\&A,
  403, 993

\bibitem[{{Knapp} {et~al.}(1998){Knapp}, {Young}, {Lee}, \&
  {Jorissen}}]{knapp1998}
{Knapp}, G.~R., {Young}, K., {Lee}, E., \& {Jorissen}, A. 1998, ApJS, 117, 209

\bibitem[{{Kwan} \& {Linke}(1982)}]{kwan1982}
{Kwan}, J. \& {Linke}, R.~A. 1982, ApJ, 254, 587

\bibitem[{{Kwok} {et~al.}(1997){Kwok}, {Volk}, \& {Bidelman}}]{kwok1997}
{Kwok}, S., {Volk}, K., \& {Bidelman}, W.~P. 1997, ApJS, 112, 557

\bibitem[{Lasdon {et~al.}(1978)Lasdon, Waren, Jain, \& Ratner}]{lasdon1978}
Lasdon, L.~S., Waren, A.~D., Jain, A., \& Ratner, M. 1978, ACM Trans. Math.
  Softw., 4, 34

\bibitem[{{Lebzelter} {et~al.}(2008){Lebzelter}, {Lederer}, {Cristallo},
  {Hinkle}, {Straniero}, \& {Aringer}}]{lebzelter2008}
{Lebzelter}, T., {Lederer}, M.~T., {Cristallo}, S., {et~al.} 2008, A\&A, 486,
  511

\bibitem[{{Levesque} {et~al.}(2007){Levesque}, {Massey}, {Olsen}, \&
  {Plez}}]{levesque2007}
{Levesque}, E.~M., {Massey}, P., {Olsen}, K.~A.~G., \& {Plez}, B. 2007, ApJ,
  667, 202

\bibitem[{{Levesque} {et~al.}(2005){Levesque}, {Massey}, {Olsen}, {Plez},
  {Josselin}, {Maeder}, \& {Meynet}}]{levesque2005}
{Levesque}, E.~M., {Massey}, P., {Olsen}, K.~A.~G., {et~al.} 2005, ApJ, 628,
  973

\bibitem[{{Likkel} {et~al.}(1991){Likkel}, {Forveille}, {Omont}, \&
  {Morris}}]{likkel1991}
{Likkel}, L., {Forveille}, T., {Omont}, A., \& {Morris}, M. 1991, A\&A, 246,
  153

\bibitem[{{Loup} {et~al.}(1993){Loup}, {Forveille}, {Omont}, \&
  {Paul}}]{loup1993}
{Loup}, C., {Forveille}, T., {Omont}, A., \& {Paul}, J.~F. 1993, A\&AS, 99, 291

\bibitem[{{Margulis} {et~al.}(1990){Margulis}, {van Blerkom}, {Snell}, \&
  {Kleinmann}}]{margulis1990}
{Margulis}, M., {van Blerkom}, D.~J., {Snell}, R.~L., \& {Kleinmann}, S.~G.
  1990, ApJ, 361, 673

\bibitem[{{Marigo}(2002)}]{marigo2002}
{Marigo}, P. 2002, A\&A, 387, 507

\bibitem[{{Marigo} {et~al.}(2008){Marigo}, {Girardi}, {Bressan}, {Groenewegen},
  {Silva}, \& {Granato}}]{marigo2008}
{Marigo}, P., {Girardi}, L., {Bressan}, A., {et~al.} 2008, A\&A, 482, 883

\bibitem[{{Massey} {et~al.}(2006){Massey}, {Levesque}, \& {Plez}}]{massey2006}
{Massey}, P., {Levesque}, E.~M., \& {Plez}, B. 2006, ApJ, 646, 1203

\bibitem[{{Massey} {et~al.}(2008){Massey}, {Levesque}, {Plez}, \&
  {Olsen}}]{massey2008}
{Massey}, P., {Levesque}, E.~M., {Plez}, B., \& {Olsen}, K.~A.~G. 2008, in IAU
  Symposium, Vol. 250, IAU Symposium, 97--110

\bibitem[{{Matsuura} {et~al.}(2002){Matsuura}, {Yamamura}, {Zijlstra}, \&
  {Bedding}}]{matsuura2002}
{Matsuura}, M., {Yamamura}, I., {Zijlstra}, A.~A., \& {Bedding}, T.~R. 2002,
  A\&A, 387, 1022

\bibitem[{{Mauron} \& {Huggins}(2000)}]{mauron2000}
{Mauron}, N. \& {Huggins}, P.~J. 2000, A\&A, 359, 707

\bibitem[{{Mauron} \& {Huggins}(2006)}]{mauron2006}
---. 2006, A\&A, 452, 257

\bibitem[{{Milam} {et~al.}(2009){Milam}, {Woolf}, \& {Ziurys}}]{milam2009}
{Milam}, S.~N., {Woolf}, N.~J., \& {Ziurys}, L.~M. 2009, ApJ, 690, 837

\bibitem[{{Neri} {et~al.}(1998){Neri}, {Kahane}, {Lucas}, {Bujarrabal}, \&
  {Loup}}]{neri1998}
{Neri}, R., {Kahane}, C., {Lucas}, R., {Bujarrabal}, V., \& {Loup}, C. 1998,
  A\&AS, 130, 1

\bibitem[{{Neufeld} \& {Kaufman}(1993)}]{neufeld1993}
{Neufeld}, D.~A. \& {Kaufman}, M.~J. 1993, ApJ, 418, 263

\bibitem[{{Nyman} {et~al.}(1992){Nyman}, {Booth}, {Carlstrom}, {Habing},
  {Heske}, {Sahai}, {Stark}, {van der Veen}, \& {Winnberg}}]{nyman1992}
{Nyman}, L.-A., {Booth}, R.~S., {Carlstrom}, U., {et~al.} 1992, A\&AS, 93, 121

\bibitem[{{Ohnaka} \& {Tsuji}(1999)}]{ohnaka1999}
{Ohnaka}, K. \& {Tsuji}, T. 1999, A\&A, 345, 233

\bibitem[{{Olivier} {et~al.}(2001){Olivier}, {Whitelock}, \&
  {Marang}}]{olivier2001}
{Olivier}, E.~A., {Whitelock}, P., \& {Marang}, F. 2001, \mnras, 326, 490

\bibitem[{{Olofsson} {et~al.}(1996){Olofsson}, {Bergman}, {Eriksson}, \&
  {Gustafsson}}]{olofsson1996}
{Olofsson}, H., {Bergman}, P., {Eriksson}, K., \& {Gustafsson}, B. 1996, A\&A,
  311, 587

\bibitem[{{Olofsson} {et~al.}(1993){Olofsson}, {Eriksson}, {Gustafsson}, \&
  {Carlstrom}}]{olofsson1993}
{Olofsson}, H., {Eriksson}, K., {Gustafsson}, B., \& {Carlstrom}, U. 1993,
  ApJS, 87, 267

\bibitem[{{Oudmaijer} {et~al.}(2008){Oudmaijer}, {Davies}, {de Wit}, \&
  {Patel}}]{oudmaijer2008}
{Oudmaijer}, R., {Davies}, B., {de Wit}, W.-J., \& {Patel}, M. 2008, ArXiv
  e-prints

\bibitem[{{Ramstedt} {et~al.}(2006){Ramstedt}, {Sch{\"o}ier}, {Olofsson}, \&
  {Lundgren}}]{ramstedt2006}
{Ramstedt}, S., {Sch{\"o}ier}, F.~L., {Olofsson}, H., \& {Lundgren}, A.~A.
  2006, A\&A, 454, L103

\bibitem[{{Ramstedt} {et~al.}(2008){Ramstedt}, {Sch{\"o}ier}, {Olofsson}, \&
  {Lundgren}}]{ramstedt2008}
---. 2008, A\&A, 487, 645

\bibitem[{{Sahai} {et~al.}(2003){Sahai}, {Morris}, {Knapp}, {Young}, \&
  {Barnbaum}}]{sahai2003}
{Sahai}, R., {Morris}, M., {Knapp}, G.~R., {Young}, K., \& {Barnbaum}, C. 2003,
  Nature, 426, 261

\bibitem[{{Samus} {et~al.}(2004){Samus}, {Durlevich}, \& {et al.}}]{gcvs}
{Samus}, N.~N., {Durlevich}, O.~V., \& {et al.} 2004, VizieR Online Data
  Catalog, 2250, 0

\bibitem[{{Schlegel} {et~al.}(1998){Schlegel}, {Finkbeiner}, \&
  {Davis}}]{schlegel1998}
{Schlegel}, D.~J., {Finkbeiner}, D.~P., \& {Davis}, M. 1998, ApJ, 500, 525

\bibitem[{{Sch{\"o}ier} {et~al.}(2007){Sch{\"o}ier}, {Bast}, {Olofsson}, \&
  {Lindqvist}}]{schoeier2007}
{Sch{\"o}ier}, F.~L., {Bast}, J., {Olofsson}, H., \& {Lindqvist}, M. 2007,
  A\&A, 473, 871

\bibitem[{{Sch{\"o}ier} {et~al.}(2005){Sch{\"o}ier}, {Lindqvist}, \&
  {Olofsson}}]{schoeier2005}
{Sch{\"o}ier}, F.~L., {Lindqvist}, M., \& {Olofsson}, H. 2005, A\&A, 436, 633

\bibitem[{{Sch{\"o}ier} \& {Olofsson}(2000)}]{schoeier2000}
{Sch{\"o}ier}, F.~L. \& {Olofsson}, H. 2000, A\&A, 359, 586

\bibitem[{{Sch{\"o}ier} \& {Olofsson}(2001)}]{schoeier2001}
---. 2001, A\&A, 368, 969

\bibitem[{{Sch{\"o}nberg} \& {Hempe}(1986)}]{schoenberg1986}
{Sch{\"o}nberg}, K. \& {Hempe}, K. 1986, A\&A, 163, 151

\bibitem[{{Skinner} {et~al.}(1999){Skinner}, {Justtanont}, {Tielens}, {Betz},
  {Boreiko}, \& {Baas}}]{skinner1999}
{Skinner}, C.~J., {Justtanont}, K., {Tielens}, A.~G.~G.~M., {et~al.} 1999,
  \mnras, 302, 293

\bibitem[{{Sloan} \& {Price}(1998)}]{sloan1998}
{Sloan}, G.~C. \& {Price}, S.~D. 1998, ApJS, 119, 141

\bibitem[{{Slootmaker} {et~al.}(1985){Slootmaker}, {Habing}, \&
  {Herman}}]{slootmaker1985}
{Slootmaker}, A., {Habing}, H.~J., \& {Herman}, J. 1985, A\&AS, 59, 465

\bibitem[{{Sopka} {et~al.}(1989){Sopka}, {Olofsson}, {Johansson}, {Nguyen}, \&
  {Zuckerman}}]{sopka1989}
{Sopka}, R.~J., {Olofsson}, H., {Johansson}, L.~E.~B., {Nguyen}, Q.-R., \&
  {Zuckerman}, B. 1989, A\&A, 210, 78

\bibitem[{{Speck} {et~al.}(2000){Speck}, {Barlow}, {Sylvester}, \&
  {Hofmeister}}]{speck2000}
{Speck}, A.~K., {Barlow}, M.~J., {Sylvester}, R.~J., \& {Hofmeister}, A.~M.
  2000, A\&AS, 146, 437

\bibitem[{{Stanek} {et~al.}(1995){Stanek}, {Knapp}, {Young}, \&
  {Phillips}}]{stanek1995}
{Stanek}, K.~Z., {Knapp}, G.~R., {Young}, K., \& {Phillips}, T.~G. 1995, ApJS,
  100, 169

\bibitem[{{Su{\'a}rez} {et~al.}(2006){Su{\'a}rez}, {Garc{\'{\i}}a-Lario},
  {Manchado}, {Manteiga}, {Ulla}, \& {Pottasch}}]{suarez2006}
{Su{\'a}rez}, O., {Garc{\'{\i}}a-Lario}, P., {Manchado}, A., {et~al.} 2006,
  A\&A, 458, 173

\bibitem[{{Teyssier} {et~al.}(2006){Teyssier}, {Hernandez}, {Bujarrabal},
  {Yoshida}, \& {Phillips}}]{teyssier2006}
{Teyssier}, D., {Hernandez}, R., {Bujarrabal}, V., {Yoshida}, H., \&
  {Phillips}, T.~G. 2006, A\&A, 450, 167

\bibitem[{{van Langevelde} {et~al.}(1990){van Langevelde}, {van der Heiden}, \&
  {van Schooneveld}}]{vanlangevelde1990}
{van Langevelde}, H.~J., {van der Heiden}, R., \& {van Schooneveld}, C. 1990,
  A\&A, 239, 193

\bibitem[{{van Zadelhoff} {et~al.}(2002){van Zadelhoff}, {Dullemond}, {van der
  Tak}, {Yates}, {Doty}, {Ossenkopf}, {Hogerheijde}, {Juvela}, {Wiesemeyer}, \&
  {Sch{\"o}ier}}]{vanzadelhoff2002}
{van Zadelhoff}, G., {Dullemond}, C.~P., {van der Tak}, F.~F.~S., {et~al.}
  2002, A\&A, 395, 373

\bibitem[{{Vassiliadis} \& {Wood}(1993)}]{vassiliadis1993}
{Vassiliadis}, E. \& {Wood}, P.~R. 1993, ApJ, 413, 641

\bibitem[{{Verhoelst} {et~al.}(2009){Verhoelst}, {van der Zypen}, {Hony},
  {Decin}, {Cami}, \& {Eriksson}}]{verhoelst2009}
{Verhoelst}, T., {van der Zypen}, N., {Hony}, S., {et~al.} 2009, A\&A, 498, 127

\bibitem[{{Wang} {et~al.}(1994){Wang}, {Jaffe}, {Graf}, \& {Evans}}]{wang1994}
{Wang}, Y., {Jaffe}, D.~T., {Graf}, U.~U., \& {Evans}, II, N.~J. 1994, ApJS,
  95, 503

\bibitem[{{Wannier} \& {Sahai}(1986)}]{wannier1986}
{Wannier}, P.~G. \& {Sahai}, R. 1986, ApJ, 311, 335

\bibitem[{{Whitelock} \& {Feast}(2000)}]{whitelock2000}
{Whitelock}, P. \& {Feast}, M. 2000, \mnras, 319, 759

\bibitem[{{Whitelock} {et~al.}(1991){Whitelock}, {Feast}, \&
  {Catchpole}}]{whitelock1991}
{Whitelock}, P., {Feast}, M., \& {Catchpole}, R. 1991, \mnras, 248, 276

\bibitem[{{Whitelock} {et~al.}(2000){Whitelock}, {Marang}, \&
  {Feast}}]{whitelock2000_bck}
{Whitelock}, P., {Marang}, F., \& {Feast}, M. 2000, \mnras, 319, 728

\bibitem[{{Whitelock} \& {Catchpole}(1985)}]{whitelock1985}
{Whitelock}, P.~A. \& {Catchpole}, R.~M. 1985, \mnras, 212, 873

\bibitem[{{Winters} {et~al.}(2000){Winters}, {Keady}, {Gauger}, \&
  {Sada}}]{winters2000}
{Winters}, J.~M., {Keady}, J.~J., {Gauger}, A., \& {Sada}, P.~V. 2000, A\&A,
  359, 651

\bibitem[{{Winters} {et~al.}(2007){Winters}, {Le Bertre}, {Pety}, \&
  {Neri}}]{winters2007}
{Winters}, J.~M., {Le Bertre}, T., {Pety}, J., \& {Neri}, R. 2007, A\&A, 475,
  559

\bibitem[{{Wood} \& {Zarro}(1981)}]{wood1981}
{Wood}, P.~R. \& {Zarro}, D.~M. 1981, ApJ, 247, 247

\bibitem[{{Ye{\c s}ilyaprak} \& {Aslan}(2004)}]{yesilyaprak2004}
{Ye{\c s}ilyaprak}, C. \& {Aslan}, Z. 2004, \mnras, 355, 601

\bibitem[{{Ziurys} {et~al.}(2007){Ziurys}, {Milam}, {Apponi}, \&
  {Woolf}}]{ziurys2007}
{Ziurys}, L.~M., {Milam}, S.~N., {Apponi}, A.~J., \& {Woolf}, N.~J. 2007, Nature,
  447, 1094

\bibitem[{{Zuckerman} \& {Dyck}(1986)}]{zuckerman1986}
{Zuckerman}, B. \& {Dyck}, H.~M. 1986, ApJ, 304, 394

\end{thebibliography}
\end{document}